\PassOptionsToPackage{unicode=true}{hyperref} 
\PassOptionsToPackage{hyphens}{url}
\documentclass[12pt,a4paper,]{article}
\usepackage{lmodern}
\usepackage{amssymb,amsmath}
\usepackage{ifxetex,ifluatex}
\usepackage{fixltx2e} 
\ifnum 0\ifxetex 1\fi\ifluatex 1\fi=0 
  \usepackage[T1]{fontenc}
  \usepackage[utf8]{inputenc}
  \usepackage{textcomp} 
\else 
  \usepackage{unicode-math}
  \defaultfontfeatures{Ligatures=TeX,Scale=MatchLowercase}
\fi
\IfFileExists{upquote.sty}{\usepackage{upquote}}{}
\IfFileExists{microtype.sty}{%
\usepackage[]{microtype}
\UseMicrotypeSet[protrusion]{basicmath} 
}{}
\IfFileExists{parskip.sty}{%
\usepackage{parskip}
}{
\setlength{\parindent}{0pt}
\setlength{\parskip}{6pt plus 2pt minus 1pt}
}
\usepackage{hyperref}
\hypersetup{
            pdftitle={Judging a book by its cover: how much of REF `research quality' is really `journal prestige'?},
            pdfauthor={David Antony Selby; David Firth},
            pdfborder={0 0 0},
            breaklinks=true}
\usepackage[margin=1in]{geometry}
\usepackage{longtable,booktabs}
\IfFileExists{footnote.sty}{\usepackage{footnote}\makesavenoteenv{longtable}}{}
\usepackage{graphicx,grffile}
\makeatletter
\def\maxwidth{\ifdim\Gin@nat@width>\linewidth\linewidth\else\Gin@nat@width\fi}
\def\maxheight{\ifdim\Gin@nat@height>\textheight\textheight\else\Gin@nat@height\fi}
\makeatother
\setkeys{Gin}{width=\maxwidth,height=\maxheight,keepaspectratio}
\providecommand{\tightlist}{%
  \setlength{\itemsep}{0pt}\setlength{\parskip}{0pt}}
\ifx\paragraph\undefined\else
\let\oldparagraph\paragraph
\renewcommand{\paragraph}[1]{\oldparagraph{#1}\mbox{}}
\fi
\ifx\subparagraph\undefined\else
\let\oldsubparagraph\subparagraph
\renewcommand{\subparagraph}[1]{\oldsubparagraph{#1}\mbox{}}
\fi

\makeatletter
\def\fps@figure{htbp}
\makeatother

\usepackage{subfig}
\usepackage[color=yellow]{todonotes}
\usepackage{booktabs}
\usepackage{longtable}
\usepackage{array}
\usepackage{multirow}
\usepackage{wrapfig}
\usepackage{float}
\usepackage{colortbl}
\usepackage{pdflscape}
\usepackage{tabu}
\usepackage{threeparttable}
\usepackage{threeparttablex}
\usepackage[normalem]{ulem}
\usepackage{makecell}
\usepackage{xcolor}

\title{Judging a book by its cover: how much of REF `research quality' is really `journal prestige'?}
\author{David Antony Selby \and David Firth}
\date{July 2020}

\begin{document}
\maketitle
\begin{abstract}
\noindent The Research Excellence Framework (REF) is a periodic UK-wide assessment of the quality of published research in universities.
The most recent REF was in 2014, and the next will be in 2021.
The published results of REF2014 include a categorical `quality profile' for each unit of assessment (typically a university department), reporting what percentage of the unit's REF-submitted research outputs were assessed as being at each of four quality levels (labelled 4*, 3*, 2* and 1*).
Also in the public domain are the original submissions made to REF2014, which include --- for each unit of assessment --- publication details of the REF-submitted research outputs.

In this work, we address the question: to what extent can a REF quality profile for research outputs be attributed to the journals in which (most of) those outputs were published?

The data are the published submissions and results from REF2014.
The main statistical challenge comes from the fact that REF quality profiles are available only at the aggregated level of whole units of assessment: the REF panel's assessment of each individual research output is not made public.
Our research question is thus an `ecological inference' problem, which demands special care in model formulation and methodology.
The analysis is based on logit models in which journal-specific parameters are regularized via prior `pseudo-data'.
We develop a lack-of-fit measure for the extent to which REF scores appear to depend on publication venues rather than research quality or institution-level differences.
Results are presented for several research fields.
\end{abstract}

\hypertarget{introduction}{%
\section{Introduction}\label{introduction}}

The Research Excellence Framework (REF; successor to the Research Assessment Exercise, or RAE) is the method used by UK funding bodies to evaluate the quality of research.
The last REF took place in 2014 and the next one is currently scheduled for 2021.
Panels of experts rate universities and research institutions in three categories: impact outside academia, research environment and quality of outputs, based on written submissions.
In the sciences and some other fields, submissions are more likely to comprise academic journal articles than books or reports (Wilsdon et al. 2015; Marques et al. 2017).

Expert panels can judge a submission to be `world-leading' (4*) `internationally excellent' (3*), `recognised internationally' (2*), `recognised nationally' (1*) or unclassified.
Results \href{https://www.ref.ac.uk/2014}{published online} describe, for each subject area (`unit of assessment') the proportion of each institution's outputs that were assigned to each of these categories.

Though it is publicly known which works were submitted for assessment, the ratings are only published in aggregate, by institution and subject: it is not disclosed which rating was assigned to which paper.
Thus, it is not obvious what constitutes a `4* paper' or which authors wrote them.
However, rumours have long circulated about lists of `4* journals' that peer review panels might use to help them determine the quality of articles (Oswald 2007).

Given that the purposes of the REF are explicitly `{[}to{]} establish reputational yardsticks' (i.e.~rank academic departments) and `to inform the selective allocation of funding for research' (REF web site 2019), it is not surprising that it has had an effect on institutional behaviour, allegedly increasing the number of staff hired on short-term contracts that coincide with the assessment period (Jump 2013), changing the way departments submit members of staff and publications for evaluation (Marques et al. 2017) and increasing productivity just before the deadline (Groen-Xu et al. 2017).

The popularity of journal-level citation metrics such as the impact factor raises the question: might some expert panels be influenced (consciously or otherwise) by a journal's reputation or citation count when judging an individual paper?

In this paper, we investigate the extent to which research institutions' REF ratings (for outputs) might be attributed to the journals in which their outputs were published.
Paper-level ratings are missing, but the margins---institutions' REF profiles and the numbers of articles they submitted from each journal---are known, so the research question becomes an `ecological inference' problem.
Using both frequentist and Bayesian approaches, we will estimate latent `quality' scores for journals, and then quantify the variation in REF results that is explained by these scores.
We also compare these scores with published journal citation metrics.

Initially, we demonstrate the methodology on the field of economics, a relatively small and well-defined discipline, which mostly publishes its outputs in academic journals and has a well-established `Top Five' journals that act as a baseline.
Results will then be compared with several other, larger academic fields whose REF-submitted outputs are also mostly in the form of journal articles.

\hypertarget{background}{%
\section{Background}\label{background}}

\hypertarget{modelling-research-assessments}{%
\subsection{Modelling research assessments}\label{modelling-research-assessments}}

Koya and Chowdhury (2017) suggested that there is, for some subject areas, a correlation between journal rankings and REF performance.
Their approach computed a `monetary value' (funding allocation) for each research output as rated in the REF, using a similar method to that described in an earlier blog post by Reed and Kerridge (2017).

Let \(F\) be the total amount of funding awarded to an institution based on the REF,
let \(n_3\) and \(n_4\) be the number of 3* and 4* outputs and
let \(x_3\) and \(x_4\) be the respective monetary value of an output with each rating.
According to the then Higher Education Funding Council for England (HEFCE), a 4* output is worth four times as much as a 3* output (Else 2015), so \(x_4 = 4x_3\).
The numbers of outputs are known.
Then \(F = n_3 x_3 + n_4 x_4 = n_3 x_3 + 4 n_4 x_3,\) from which we obtain
\[x_3 = \frac{F}{n_3 + 4n_4},\]
for a given institution and subject area.

Consider the example of general engineering at the University of Cambridge.
It was awarded \(F =\) £5,328,295 in 2015--16 as a result of its outputs submitted to the 2014 REF\footnote{according to the HEFCE 2015--16 funding allocation tables for research}.
Of the submitted outputs, 37.4\% were rated 4* and 55.8\% at 3*, for 177.2 full-time equivalent staff.
Each staff member was allowed up to four submissions, and funding was allocated assuming that staff submitted this maximum, even if they did not.
So the theoretical (not actual) number of outputs was \(4 \times 177.2 = 708.8\).
Thus \(n_3 = 708.8 \times 55.8\% = 395.5104\) and \(n_4 = 708.8 \times 37.4\% = 265.0912\), from which we obtain \(x_3 = \pounds 5328295/1455.875 =\) £3,659.86 and \(x_4 =\) £14,639.43.

From here, Koya and Chowdhury (2017) studied the relationship between the distribution of an institution's REF scores with the venues in which the outputs were published.
REF results do not reveal which article/submission received each rating; the data are only published in aggregate.
Koya and Chowdhury (2017) ``identified how many of the submitted articles were in top quartile \emph{{[}sic{]}} journals'' based on impact factors published in the 2013 edition of Thomson Reuters'\footnote{Now operated by Clarivate Analytics} \emph{Journal Citation Reports}, and compared this proportion with the percentages of articles awarded 4* and 3* ratings\footnote{As pointed out by Hill (2017), this means only a subset of journal articles are being compared with the full range of outputs submitted to the REF---not a like-for-like comparison.}.
Positive correlations, where found, were weak and only present in some subject areas.
Surprisingly, Koya and Chowdhury (2017) did not directly compare the computed `monetary value' of research outputs with the corresponding bibliometric indicators for each institution.

Wilsdon et al. (2015 Section~9.1) commissioned HEFCE to perform a more detailed study of the relationship between bibliometric indicators and REF scores, with privileged access to ratings at the level of the individual outputs.
That analysis found low (\(<0.5\)) positive correlations between citation metrics and 4* outputs, but with stronger relationships for some fields such as medicine, biology, chemistry, physics and economics.
The strongest predictors were full-text clicks (on Scopus), number of authors, citation count (according to Google Scholar), SJR (a Scopus-published journal metric based on PageRank score), source-normalized impact per paper (a another Scopus metric, similar to a weighted impact factor), tweets, and downloads from the web site \emph{Science Direct}.

In the field of Art and Design, Mansfield (2016) ranked journals according to their popularity in REF submissions, but did not attempt to infer star ratings for the publications.

Stockhammer, Dammerer, and Kapur (2017) investigated the `grade point average'---the average star rating---of each institution in the 2014 REF, modelling it as a linear function of either the SCImago Journal Rank citation score (SJR) or of journal ratings assigned by the \href{https://charteredabs.org}{Chartered Association of Business Schools}.
That analysis, applied to the fields of economics, found a coefficient of determination of up to \(R^2 = 89\%\), with the 2014 log-SJR score having a statistically significant effect under their model.

Italy's research assessment exercise, the \emph{Valutazione Triennale della Ricerca} (triennial research evaluation) began in 2003 with a similar remit to the UK's RAE/REF and was initially `fully based on peer review'.
Franceschet and Costantini (2011) found positive correlations between the peer review assessments and citation metrics, but the strength of the correlation varied between fields, and was particularly weak for journal impact factor.
The then-recently proposed \(h\)-index (Hirsch 2005) provided a better approximation.

From 2004, the \emph{Valutazione della Qualità della Ricerca} (research quality evaluation; VQR) introduced a `dual system of evaluation' using a combination of peer review and bibliometrics.
The Italian National Agency for the Evaluation of the University and Research Systems (ANVUR) compared the results from each approach and found a `more than adequate concordance', apparently justifying the decision to use bibliometrics.
However, this conclusion has been strongly challenged by Baccini and Nicolao (2016), who insist the methodology was `fatally flawed' and undermines the results for the field of economics and statistics in particular.

Between the RAE2008 and REF2014, Mryglod et al. (2015a) compared departmental \(h\)-indices with performance in the RAE, finding a correlation between \(h\)-index and certain grade-point averages of RAE results.
Using this relationship, they made predictions for the upcoming REF2014 for several institutions and fields.
However, in a follow-up after the REF2014 results were published, Mryglod et al. (2015b) reported the predictions ``failed to anticipate with any accuracy either overall REF outcomes or movements of individual institutions in the rankings relative to their positions in the previous Research Assessment Exercise''.
Thus care should be taken in trying to predict one research assessment from the results of another that took place years before.

Our research is not the first attempt at producing a journal ranking from REF results for economics, let alone for academic fields in general.
Hole (2017) used a greedy iterative algorithm to assign star ratings to individual papers (assuming these were entirely dependent on the journals in they appeared) minimizing the squared error in predicted ratings for institutions,
\[Q = \sum_{i=1}^I \sum_{r=1}^4 N_i (p_{ir} - \hat p_{ir})^2,\]
where \(N_i\) is the number of submissions from each institution, \(p_{ir}\) is the observed proportion of \(r\)-star submissions from that institution and \(\hat p_{ir}\) is the predicted proportion, based on the imputed ratings.
The algorithm first assigns an arbitrary star-rating \(r\) to each journal, calculates the objective function \(Q\), then iterates over the list of journals, changing each journal's star rating to that which would decrease \(Q\) the most, terminating when a full pass over all journals produces a change in \(Q\) smaller than a pre-specified threshold.
The analysis of Hole (2017) excluded journals with fewer than five submissions in the REF.
Since this would result in the number of submissions no longer adding up to the total number of ratings, they assigned arbitrary ranks to these left-out journals.
The results had a correlation of approximately \(\rho = 0.5\) with previously-published economics journal rankings.

More recently, Balbuena (2018) adopted a machine learning approach, using a Bayesian additive regression tree model to predict grade point average from a range of institutional covariates, including the number of attributed documents indexed in the \emph{Web of Science} and the proportional intake of students from state schools.
However this analysis focussed more on possible inequities in distribution of funding, rather than investigating an explicit journal identity effect.

Yan (2017) used a Metropolis-within-Gibbs sampling regime to fit an ordinal response model to Economics \& Econometrics outputs for REF2014.
Whilst broadly similar to our approach, their framework is based on a proportional-odds cumulative probit model, which assumes a common set of thresholds between star ratings for all journals.
In other words, the increase in difficulty of attaining a 4* rating over a 3* one is the same for every journal.
Our analysis fits models for several different subjects and finds that this assumption does not hold, even for Economics \& Econometrics.

\hypertarget{ecological-inference}{%
\subsection{Ecological inference}\label{ecological-inference}}

The previous section provided examples of limited analyses comparing some citation indices and other journal- or institution-level covariates with REF results, and of approaches to produce journal rankings from institutional scores.
However, to our knowledge, no principled \emph{statistical} analyses (that is, with quantified uncertainty) of the relationship between journal identities and UK research assessment have been published.
Moreover, modelling REF ratings as a function of citation metrics is problematic; criticisms abound of certain indicators under inspection---impact factor and its variants, as well as `alternative metrics' such as tweets and download counts (e.g.~Colquhoun and Plested 2014; MacRoberts and MacRoberts 2017).
Instead of using a flawed and imprecise proxy such as a citation metric to analyse the relationship between publications and research assessment, one might consider modelling published REF results against the actual journal identities instead.
The problem with this approach is that HEFCE (or since April 2018 its successor, Research England) will never publish the individual ratings given to submissions in the REF; indeed they were destroyed upon completion of the research assessment (REF 2015).

We are therefore left in a quandary: how do we model the effect of journals on star ratings, if we don't know which journal articles received which ratings?
What if we wanted to try to infer these publication-level ratings?
This would allow us to construct a ranking of \emph{journals}, not just institutions, from the REF results, similar to the work by Hole (2017).
Moreover we might attempt to answer the question: is an institution's REF rating simply a function of the journals in which it published?
Were that to be the case, it would suggest that the REF is directly measuring prestige rather than quality---a common criticism of citation indices.
On the other hand, if an institution's REF score is \emph{more} than the sum of its output journals then it might be used as evidence against using journal-level metrics to assess research quality.

However, as already mentioned, the REF scores are aggregated by institution, not by journal.
Journal-level scores must therefore be imputed rather than observed.
Such a task---inferring individual-level properties from aggregate data---is known as \emph{ecological inference} or \emph{ecological regression} (Goodman 1953), typically applied to estimate voting behaviour in a secret ballot, when exit polls are infeasible or unreliable.
Examples include modelling voter transitions between parties (Brown and Payne 1986) and estimating who voted for the Nazi Party in Weimar Germany (Rosen et al. 2001).
A detailed review of the topic is provided by Wakefield (2005).
The following is a brief summary.

Sociologists and political scientists often use the term `ecological inference' to refer to inference on voting populations---for example, voter transitions between elections in a two-party system, or turnout for two demographics.

Consider an election where, to comply with civil rights legislation, authorities in the US desire to compare turnout amongst black and white voters.
Suppose for a given electoral district (constituency) \(i\), the demographic makeup is known with proportion \(X_i\) of the population black and the remainder white.
Overall voter turnout, \(T_i\), is observed for a particular election but the ballot is secret, so turnout among blacks and whites, respectively \(\beta_i^b\) and \(\beta_i^w\), are unknown.
These data yield the following \(2 \times 2\) table of proportions.

\begin{longtable}[]{@{}llll@{}}
\caption{\label{tab:voters}Observed and unobserved proportions for a two-dimensional voter turnout model}\tabularnewline
\toprule
& Vote & Not vote &\tabularnewline
\midrule
\endfirsthead
\toprule
& Vote & Not vote &\tabularnewline
\midrule
\endhead
Black & \(\beta_i^b\) & \(1 - \beta_i^b\) & \(X_i\)\tabularnewline
White & \(\beta_i^w\) & \(1 - \beta_i^w\) & \(1 - X_i\)\tabularnewline
& \(T_i\) & \(1 - T_i\) &\tabularnewline
\bottomrule
\end{longtable}

At first glance, it may not appear that one can really glean any information about individuals only from the margins.
Via the \emph{method of bounds} however, we can obtain deterministic bounds on (at least one of) the parameters: black turnout \(\beta_i^b\) must be greater than \(\frac{T_i - (1-X_i)}{X_i}\) and smaller than \(\frac{T_i}{X_i}\), whilst
white turnout \(\beta_i^w\) must be between \(\frac{T_i - X_i}{1 - X_i}\) and \(\frac{T_i}{1-X_i}\), to ensure they are valid proportions that add up to one (Duncan and Davis 1953).
For example, if a district's population were 70\% black and overall turnout were 40\%, then black turnout must be in the range (14\%, 57\%), but white turnout could still be anywhere in (0, 100\%).

Unlike this limited deterministic approach to the ecological inference problem, \emph{ecological regression} or \emph{Goodman regression} (1953, 1959) is one of the first \emph{statistical} solutions.
Using the identity
\[T_i = X_i \beta_i^b + (1 - X_i) \beta_i^w,\]
one may construct a simple linear regression model of turnout on racial proportions:
\[\mathbb{E}[T_i|X_i] = \alpha + \beta X_i,\]
where \(\alpha = \beta_i^w\) and \(\beta = \beta_i^b - \beta_w\).
A notable criticism is that these voting propensities are assumed to be homogeneous over districts, regardless of the racial mix in each area.
Moreover, least squares does not constrain these parameters to lie within the bounds described above, or even between zero and one (Wakefield 2005).

Brown and Payne (1986) proposed modelling voter turnout using a convolution of Dirichlet--multinomial distributions, with the response approximated by a multivariate normal distribution.
However, this model is sensitive to the choice of prior (Wakefield 2005).
More recently, King (2013) combined the method of bounds with a pseudo-`likelihood' function---equivalent to an asymptotic form of the binomial distribution---and imposed a truncated bivariate normal distribution to tighten the bounds.
This approach describes itself as `a solution to the ecological inference problem', however this claim was criticized as overly optimistic (Cho 1998; Freedman et al. 1999).

Since then, King, Rosen, and Tanner (1999) proposed a different solution in the form of an hierarchical Dirichlet--multinomial model where the unobserved probabilities (the voter turnouts by ethnicity) are beta-distributed latent random variables.
For a constituency/district \(i\) with total voting-age population \(N_i\) and observed voter turnout count \(Y_i = N_i T_i\) the hierarchical model takes the form
\[Y_i \sim \operatorname{Binomial}(N_i, T_i),\]
where the marginal probability of voter turnout in constituency \(i\) is
\[T_i = \sum_{j=1}^J x_{ij} \beta_i^j,\]
with \(x_{ij}\) denoting the proportion of people of ethnicity \(j\) in constituency \(i\), and where the prior constituency-level probabilities of voter turnout, by ethnic group, are
\[\beta_i^j \vert a_j, b_j \, \overset{\text{iid}_i}{\sim} \, \operatorname{Beta}(a_j, b_j)\]
with hyper priors
\[a_j, b_j \overset{\text{iid}_i} \sim \operatorname{Exp}(\lambda)\]
and the default hyper-parameter setting \(\lambda = 0.5\).

In the two-dimensional (black--white voter turnout) case described above, \(J=2\) and the middle level is
\[T_i = X_i \beta_i^b + (1 - X_i) \beta_i^w.\]
The model generalizes to \(J>2\) ethnic groups (or journals, in our case) and can be further extended to multiple outcomes (beyond binary `vote or not') by replacing the beta--binomial distribution pair with a Dirichlet--multinomial (Rosen et al. 2001).

In our view, the top level of King's hierarchical model possibly adds an unnecessary random component, for the total turnout should simply be a deterministic, weighted sum of the turnout among each ethnic group.
The election result is not an approximation of the counted votes: it \emph{is} the counted votes.
All that is necessary is for the \(\beta\)s to be constrained so that the sum over ethnic groups of voters adds up to the observed overall turnout.
This is perhaps more easily said than done, however.

More recent approaches to ecological inference make use of \emph{distribution regression}, by treating the makeup of each electoral district as a probability distribution (Flaxman, Wang, and Smola 2015; Szabó et al. 2014).
The basic idea is to project the distributions into a feature space, then fit a regularized regression model, such as kernel ridge regression, using this embedding.
Flaxman, Wang, and Smola (2015) used this technique to combine demographic and spatial information and infer the groups who voted for Barack Obama in the 2012 US presidential elections, and again for the 2016 elections (Flaxman et al. 2016).

Rosenman and Viswanathan (2018) derived a `heteroscedastic Gaussian' approximation to the Poisson binomial log-likelihood, via a central limit theorem, and later applied this to a large voter transition model (Rosenman 2019), which they term a \emph{Poisson binomial generalized linear model}.
Unlike the presidential election studies by Flaxman et al., which used Bayesian techniques, the Poisson binomial GLM is `purely frequentist'.
This offers the advantages of `simpler fitting procedures, straightforward estimation of individual-level probabilities, and greater model interpretability' at the expense of reduced flexibility (Rosenman 2019).

\hypertarget{model}{%
\section{Model}\label{model}}

The REF ratings received by institutions on their outputs could be assumed to be drawn from a Poisson binomial distribution (Poisson 1837), which describes the probability of obtaining \(K\) successes in \(n\) independent but non-identically distributed Bernoulli trials, with probability mass function
\[\operatorname{Pr}(K = k) = \sum_{A \in F_k} \prod_{i \in A} \pi_i \prod_{j \in A^c} (1 - \pi_j),\]
where \(F_k\) is the set of all subsets of \(k\) integers that can be selected from \(\{1, 2, \dots, n\}\), for \(n\) the number of Bernoulli trials and \(\pi_i\) the success probability of the \(i\)\textsuperscript{th} trial (Wang 1993).
For our purposes, \(\pi_i\) represents the probability that an output \(i\) was awarded a particular rating, for example 4*.

The Poisson binomial is a special case of the aggregated compound multinomial model used by Brown and Payne (1986).
That paper describes a Dirichlet-multinomial (`compound multinomial') model for the unobserved numbers of voters who switched between each of the major parties from one election to another.
In their notation, each election featured the same set of political parties.
The model estimates the probability, \(p_{ijk}\), that a voter for party \(i\) in constituency \(k\) becomes a voter for party \(j\).

Our analogy is rather different: there are \(J\) parties at the first election, representing the journals in which the articles are published, but only two parties at the next election: `4*' and `not 4*'.
Voters are articles, and constituencies are academic institutions.
We model the probability that an article published in a particular journal is awarded a 4* rating, or not.

Let \(x_{ij}\) denote the (known) number of articles published by institution \(i\) in journal \(j\).
Let \(y_{ij}\) denote the (unknown) number of such articles that attained a 4* rating in the REF, with \(0 \leq y_{ij} \leq x_{ij}\) for all \(i = 1, \dots, I\) and \(j = 1, \dots, J\).
Let \(y_i = \sum_j y_{ij}\) denote the published number of 4* ratings awarded to each institution and let \(x_j = \sum_i x_{ij}\) denote the total number of articles submitted from each journal.
Then the marginal totals, \(\mathbf{Y} = (Y_1, \dots, Y_I)\), are aggregated compound multinomial (Dirichlet-Poisson-binomial) random variables with expectation
\[\mathbf{E}(\mathbf{Y}) = \mathbf{P}^T\mathbf{x}\]
and covariance
\[\operatorname{cov}(\mathbf{Y}) = \operatorname{diag}(\mathbf{P}^T\mathbf{w}) - \mathbf{P}^T\operatorname{diag}(\mathbf{w})\mathbf{P},\]
where \(\mathbf{P}\) is the \(J\)-vector of journal success probabilities\footnote{More generally, \(\mathbf{P}\) is a \emph{matrix} of multinomial probabilities}, \(\mathbf{x}=(x_1, \dots, x_J)\) is a vector of the number of articles in each journal and \(\mathbf{w}\) is a \(J\)-vector of weights \(w_j = x_j(x_j + \alpha_j)/(1 + \alpha_j)\).
Brown and Payne (1986) note that `election data involve more variability than a multinomial would suggest' and add the \(\alpha\) vector of \(J\) precision/dispersion parameters to account for this.

The variance of a Poisson binomial-distributed random variable is
\[
\operatorname{Var}(Y_i)=\sum_{j}(1-\pi_{j})\pi_{j} = \sum_j (\pi_j - \pi_j^2),
\]
which differs from the variance of the aggregate compound multinomial model only by the \(w_j\) term.
We notice that as \(\alpha_j\) grows large, then (dropping the subscripts for the moment)
\[\lim_{\alpha\to\infty} w
= \lim_{\alpha\to\infty} x\left(\frac{x}{1 + \alpha} + \frac{1}{\frac1\alpha + 1}\right)
= x,\]
and since \(\alpha=\infty\) corresponds to the (non-compound) aggregate multinomial distribution (Brown and Payne 1986), we can see the Poisson binomial and aggregated multinomial models are equivalent.

If we consider every paper grading to constitute an independent trial, with success probability dependent on the journal in which it is published (but not the institution or any paper-level characteristic), then for each institution \(i\), the number of 4* ratings received is distributed
\begin{equation}
Y_i \sim \text{Poisson-Binomial}(\boldsymbol\pi),
\label{eq:model}
\end{equation}
where \(\boldsymbol\pi = (\pi_1, \dots, \pi_J)\) is the vector of journal probabilities\footnote{In practice, each element \(\pi_j\) is repeated \(x_{ij}\) times, representing repeated trials for the number of articles in journal \(j\) that were submitted by institution \(i\) to the REF.}.
Here the success probabilities are not identical for every publication, but they do coincide wherever two submissions are published in the same journal; if an institution submits two or more articles from the same journal then each of these articles is regarded as a separate independent trial.
Of course, independence might be an heroic assumption here, but \emph{ideally} one would hope the REF panels consider each article on its own merits rather than ranking them against one another.

We can fit the model twice: firstly with `success' defined as 4* ratings, and secondly with success defined as 3* \emph{or} 4* ratings, i.e.~3* or better.
As the star ratings are ordinal responses (4* is better than 3*, which is better than 2* and so on), it seems reasonable to assume cumulative odds, and infer the probability of 3* from the estimated probabilities of 4* and of 3*-or-better.
Thus a journal's probability of obtaining a 3* rating is assumed to be \[\pi_j^3 = \pi_j^{34} - \pi_j^4,\]
where we introduce superscript notation: \(\pi_j^3\), \(\pi_j^4\) and \(\pi_j^{34}\) respectively denote journal \(j\)'s probability of accruing 3*, 4* and \(\geq 3^*\) ratings.

This separate fitting of the model for the 4* and 3*-or-4* is a key difference from the work of Yan (2017), which used a cumulative odds model with common thresholds for every journal.
That is, under Yan's model, \(\operatorname{probit} (\pi_j^{34}) - \operatorname{probit} (\pi_j^4) = c\), a constant offset that does not depend on the journal \(j\).
Our approach replaces \(c\) with \(c_j\), a difference that can might be distinct for every journal.

On top of the likelihood \eqref{eq:model} we impose a prior on the journal success parameters,
\begin{equation}
  \pi_j \sim \text{Beta}\bigl( \gamma \mu, \gamma(1 - \mu) \bigr)
  \label{eq:prior}
\end{equation}
for each journal \(j\), such that the mean probability of success is \(\mu\), and \(\gamma\) is a regularizing concentration parameter.
On top of these we impose hyper-priors
\begin{equation}
\begin{aligned}
    \mu &\sim \text{Uniform}(0, 1)
    \\
    \gamma &\sim \text{Gamma}(\tfrac1{10}, \tfrac1{20}),
\end{aligned}
  \label{eq:hyperprior}
\end{equation}
where the given hyper-parameters of the gamma distribution are the shape and rate, respectively---corresponding to a mean of 2 and variance 40.
In principle one could set these manually, for instance setting \(\mu\) equal to the empirical mean institutional profile, but we shall try to learn them from the data.

There are more differences between institutions than just the journals in which they publish, so to check for aggregation bias, we extend the model \eqref{eq:model} such that an article success depends not just on the journal parameter, but on an institutional covariate linked to the REF Environment profiles.
In this way we might hope to detect any institutions that perform better or worse in output scores due to the quality of their research environment rather than on the journals in which they publish.
Thus the success probability of an article from institution \(i\) in journal \(j\) is
\begin{equation}
  \text{log-odds}(4* | i, j) = \operatorname{logit} \pi_j + \alpha ~ \text{envir}_i
  \label{eq:envir}
\end{equation}
where \(\alpha\) is a parameter to be estimated and \(\text{envir}_i\) is the proportion of `Environment' in institution \(i\) rated 4* (centred by subtracting the mean).
If the \(\alpha\) is near zero, then we might conclude that output profiles depend more on the journals than on the unique characteristics of each research institution.

\hypertarget{methods}{%
\section{Methods}\label{methods}}

We will employ two different methods to estimate the parameters of the model.
Firstly, a Bayesian Monte Carlo method, and secondly a maximum likelihood approach using an expectation--maximization algorithm.
This section describes the details behind each technique.

\hypertarget{hamiltonian-monte-carlo}{%
\subsection{Hamiltonian Monte Carlo}\label{hamiltonian-monte-carlo}}

Owing to the limited computational power available at the time, Brown and Payne (1986) employed a normal approximation to the Poisson binomial model to estimate the unknown coefficients.
The Poisson binomial distribution can also be approximated by a Poisson distribution, though the performance of this approximation is poor when the number of trials is large (Hong 2013).

Advances in computation capacity allow us to consider a couple of different approaches of fitting a Poisson binomial model.
The first would be to employ the probabilistic programming language Stan (Carpenter et al. 2017; Stan Development Team 2018) to sample from the posterior distribution via Hamiltonian Monte Carlo (also known as hybrid Monte Carlo or HMC; Duane et al. 1987).
Because enumerating all possible sets of integers \(F_k\) is computationally infeasible, instead one can program a routine to compute the mass by enumerating with a recursive formula (Shah 1973)
\[\operatorname{Pr}(K = k) = \begin{cases}
\prod_{j=1}^n (1 - p_j)    &    k = 0    \\
\frac1k\sum\limits_{j=1}^k(-1)^{j-1} \operatorname{Pr}(K = k-j) \sum\limits_{l=1}^n\left(\frac{p_l}{1 - p_l}\right)^j    &    k > 0,
\end{cases}\]
however this may not be numerically stable for large \(n\) (Hong 2013) unless computed on the logarithmic scale.

Using such a dynamic programming algorithm on the logarithmic scale, we fit the model in Stan and report the results in Section~\ref{results}.
As a robustness check, we also consider a maximum likelihood approach, described in the next subsection.

\hypertarget{expectationmaximization-algorithm}{%
\subsection{Expectation--maximization algorithm}\label{expectationmaximization-algorithm}}

The expectation--maximization (EM) algorithm can provide alternative maximum likelihood point estimates, albeit without any covariance estimate as a measure of uncertainty.

The EM algorithm makes use of the \emph{extended multivariate hypergeometric distribution}.
Recall the more familiar hypergeometric distribution describes the probability that, given an urn of \(N\) balls, \(K\) of them white and \(N-K\) black, that if we draw \(n\) balls at random then \(k\) of them are white.
The \emph{extended} hypergeometric distribution, also known as Fisher's \emph{noncentral} hypergeometric distribution, extends this scenario to non-uniform sampling---where the white balls are more likely to be drawn than black ones because of differences in size or weight.

A multivariate hypergeometric distribution generalizes to a situation where there are more than two colours of balls and describes the probability of picking a particular mixture of colours.
Hence, an extended multivariate hypergeometric distribution describes the probability of picking a certain mix from an urn of balls whose weights are not all equal (McCullagh and Nelder 1989, 260--261).
For dimension \(d\) different colours, the probability mass function for drawing a mixture \(\mathbf{x} = (x_1,\dots,x_d)\) of \(n\) balls is
\[f(\mathbf{x}; n, \mathbf{m}, \boldsymbol\omega) = \frac{1}{P_0} \prod_{i=1}^d {m_i \choose x_i} \omega_i^{x_i}\]
where \(\mathbf{m}=(m_1, \dots, m_d)\) is the number of each colour of balls in the urn and \(\boldsymbol\omega = (\omega_1, \dots, \omega_d)\) are their respective weights.
The denominator \(P_0\) is
\[P_0 = \sum_{\mathbf{y}\in \mathcal S} \prod_{i=1}^d {m_i \choose y_i} \omega_i^{y_i}\]
with \(\mathcal S\) denoting the set of all possible non-negative integer \(d\)-vectors \(\mathbf{y} = (y_1, \dots, y_n)\) where \(\sum_{i=1}^d y_i = n\).

If the article-level ratings were known, we could fit a Rasch-like generalized linear model of the following form to estimate journal effects:
\begin{equation}
\text{logit}~\mathbb{E}[\operatorname{Pr}(4^*|i, j)] = \mu + \alpha z_i + \beta_j,
\label{eq:rasch}
\end{equation}
for a paper by institution \(i\) in journal \(j\), where parameter \(\mu\), analogously to \eqref{eq:model}, acts as a `grand mean' intercept term, here in the logit space, and where \(z_i\) is an indicator variable for a `pseudo-institution'.
The latter submits an equal number of 4* and not-4* papers from every journal, acting as a regularizing prior on the strength of the journal effects to avoid overfitting.
The more pseudo-papers submitted, the stronger the effect of the regularization.
The maximum likelihood estimate for such an artificially augmented dataset is equivalent to the posterior mode with a conjugate Bayes model (Jannarone, Yu, and Laughlin 1990).
The optimum strength of regularization (i.e.~the number of pseudo-articles augmenting the data) is determined via cross-validation, described at the end of this section.

We adopt an expectation--maximization procedure as follows.

\begin{enumerate}
\def\labelenumi{\arabic{enumi}.}
\item
  Initialize the weights of the noncentral multivariate hypergeometric distribution.
  That is, randomly generate a probability for each journal--institution that corresponding outputs will be 4*-rated in the REF. (We use a logit-normal distribution for this.)
\item
  Compute the (approximate) expectation of the noncentral multivariate hypergeometric distribution with these odds (for this, we use R package \texttt{BiasedUrn} by Fog 2015). This vector forms an imputation of the latent individual-level ratings.
\item
  Fit the model described in Equation~\eqref{eq:rasch}. Extract coefficients from this model to get new odds for the noncentral multivariate hypergeometric distribution.
\item
  Repeat steps 2--3 until convergence.
\end{enumerate}

To obtain new odds form the logistic regression model for the noncentral hypergeometric distribution, we simply use the relation
\[\text{odds}(4^* | j) = \exp(\hat\mu + \hat\beta_j)\]
for all journals \(j\), where \(\hat\mu\) and \(\hat\beta_j\) are the estimated parameters from the previous EM step.

Our chosen prior for this model is essentially uninformative on the expected journal ranking.
In principle, one could attempt to elicit distributions for the relative strengths of the journals, or (by asking someone who might have served on REF/RAE expert panels in the past) the probability that papers in a certain journal might accrue 4* ratings.
However, such an approach is not very scalable to large numbers of journals or fields, so we do not adopt it here.

The cross-validation procedure works as follows.

\begin{enumerate}
\def\labelenumi{\arabic{enumi}.}
\tightlist
\item
  Randomly divide the institutions into (say) 10 groups.
\item
  For each group:

  \begin{enumerate}
  \def\labelenumii{\alph{enumii}.}
  \tightlist
  \item
    Run the above expectation--maximization algorithm on data from the other 9 groups.
  \item
    Use the estimated journal parameters to predict the institutional results for the held-out group.
  \item
    Compute the index of dissimilarity between the predicted and actual institutional results.
  \end{enumerate}
\item
  Repeat steps 1--2 for different levels of regularization.
\end{enumerate}

We seek the parameter that minimizes the index of dissimilarity (described in the next section) between the predicted institutional scores and the actual scores of the held-out institutions.

\hypertarget{diagnostics-and-summary-statistics}{%
\subsection{Diagnostics and summary statistics}\label{diagnostics-and-summary-statistics}}

To obtain a `prediction' or fitted value from the Poisson binomial model, we take the posterior median of the journal probability estimates \(p_j\) and take them to be the proportion of the time that articles in those journals were awarded 4*.

That is, we compute
\[\hat y_i^4 = \sum_{j=1}^J n_{ij} \hat \pi_j^4\]
from the model fitted to 4* outputs and
\[\hat y_i^{34} = \sum_{j=1}^J n_{ij} \hat \pi_j^{34}\]
from the same model fitted to a dataset of 3* \emph{or} 4* outputs, where \(y_i^{34}\) denotes the number of an institution's outputs rated 3* or better (i.e.~3* \emph{or} 4*), \(\pi_j^{34}\) represents the probability that articles in journal \(j\) are awarded 3* or better, and \(n_{ij}\) denotes the number of articles from institution \(i\) in journal \(j\).
Hence we can compute the predicted number of 3* outputs,
\[\hat y_i^3 = \hat y_i^{34} - \hat y_i^4,\]
for each institution \(i = 1, \dots, I\).

Recall that our main aim is to answer the question: to what extent are REF output profiles a function of journal identities?
In other words: given the journals in which an institution published its submissions, can we predict that institution's REF score?

To determine the quality of fit of the Poisson binomial model we adopt the index of dissimilarity (Duncan and Duncan 1955; Kuha and Firth 2011), which here represents the proportion of an institution's articles predicted a different rating to that observed in the REF.
It is computed using the formula
\[\Delta = \frac1{2N} \sum_i \bigl( |y_i^4 - \hat y_i^4| + |y_i^3 - \hat y_i^3| + |y_i^4 + y_i^3 - \hat y_i^4 - \hat y_i^3| \bigr),\]
where \(N = \sum_i \sum_j n_{ij}\), the total number of submitted outputs.

From the index of dissimilarity we propose another metric, the \emph{redistribution of monetary reward}, based on the notion that a 4* output is worth four times as much in research funding as a 3* output, and outputs rated 2* or lower accrue no direct funding at all (see e.g.~Koya and Chowdhury 2017).
This metric describes the fraction of total monetary reward that would move between institutions if the estimated REF profiles \((\hat y_i^4, \hat y_i^3)\) were used instead of the observed profiles \((y_i^4, y_i^3)\), and is measured by
\[\Delta_{\pounds} = {{\frac12 \sum_i m_i | r_4 (p_i^4 - \hat p_i^4) + r_3 (p_i^3 - \hat p_i^3)|} \over {\sum_i m_i (r_4 p_i^4 + r_3 p_i^3)}},\]
where \(m_i\) is the number of full-time equivalent (FTE) staff submitted by institution \(i\) in the unit of assessment, \(\hat p_i^4 = \hat y_i^4 / \sum_j n_{ij}\), \(\hat p_i^3 = \hat y_i^3 / \sum_j n_{ij}\) and \(r_4\) and \(r_3\) are the respective monetary reward per FTE for the 4* and 3* components of output profiles, in arbitrary units with \(r_4 = 4 r_3\).
(Implicitly, terms for 2*, 1* and unclassified outputs can appear in the above formula, but we take \(r_2 = r_1 = r_u = 0\).)

The monetary index might even be combined with calculations of the kind by Koya and Chowdhury (2017) to compute an absolute sterling figure for the amount of funding that would move institutions in a switch from the actual REF profiles to those estimated our model.

\hypertarget{data}{%
\section{Data}\label{data}}

\hypertarget{units-of-assessment}{%
\subsection{Units of assessment}\label{units-of-assessment}}

To demonstrate the method, we will first consider the `Economics and Econometrics' unit of assessment.
We choose this particular subject area because it is small, relatively self-contained, and a high proportion of output submissions (92\%) are in the form of journal articles (rather than books, conference proceedings or other works).
One might expect (this being a statistics PhD thesis) to look at statistical science submissions first, however these fall under the umbrella of Mathematical Sciences---along with research in probability, pure and applied mathematics and mathematical physics---which is a larger and more hetereogeneous field.

From Table~\ref{tab:outputs} it is easy to see that the hard sciences (REF panels A and B) mostly submitted outputs in the form of journal articles; the arts and humanities (panel D) used other formats, and social sciences (panel C) were somewhere in between.
A notable exception to this rule is the field of Computer Science and Informatics, where the role of academic journals is often supplanted by conference proceedings.

After the initial analysis of the Economics and Econometrics sub-panel, we also examine three other REF subpanels, all from the Physical Sciences main panel, to see how they compare.
The arts (main panel D) publish too few of their outputs as journal articles for this model to be of practical relevance.

\begin{table*}[t]
\centering
\begingroup\small
\begin{tabular}{lp{0.7\linewidth}rr}
  \toprule
Panel & Unit of assessment & Outputs & Journals \\ 
  \midrule
A & Clinical Medicine & 13400 & 99.9 \\ 
   & Biological Sciences & 8608 & 99.7 \\ 
   & Public Health, Health Services and Primary Care & 4881 & 99.6 \\ 
   & Psychology, Psychiatry and Neuroscience & 9126 & 99.6 \\ 
   & Agriculture, Veterinary and Food Science & 3919 & 99.1 \\ 
   & Allied Health Professions, Dentistry, Nursing and Pharmacy & 10358 & 98.9 \\ 
  B & Chemistry & 4698 & 99.8 \\ 
   & Earth Systems and Environmental Sciences & 5249 & 99.1 \\ 
   & Aeronautical, Mechanical, Chemical and Manufacturing Engineering & 4143 & 99.0 \\ 
   & Electrical and Electronic Engineering, Metallurgy and Materials & 4025 & 98.9 \\ 
   & Physics & 6446 & 98.9 \\ 
   & General Engineering & 8679 & 98.4 \\ 
   & Civil and Construction Engineering & 1384 & 97.4 \\ 
   & Mathematical Sciences & 6994 & 96.2 \\ 
   & Computer Science and Informatics & 7651 & 72.6 \\ 
  C & Sport and Exercise Sciences, Leisure and Tourism & 2757 & 96.8 \\ 
   & Business and Management Studies & 12202 & 95.6 \\ 
   & Economics and Econometrics & 2600 & 91.8 \\ 
   & Geography, Environmental Studies and Archaeology & 6017 & 82.6 \\ 
   & Education & 5519 & 78.3 \\ 
   & Architecture, Built Environment and Planning & 3781 & 77.6 \\ 
   & Social Work and Social Policy & 4784 & 77.4 \\ 
   & Sociology & 2630 & 76.1 \\ 
   & Politics and International Studies & 4365 & 70.6 \\ 
   & Anthropology and Development Studies & 2013 & 67.3 \\ 
   & Law & 5522 & 62.5 \\ 
  D & Philosophy & 2173 & 61.8 \\ 
   & Area Studies & 1724 & 56.6 \\ 
   & Communication, Cultural and Media Studies, Library and Information Management & 3517 & 52.5 \\ 
   & Modern Languages and Linguistics & 4932 & 48.3 \\ 
   & History & 6431 & 44.0 \\ 
   & Theology and Religious Studies & 1558 & 37.2 \\ 
   & English Language and Literature & 6923 & 35.7 \\ 
   & Music, Drama, Dance and Performing Arts & 4246 & 29.8 \\ 
   & Classics & 1386 & 28.9 \\ 
   & Art and Design: History, Practice and Theory & 6321 & 26.2 \\ 
   \bottomrule
\end{tabular}
\endgroup
\caption{Units of assessment in REF2014, the number of outputs submitted and the percentage of which that were classified as journal articles} 
\label{tab:outputs}
\end{table*}

\hypertarget{wrangling-ref2014-data}{%
\subsection{Wrangling REF2014 data}\label{wrangling-ref2014-data}}

\begin{figure}
\subfloat[number of submitted articles\label{fig:submissions-1}]{\includegraphics[width=.9\linewidth]{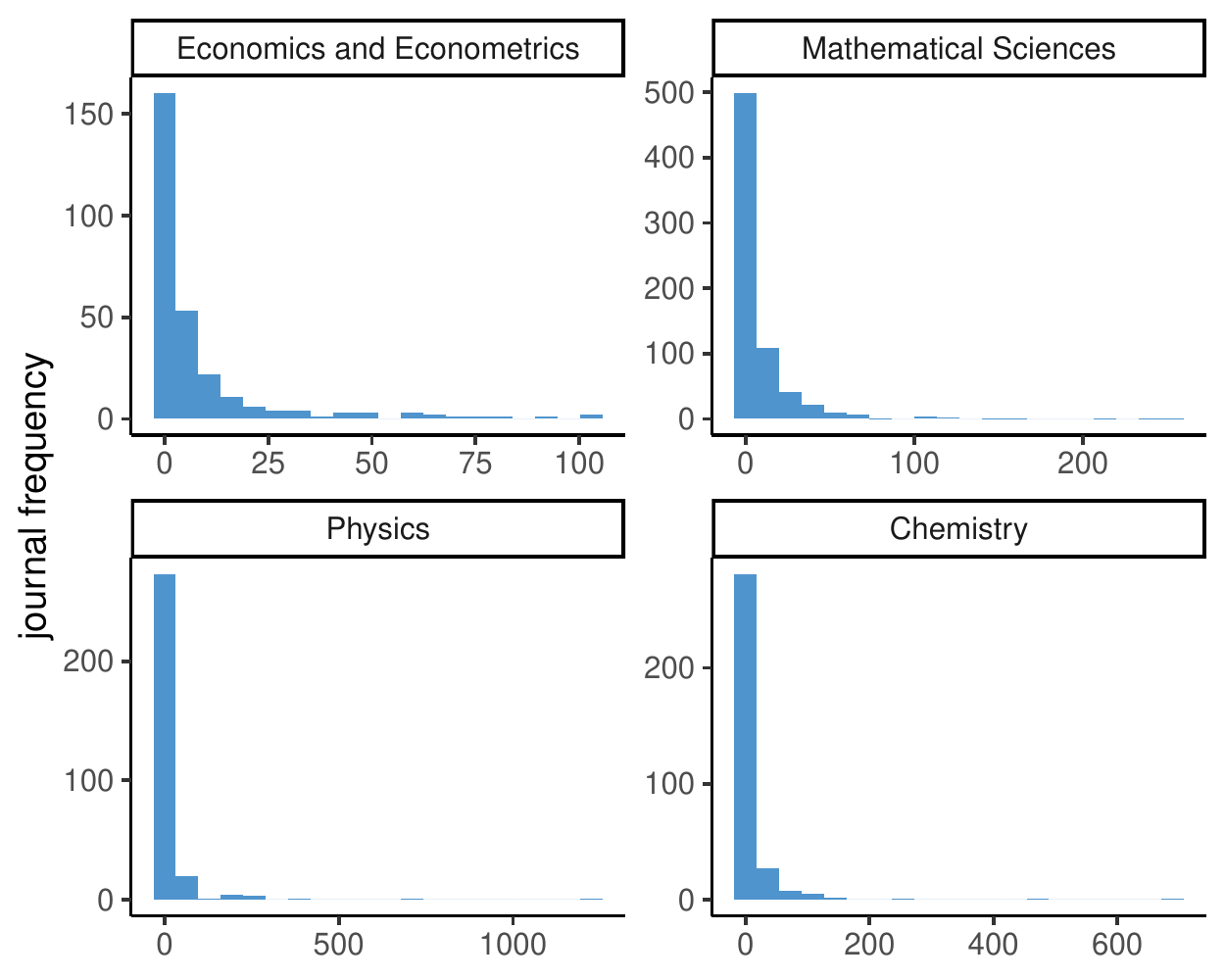} }\newline\subfloat[number of submitting institutions\label{fig:submissions-2}]{\includegraphics[width=.9\linewidth]{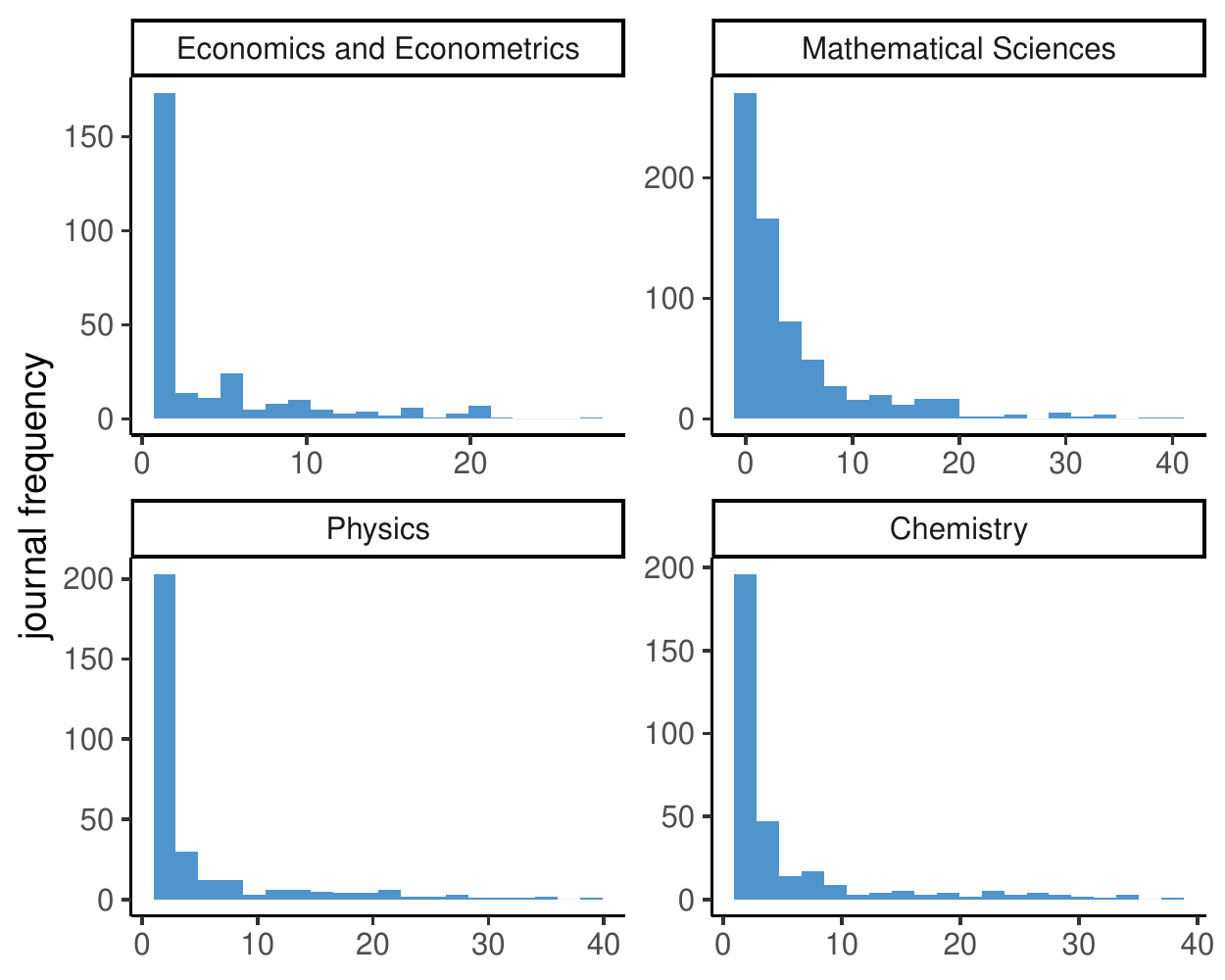} }\caption{Distribution of journal articles across journals and institutions, by unit of assessment. Some journals are much more popular than others, and not all institutions publish in the same journals}\label{fig:submissions}
\end{figure}

In the published REF2014 submissions data, outputs are explicitly categorised into types, such as journal articles, book chapters, conference proceedings, working papers and so on.
However, there is no sure-fire way to group together articles published in the same journal or book, as the titles are unstandardised, ISSNs, if provided, can vary between print and online editions and DOIs, where present, can be difficult to parse.
Labour-intensive manual tagging of the data has rather little appeal, not least because it is error-prone and does not scale well to larger future data sets.
But there is a network science solution to the problem.

We coerced the output submissions data into a long-format table comprising just a journal identifier---the ISSN, ISBN, DOI or standardised journal title (coerced to lower case, with punctuation, diacritics, spaces and leading ``the''s removed)---and a unique identifier for each output, then constructed an undirected bipartite graph between the journal identifiers and individual output identifiers.
Each connected component in this graph represents a unique journal, containing outputs with a common journal title, DOI, ISSN and/or ISBN.
Each is assigned a unique journal ID, as well as a human-readable title, the latter sampled from one of the journal title variants found in the component.

Unfortunately, this methodology on the published REF data alone assumes integrity of the published data, which was later found to be lacking.
Some administrators entered article metadata by hand, rather than retrieving it programmatically via CrossRef, as perhaps they should have done.
This inevitably introduced human error; for example one entry that should have been from the \emph{Annals of Mathematics} had the correct DOI, article and journal title, but the ISSN was that for the separate \emph{Advances in Mathematics} journal, which causes the above mini-algorithm to merge the works in \emph{Annals} and \emph{Advances} as if they came from the same journal.
In turn, the \emph{Advances in Mathematics} journal was grouped with \emph{Advances in Applied Mathematics} due to similarly careless data entry.
Further issues were caused by journal titles that were ambiguous if not completely erroneous, for example various articles published in \emph{Physical Review Letters}, \emph{Physical Review A}, \emph{Physical Review B} and so on all being given the unhelpful abbreviation \emph{PHYS REV}.

Evidently, the metadata in the published REF outputs data set cannot be trusted, except possibly the DOIs.
To remedy this, we used the R package \textbf{rcrossref} (Chamberlain et al. 2019) to access CrossRef application programming interface (API), allowing retrieval of metadata associated with the 25,000 unique DOIs for the Economics \& Econometrics, Mathematical Sciences, Chemistry and Physics submissions.
All except 22 returned results.
Of these few `invalid' DOIs, manual inspection showed the same broken DOIs to be published on publishers' own web pages (and this was reported to CrossRef) so these were not a problem with the REF data itself.
For the remaining (vast majority) of DOIs, the CrossREF API returned the titles of the articles and the names and ISSNs of the containing periodicals.

A small amount of data wrangling remained, however.
Though no single DOI yielded multiple entries in the CrossRef database, our mini clustering algorithm was still required to merge journals which have multiple titles appearing in CrossRef, for example \emph{The Review of Economic Studies} and \emph{Review of Economic Studies}.
These were able to be clustered by shared ISSNs (and we assume that CrossRef, at least, gets these correct).

This approach can easily be applied to every field with no manual or \emph{ad hoc} data processing necessary (except those articles with missing or invalid DOIs).
The distribution of outputs to journals and to institutions is illustrated in Figure~\ref{fig:submissions}, where we can see that it is quite skewed.
An uneven spread of journals between institutions is desirable for an ecological inference model; if every institution published in the same profile of journals then it would be impossible to learn any journal-level effect.

Nevertheless, estimating several hundred journal parameters from just a few dozen institution-level observations is particularly ambitious, especially when it is evident that many journals accounted for very few submitted outputs.

Ordinarily in high-dimensional data analysis, one can apply some level of regularization to the model, the exact level of regularization to be determined by, say, empirical Bayes estimation.
However, standard techniques of `soft' regularization do not seem to work very well for aggregated data like those found in our ecological inference problem.
Instead we adopt a fairly pragmatic approach: any journal containing fewer than some threshold number of articles is aggregated into a single \emph{super-journal} entitled `Other journals'.
We choose the threshold such that \emph{most} (i.e.~\(\geq 50\%\)) of the articles in the data fall into a named journal rather than an anonymous `other' journal, while hopefully also keeping the number of parameters low enough to be practical for reporting and visualization.
Conference proceedings and other non-journal outputs cannot be ignored, as the Poisson binomial model requires we account for all submitted outputs, so these publications are aggregated into their own respective `other' categories.

We apply our methodology to the Economics and Econometrics sub-panel as well as three other fields: Mathematical Sciences, Physics and Chemistry, representing three units of assessment from REF2014 main panel B.
Biological Sciences (main panel A) was also considered, but modelling this field proved too computationally intensive, possibly due to the large number of submitted outputs (8,608) and institutions (44) or the distributions thereof.
(This unit of assessment could still be analysed in future with a more efficient model fitting implementation.)

Compared to Economics and Econometrics, several times more outputs were submitted to each of these sub-panels (see Table~\ref{tab:outputs}), the vast majority of them (\(\geq 96\%\)) in the form of journal articles.

\hypertarget{economics-econometrics}{%
\subsection{Economics \& Econometrics}\label{economics-econometrics}}

Our first REF sub-panel of interest, the `Economics and Econometrics' unit of assessment, received 2600 publications from 28 institutions for the outputs submission.
Of these, 2388 were journal articles, distributed in various publications as shown in Table~\ref{tab:econJournals}.

Using a combination of CrossRef data and the clustering algorithm described in the previous section, eventually we were able automatically to assign the 2388 economics outputs into 277 unique journals.

Setting the threshold at all Economics and Econometrics journals containing \(\geq 20\) submitted articles, we obtain the distribution shown in Table~\ref{tab:econJournals}.
There are 29 named journals, representing over half of the total submissions.

\hypertarget{mathematical-sciences}{%
\subsection{Mathematical Sciences}\label{mathematical-sciences}}

After the field of economics, we study the Mathematical Sciences unit of assessment, which encompasses pure and applied mathematics, statistics and probability---though no distinction was made between these sub-fields, so the REF panel perhaps had the dubious honour of trying to assess subfields as diverse as pure mathematics and applied statistics together on the same measurement scale.

Mathematical Sciences was larger than Economics \& Econometrics, with 53 submitting research institutions.
The 6994 Mathematical Sciences outputs, of which 6731 were classified as journal articles, span some 696 unique scholarly journals.
In this case, and for the remaining three sub-panels, the larger number of articles per journal necessitates a higher threshold for named journals: we increase the minimum number of submitted articles to 30 for Mathematical Sciences, Physics and Chemistry.
This ensures that `named' journals still provide a good representation (\(\geq 50\%\) coverage) of outputs in the data, while keeping model complexity reasonably low.

Figure~\ref{fig:submissions}a suggests a similarly skewed distribution of articles across journals: many journals represented just one or two article submissions each, but a small number of mainly physical science journals had article counts in triple figures, including \emph{Journal of Fluid Mechanics} with 254 articles and \emph{Physical Review Letters} with 209.
Some sub-fields appear to have published (or at least been submitted) more prolificly than others: the biggest statistical journal submission number was from \emph{Biometrika} with 57 articles.
See Table~\ref{tab:mathsJournals} for a full breakdown.

Across institutions, the journal submissions data for Mathematical Sciences are skewed: most journals were published in by only a handful of unique institutions, but there were a small number of journals popular with nearly all of the institutions assessed by the sub-panel.
See Figure~\ref{fig:submissions}b.

\hypertarget{physics}{%
\subsection{Physics}\label{physics}}

We now turn to Physics, with 6446 REF2014 outputs, of which 6376 were journal articles in 304 unique journals, which we might expect to have some overlap with the Mathematical Sciences.
Indeed, some 100 journal titles appear in both submissions.

There were 41 different institutions who submitted to the Physics sub-panel for REF2014.

As with Mathematical Sciences, in Physics we used a cut-off of 30 articles for a publication to be `named' in the model, rather than aggregated under `Other journals'.

The distribution of journals by article count and across institutions, shown in Figure~\ref{fig:submissions}, appear largely similar to the aforementioned subjects, but the breakdown of article counts by journal in Table~\ref{tab:physicsJournals} reveals that two journals---\emph{Physical Review Letters} and \emph{Monthly Notices of the Royal Astronomical Society}---were extremely strongly represented, constituting nearly 30\% of all outputs.

\hypertarget{chemistry}{%
\subsection{Chemistry}\label{chemistry}}

Our fourth unit of assessment to model is Chemistry.
The data for this field comprise 37 institutions, who submitted 4698 outputs, of which 4688 were journal articles in 326 unique journals.

The distributions of submissions, shown in Figure~\ref{fig:submissions}, once again look similar to the other fields.
As in Physics, a couple of journals stand out for containing a very high proportion of outputs: the \emph{Journal of the American Chemical Society} and \emph{Angewandte Chemie} together represent nearly 25\% of all submitted works.

\hypertarget{results}{%
\section{Results}\label{results}}

Figure~\ref{fig:alphaDensity} represents the posterior marginal density for the parameter \(\alpha\), defined in \eqref{eq:envir} as the effect of institutions' research environments---rather than journal submissions---on the probability of their outputs attaining 4* ratings in the REF.

For Economics and Econometrics, Figure~\ref{fig:alphaDensity}a suggests there is little evidence for the environmental effect being distinct from zero, either when estimating the probabilities of journals attaining 4* or \(\geq3^*\) ratings in the REF.
The same was found also for Mathematical Sciences, Physics and Chemistry.
This simple diagnostic check---for sensitivity of the results when controlling for an institution-level covariate---provides some, albeit limited, reassurance: the results appear robust to potential effects of aggregation bias, and there is no indication from this check of anything like Simpson's paradox.

Trace plots for the Hamiltonian Monte Carlo runs are given in Figure~\ref{fig:traceplots}, and suggest good mixing of the chains for each of the parameters.

To catch any glaring errors in the results, and for a more informed interpretation of the findings (especially the implied journal rankings in each field) the initial results were presented to several senior University of Warwick academics with expertise in their respective disciplines.
This was invaluable, for example, in spotting the conspicuous absence of \emph{Annals of Mathematics} from the rankings, due to the aforementioned coding error in the REF2014 data.
With such anomalies fixed, our informal panel of experts provided useful context for the final results, presented in the following sections.

\hypertarget{economics-econometrics-1}{%
\subsection{Economics \& Econometrics}\label{economics-econometrics-1}}

\begin{figure}
\subfloat[Probability of 4*\label{fig:econLeague-1}]{\includegraphics{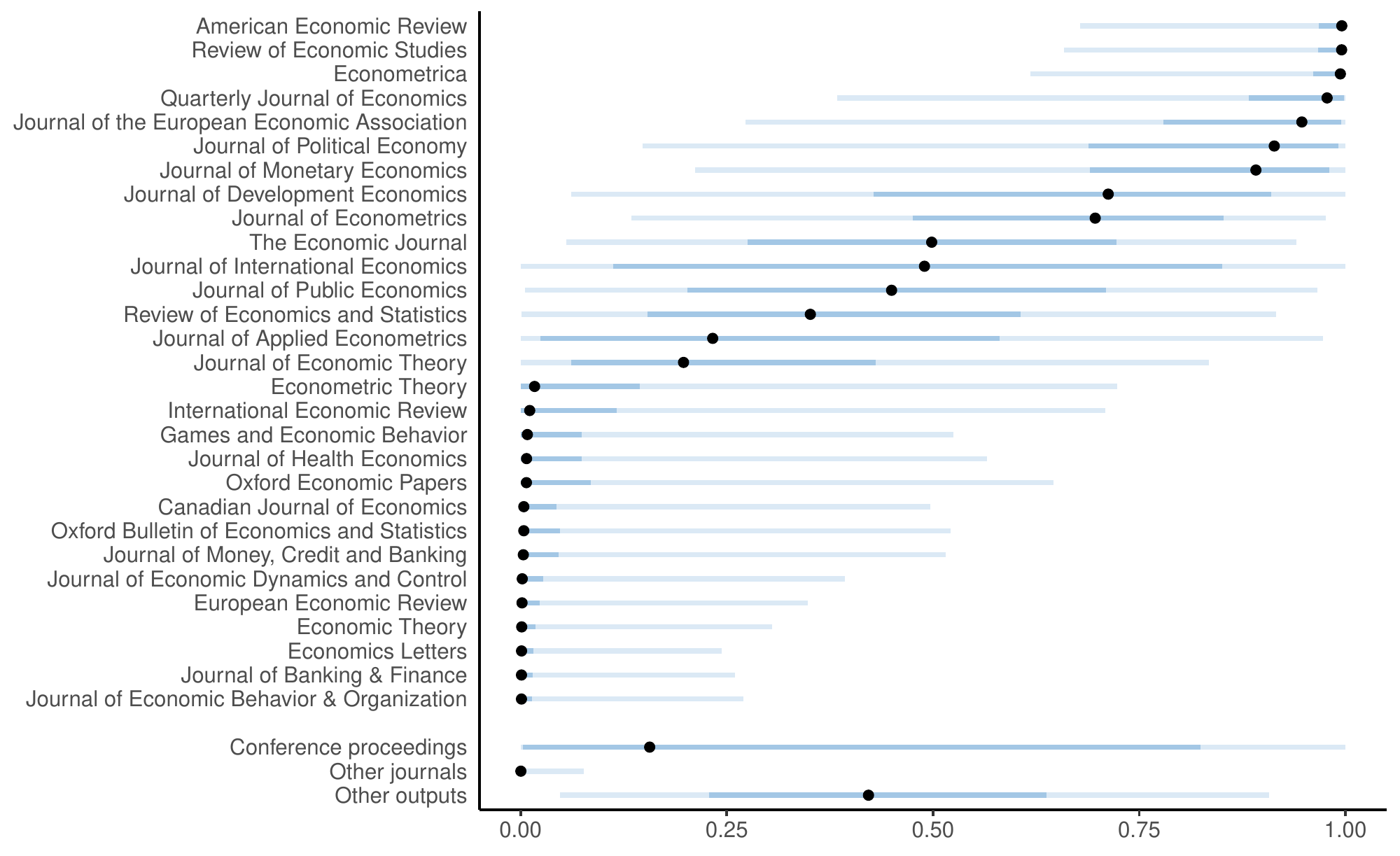} }\par\subfloat[Probability of 3* or 4*\label{fig:econLeague-2}]{\includegraphics{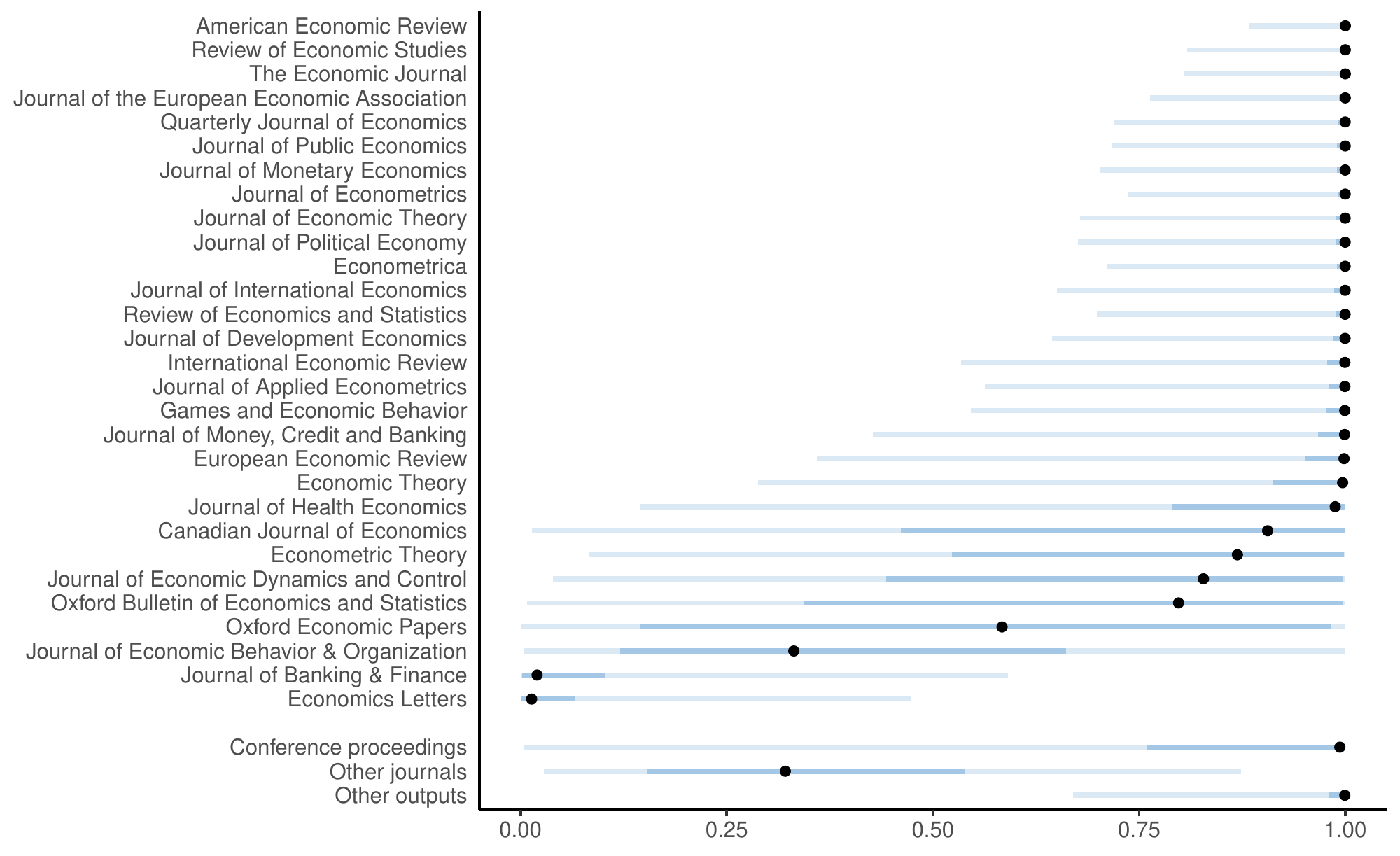} }\caption{Median estimated journal success probabilities in Economics and Econometrics. Shaded line segments represent 50\% and 95\% posterior intervals. Named journals had 20 or more articles submitted in REF2014}\label{fig:econLeague}
\end{figure}

Figure~\ref{fig:econLeague} provides a `league table', in the form of a series of box plots of the marginal posterior distributions, of the estimated Economics and Econometrics journal probabilities of attaining 4* and 3* or 4* ratings in the REF.
Strikingly, the five journals considered among economists to be the `Top Five' in their field (Heckman and Moktan 2018) are near the top of this ranking as well: namely, the \emph{American Economic Review}, \emph{Econometrica}, \emph{Quarterly Journal of Economics}, \emph{Review of Economic Studies}, and even the \emph{Journal of the Political Economy}, despite the latter only representing a handful of outputs, at 22 articles.
Looking at the probability of achieving 3* or greater (Figure~\ref{fig:econLeague}b), the top probabilities are all so close to 1 that little can be inferred from the ordering of the journals.

The 95\% posterior intervals are quite wide, especially for publications with fewer articles submitted in the REF, which is to be expected given the inherent uncertainty associated with estimating a large number of parameters from a small number of incomplete observations.

Our journal ranking has several notable omissions: the \emph{Journal of Labor Economics} and the \emph{RAND Journal of Economics} are highly respected (Sgroi 2019; Oswald 2019), as are general science journals such as \emph{Science}, \emph{Nature} and \emph{PNAS}.
However none of these journals met the minimum threshold of 20 articles submitted to the REF2014 Economics \& Econometrics sub-panel, so they do not appear as `named journals' in our results.

\begin{figure}
\centering
\includegraphics{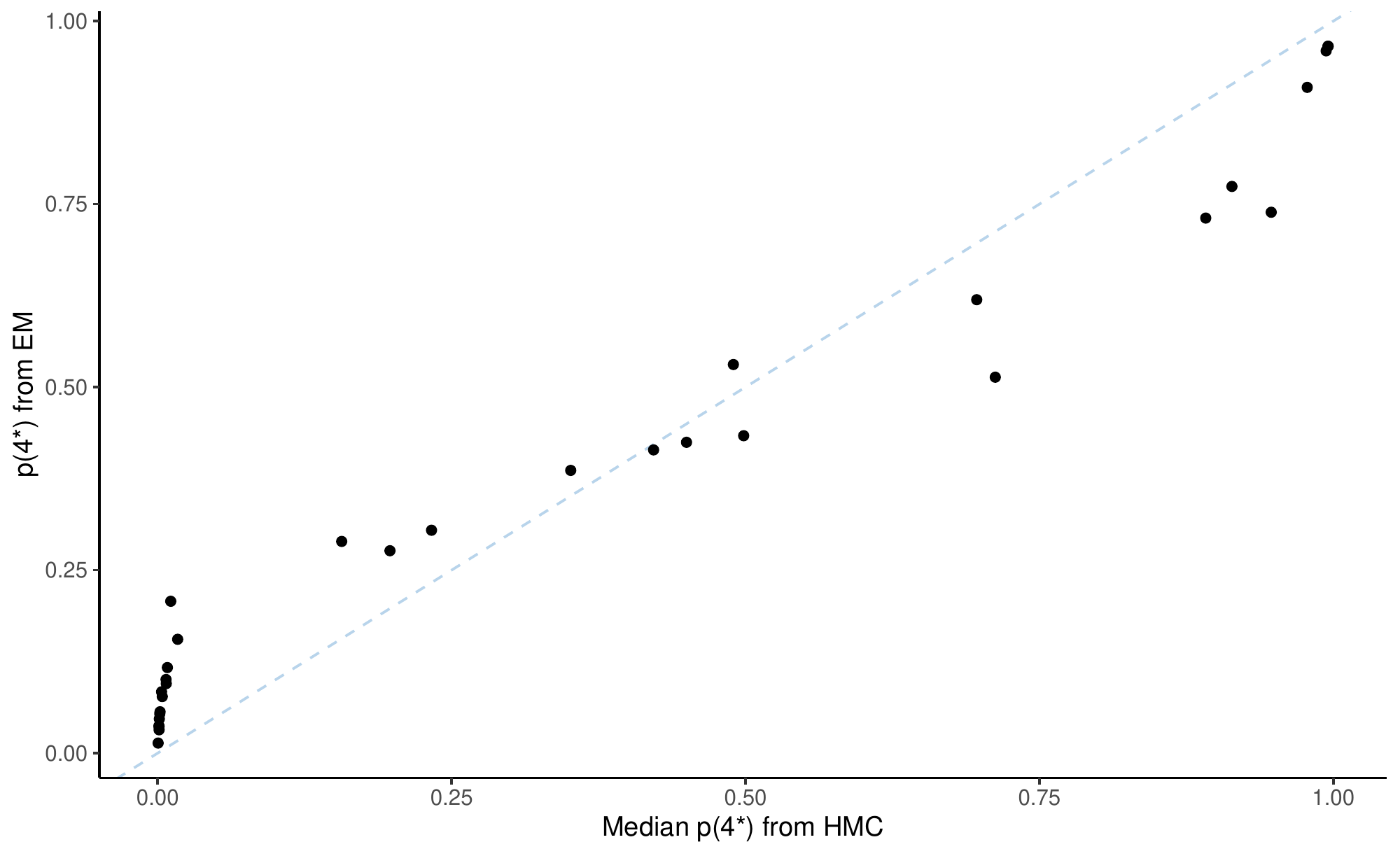}
\caption{\label{fig:EMvHMC}Maximum likelihood estimates of journal effects, \(\hat\beta_j\), versus Hamiltonian Monte Carlo estimates of journal success probabilities (on a logit scale), for Economics and Econometrics, with line of best fit}
\end{figure}

As a robustness check, Figure~\ref{fig:EMvHMC} compares Hamiltonian Monte Carlo estimated journal probabilities of attaining 4* with the respective maximum likelihood estimates computed via the expectation--maximization algorithm (with the level of pseudo-data set (arbitrarily) at one article per journal).
The maximum likelihood estimates come without any uncertainty quantification, but we can see a strong correlation in point estimates between the \(\beta_j\) estimates of Equation \eqref{eq:rasch} and the (logit) success probabilities corresponding to the same journals, so the general approach seems sound.

\begin{figure}
\subfloat[\% articles at 4*\label{fig:econPredictions-1}]{\includegraphics{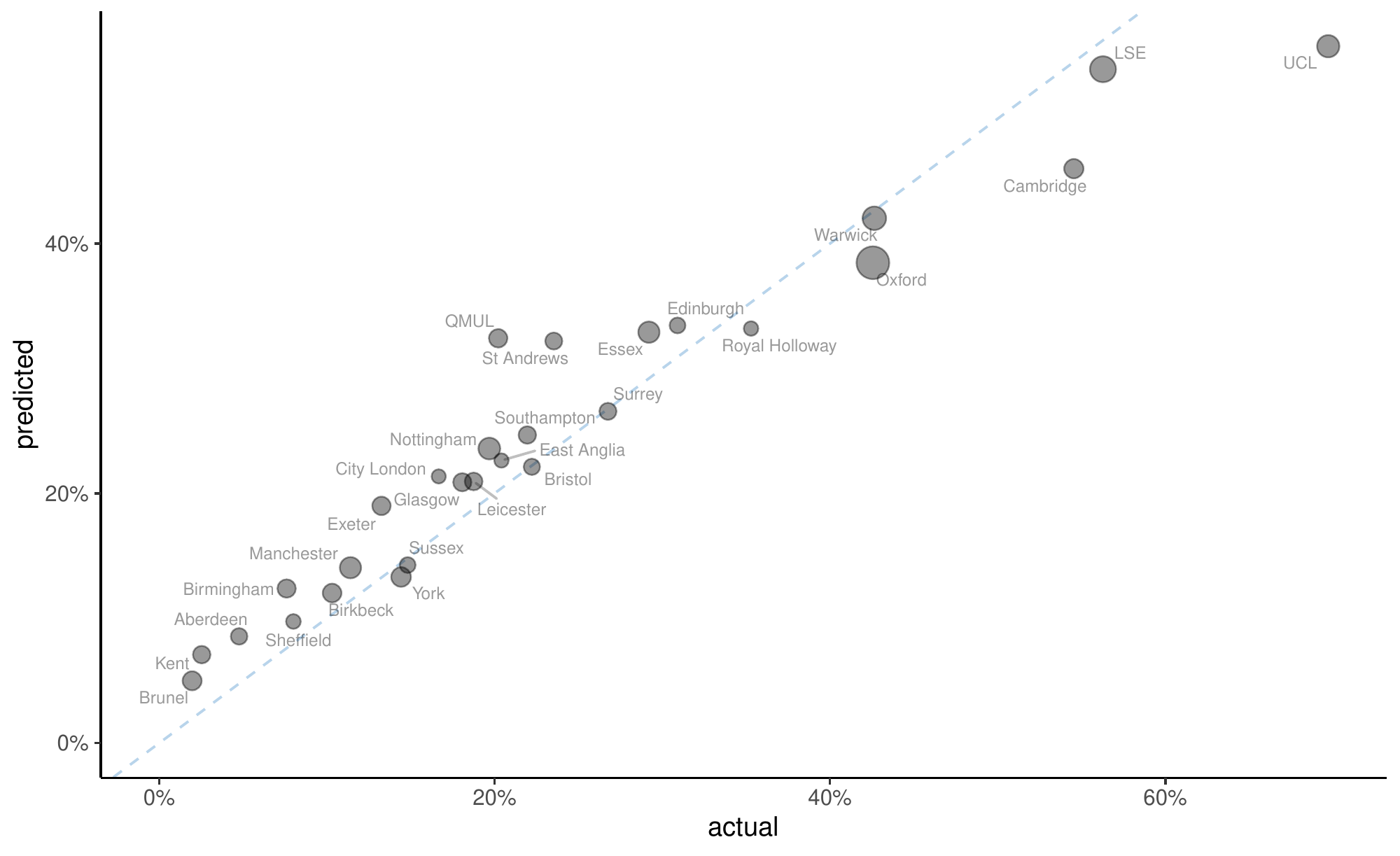} }\newline\subfloat[funding allocation\label{fig:econPredictions-2}]{\includegraphics{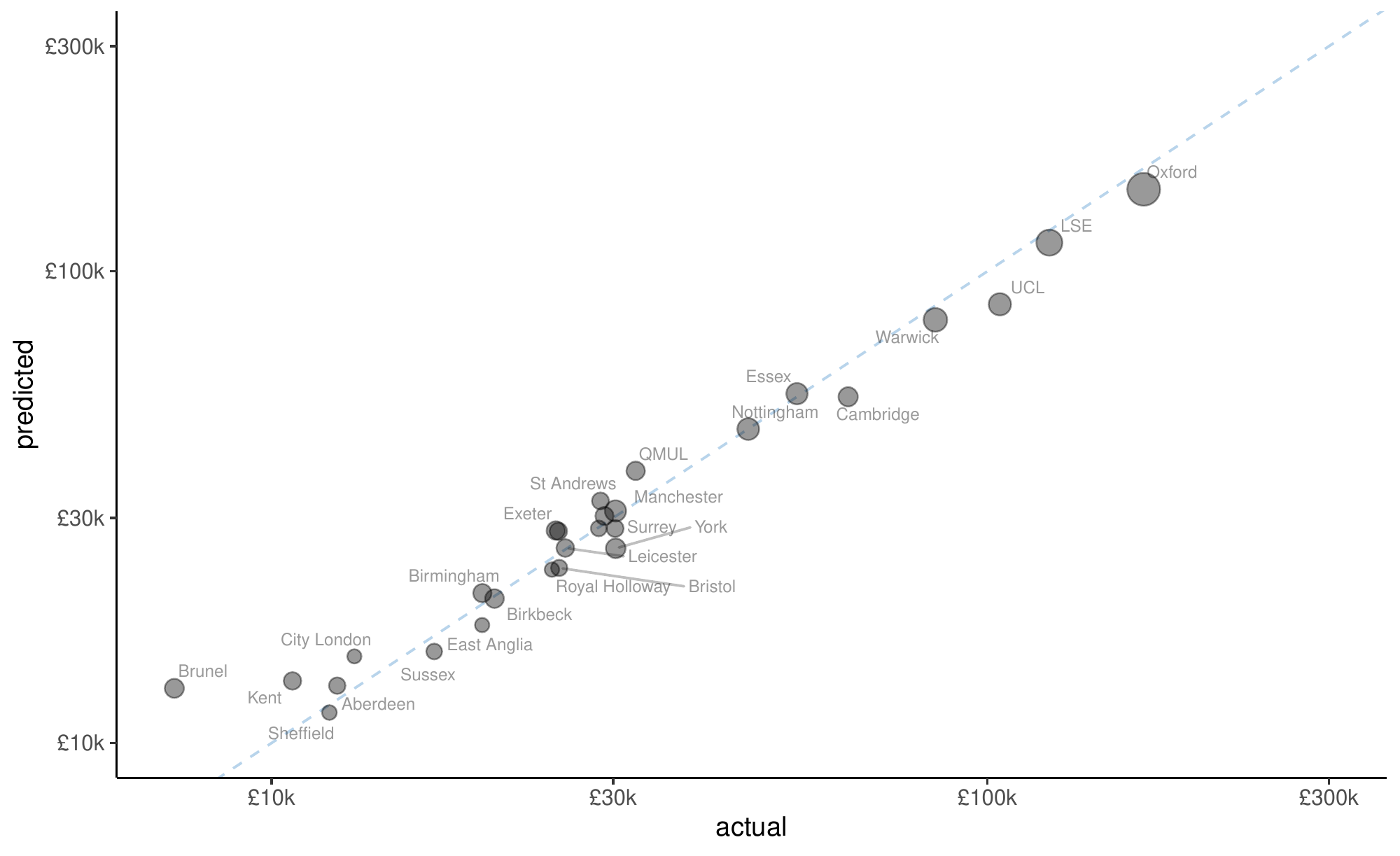} }\caption{Predictions versus observed REF2014 results for institutions submitting outputs to the Economics \& Econometrics sub-panel, with point sizes proportional to number of FTE staff}\label{fig:econPredictions}
\end{figure}

Figure~\ref{fig:econPredictions}a shows the predicted versus actual 4* output profiles for each of the institutions in Economics and Econometrics.
With the predicted 4* and 3* profiles converted into funding allocations, Figure~\ref{fig:econPredictions}b shows the resulting discrepancies between the predicted institutional funding versus that actually allocated (based on the methodology of Koya and Chowdhury 2017) based on HEFCE data.
Not all institutions are based in England, of course, so the `actual' funding figures for other nations in the UK assume that the respective research councils used similar formulae to allocate funding based on REF2014 outputs.

The quality of prediction appears reasonably good, with most points falling close to the line of \(y = x\).
Some institutions appear to have received more 4* ratings than predicted from their journal choices, notably Cambridge and UCL, and Queen Mary University London appears to have received fewer 4* ratings than suggested by the model.
Otherwise there are no noticeable outliers.

When it comes to funding, Figure~\ref{fig:econPredictions}b shows how much funding would be allocated, when combining the estimated 4* profiles with 3* ratings and the FTE headcount for each department.
The only institution with a significant discrepancy is Brunel, and this can be accounted for by the fact that most of that university's Economics and Econometrics outputs were published in less popular journals not named in Table~\ref{tab:econJournals}, so the model has less information available to predict this institution's results.

\begin{figure}
\subfloat[Index of dissimilarity\label{fig:indicesRidges-1}]{\includegraphics{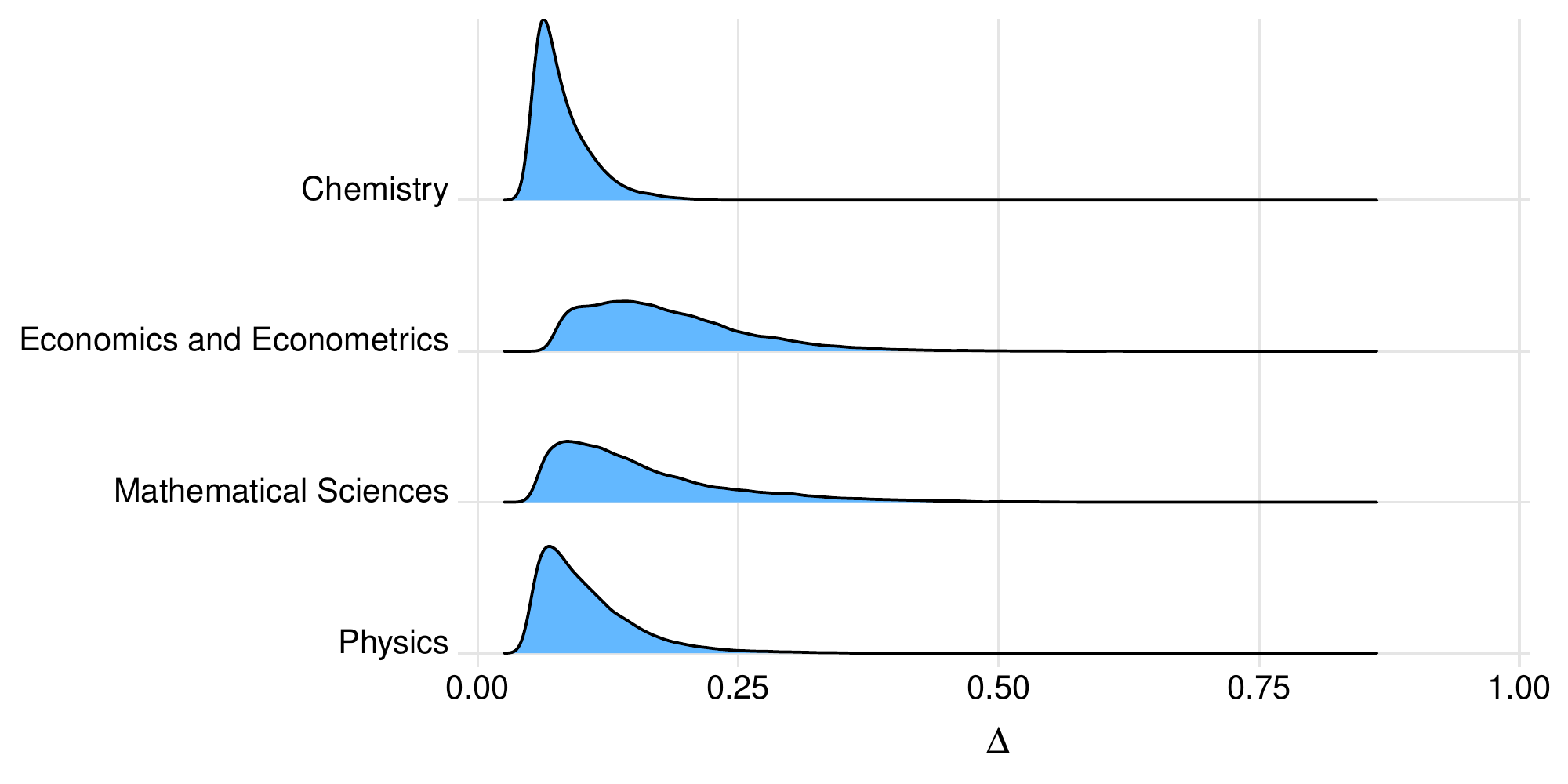} }\par\subfloat[Index of redistribution of monetary reward\label{fig:indicesRidges-2}]{\includegraphics{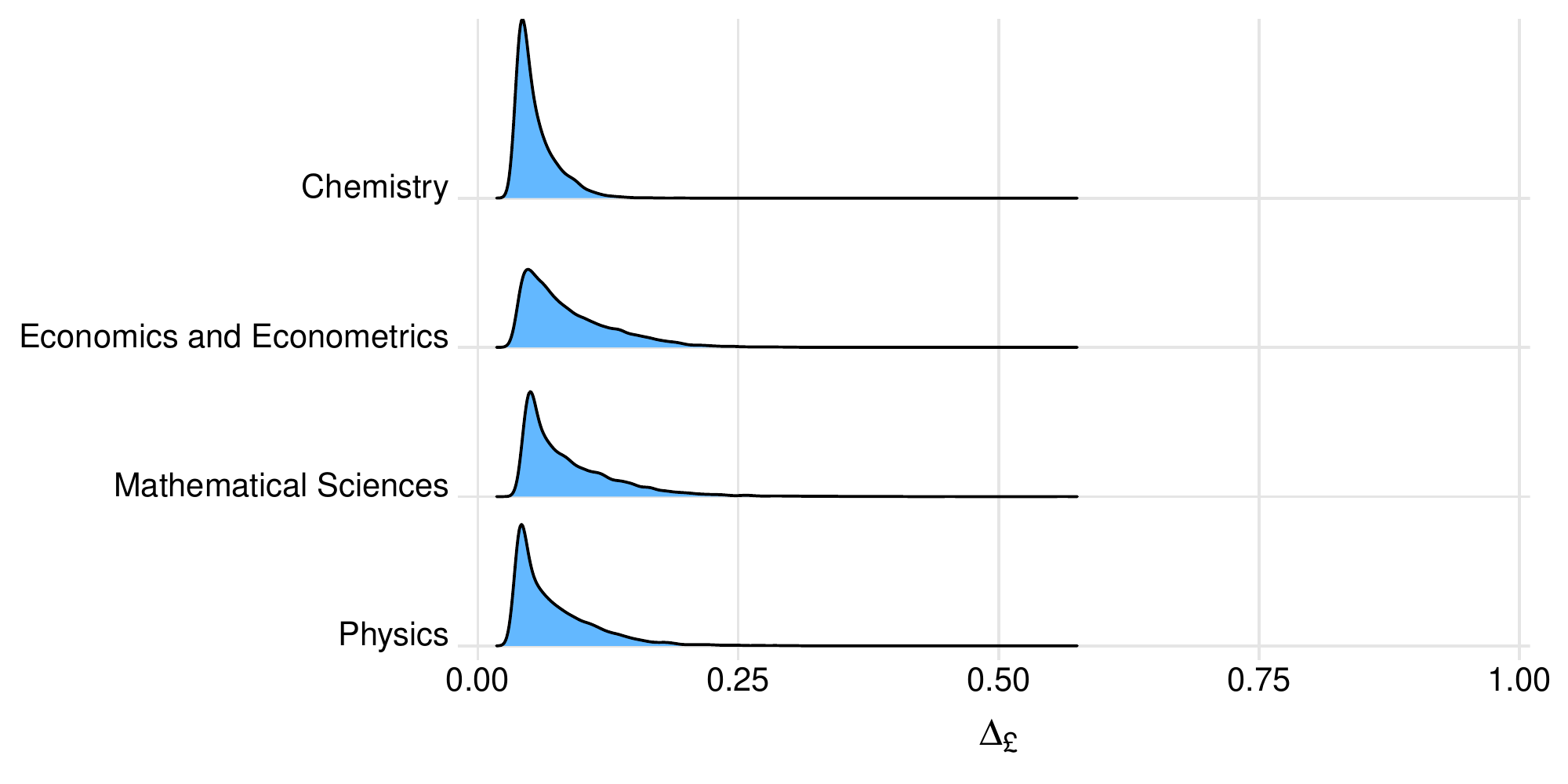} }\caption{Density plots of indices of dissimilarity and of redistribution of monetary reward, by unit of assessment}\label{fig:indicesRidges}
\end{figure}

Across the four fields, we compute the index of dissimilarity, \(\Delta\), and the index of redistribution of monetary reward, \(\Delta_\pounds\).
The distributions of these metrics are plotted in Figure~\ref{fig:indicesRidges}.
Lower numbers are better.

The median value for Economics and Econometrics is \(\Delta = 17.9\%\), that is, this proportion of articles would need to be reclassified for the estimated institutional profiles to match exactly those published in the REF.
As a metric, 82.1\% accuracy sounds like it might be quite good, but we should be careful not to draw too many conclusions from a single number.
In funding terms, that translates to \(\Delta_\pounds = 8.7\%\) of funding in Economics and Econometrics needing to be reallocated if an initial allocation was made based on the Poisson binomial model alone.
Across dozens of institutions, that represents a substantial amount of money, though.

\begin{figure}
\centering
\includegraphics{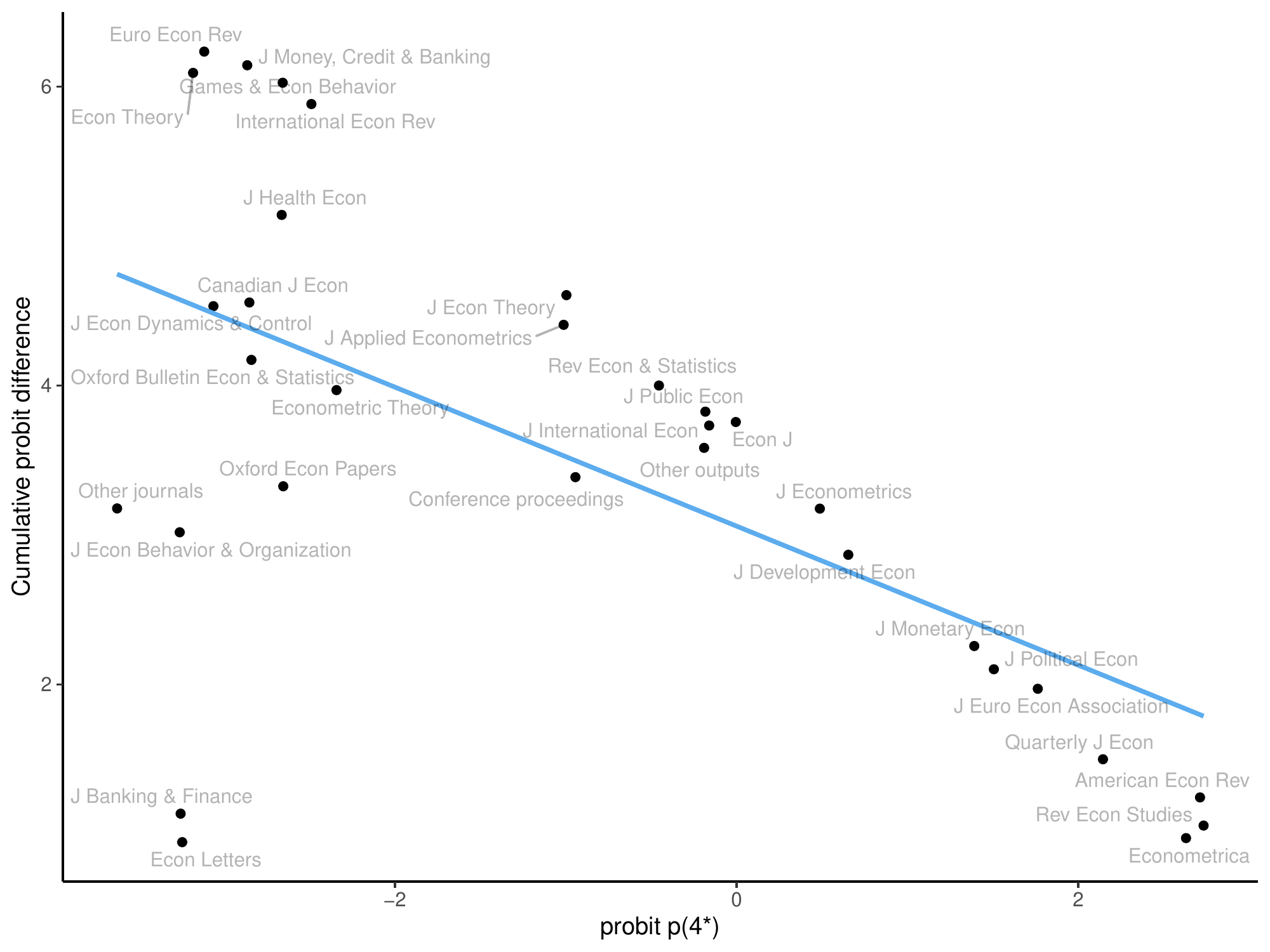}
\caption{\label{fig:Yan}Comparison of cumulative probit differences, \(c_j = \operatorname{probit}(p_j^{34}) - \operatorname{probit}(p_j^4)\), versus estimated probit probability of attaining 4*, by journal in Economics and Econometrics in REF2014, with line of best fit. A non-zero slope implies \(c_j \neq c\), that the cumulative probit difference is not constant across journals}
\end{figure}

Figure~\ref{fig:Yan} provides evidence against the model of Yan (2017), which assumed a constant cumulative probit difference between the probability of getting 4* and the probability of getting 3* or better.
It is clear that `better' journals (those more likely to attain 4*) have a smaller cumulative probit difference, suggesting that it is not much harder for an output in such an apparently high-achieving journal to get a 4* than a 3* rating, whereas for `weaker' journals, it is harder to improve from 3* to 4*.

\hypertarget{mathematical-sciences-1}{%
\subsection{Mathematical Sciences}\label{mathematical-sciences-1}}

\begin{figure}
\centering
\includegraphics{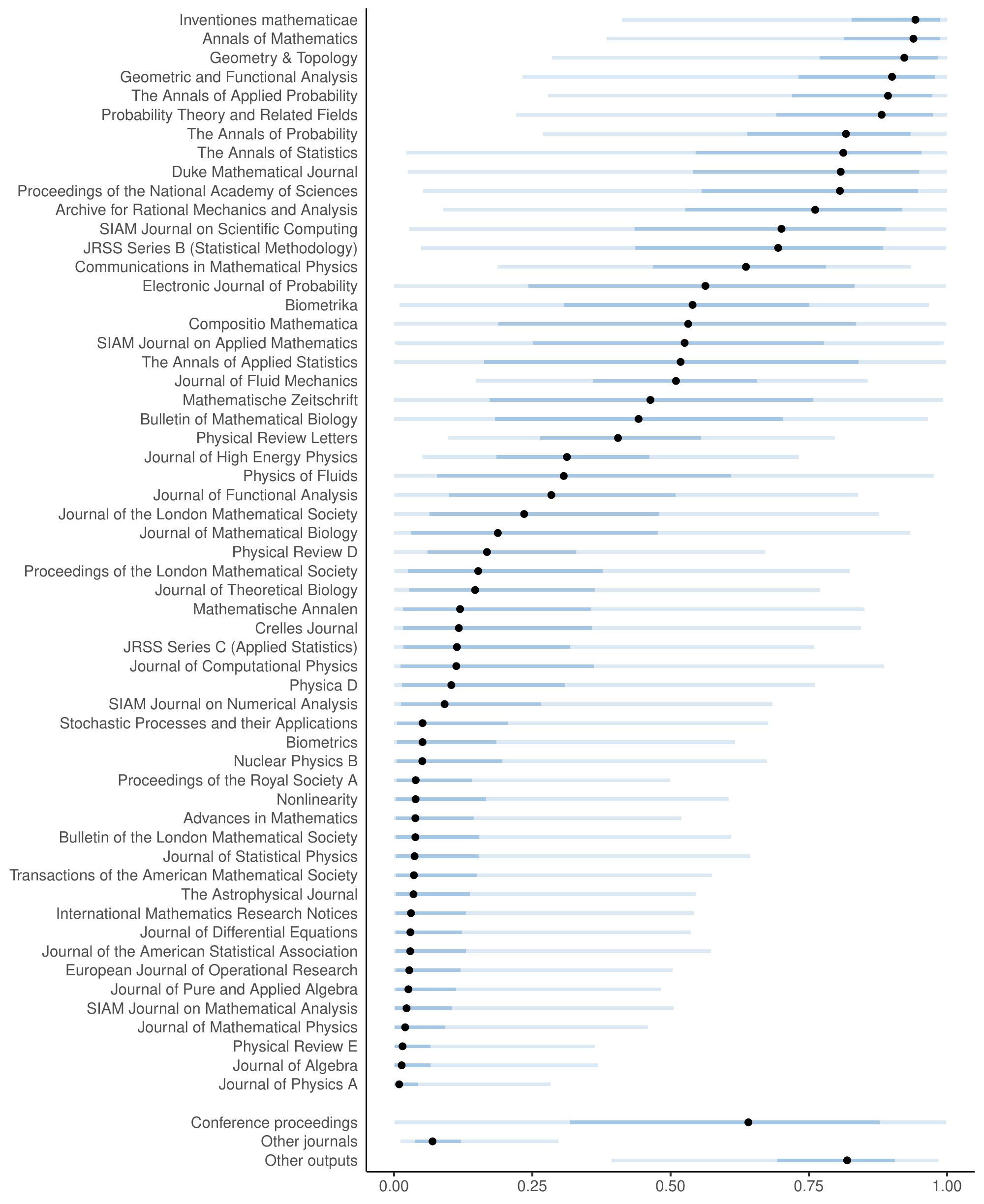}
\caption{\label{fig:mathLeague}Median estimated journal success probabilities of 4* ratings in Mathematical Sciences. Shaded line segments represent 50\% and 95\% posterior intervals. Named journals had 30 or more articles submitted in REF2014}
\end{figure}

In Mathematical Sciences, we face the problem of several partly disjointed sub-fields, such as pure mathematics, statistics, mathematical physics and mathematical biology, all falling under the same umbrella.
As a result it is harder to gauge what might be considered a group of `top' mathematical sciences journals---mathematicians might declare that statistics, for example, is merely applied mathematics and that a pure mathematics journal should lead the field (Monroe 2008) whilst statisticians might counter that statistical journals should come top because of the widespread application of statistics.
It is perhaps surprising, then, that some of the reputed top journals in statistics, \emph{Annals of Statistics}, \emph{Biometrika} and the \emph{Journal of the Royal Statistical Society: Series B} (Varin, Cattelan, and Firth 2016) are still ranked highly based on the model for attaining 4* ratings in the REF.
See Figure~\ref{fig:mathLeague}.
However, the \emph{Journal of the American Statistical Association}, also a highly-regarded statistics journal, has a low estimated probability of obtaining 4* ratings.

For the mathematicians, \emph{Inventiones Mathematicae} and \emph{Annals of Mathematics} are both highly reputed and have the highest estimated probabilities of yielding 4* ratings in the REF.
The \emph{Journal of the American Mathematical Society} and \emph{Publications Mathémathiques de l'IHÉS} are also highly regarded (Loeffler 2019), but do not appear in the results as named journals because fewer than 30 of their respective articles were submitted to the REF.

\begin{figure}
\centering
\includegraphics{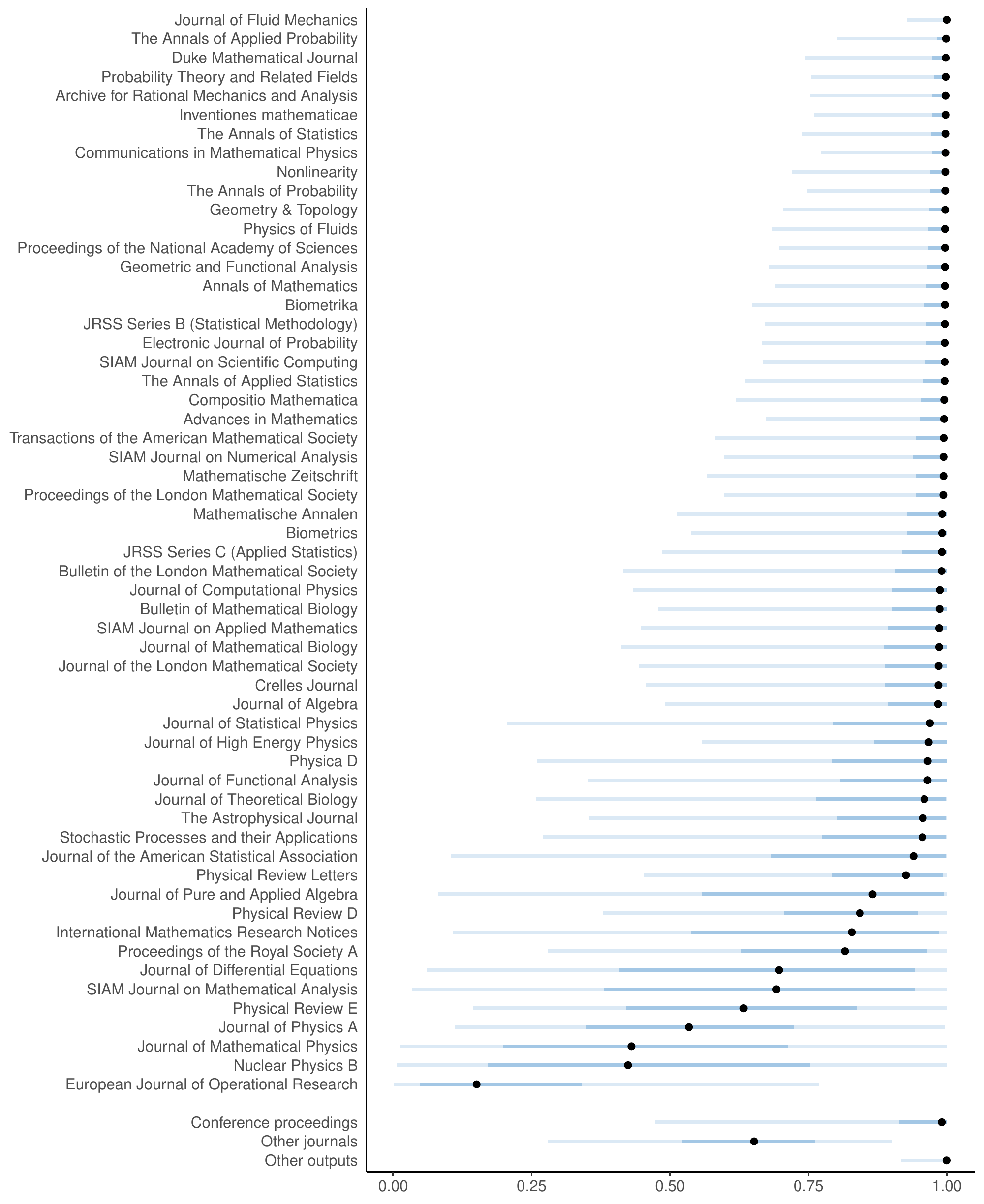}
\caption{\label{fig:mathLeague2}Median estimated journal success probabilities of 3* or 4* ratings in Mathematical Sciences. Shaded line segments represent 50\% and 95\% posterior intervals. Named journals had 30 or more articles submitted in REF2014}
\end{figure}

\begin{figure}
\subfloat[\% articles at 4*\label{fig:mathPredictions-1}]{\includegraphics{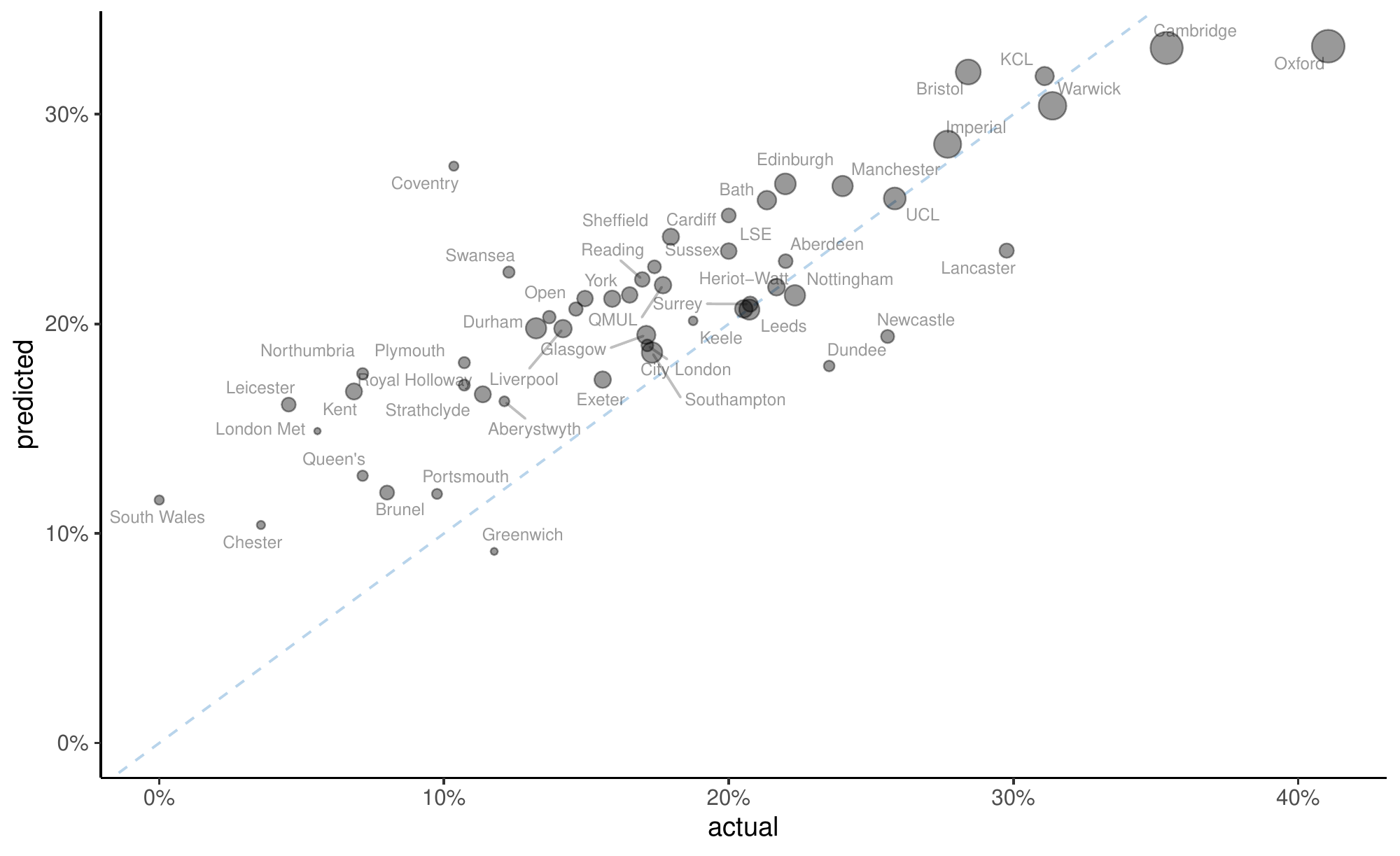} }\par\subfloat[funding allocation\label{fig:mathPredictions-2}]{\includegraphics{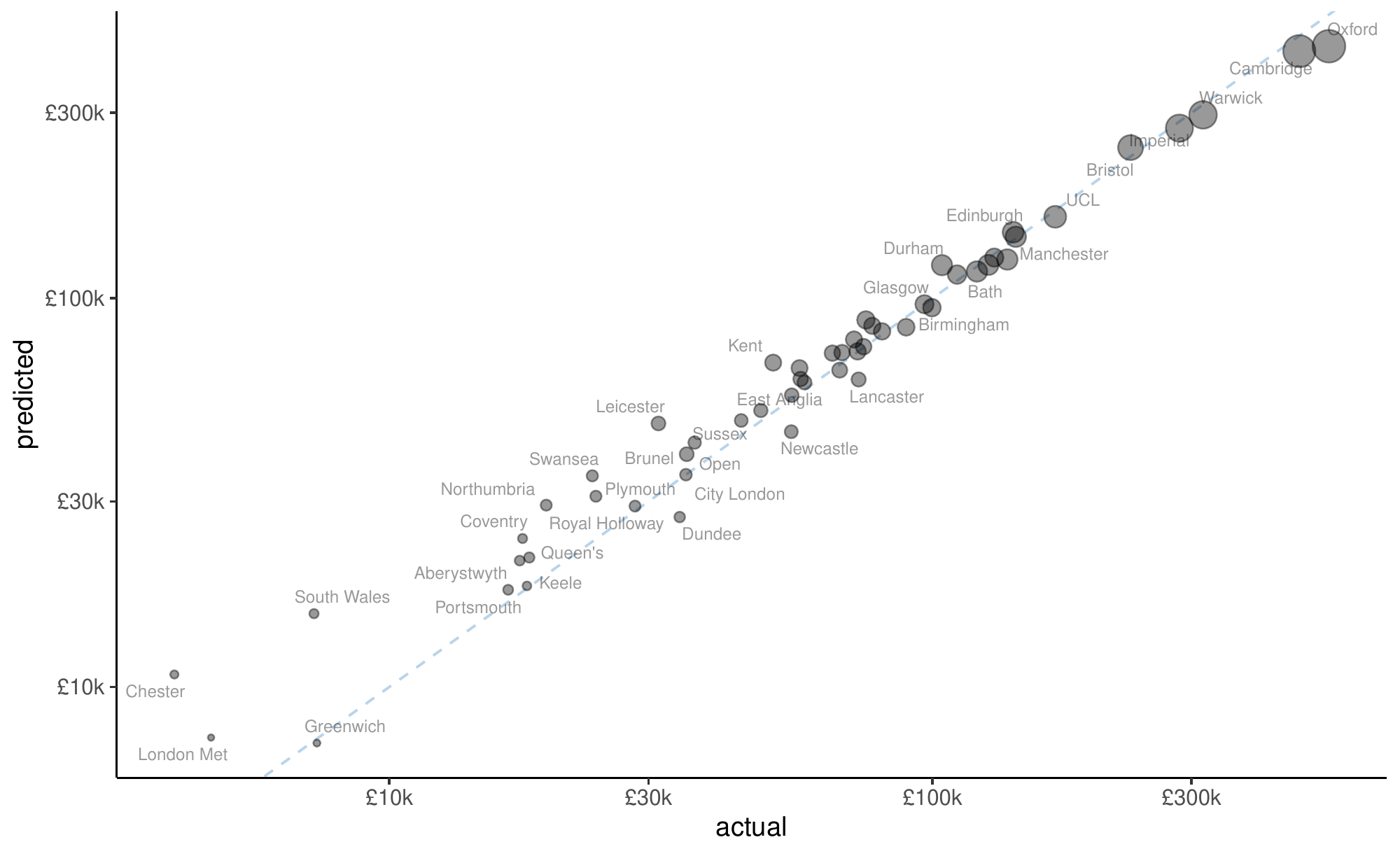} }\caption{Predictions versus observed REF2014 results for institutions submitting outputs to the Mathematical Sciences sub-panel, with point sizes proportional to number of FTE staff}\label{fig:mathPredictions}
\end{figure}

In Mathematical Sciences, the predicted versus allocated 4* ratings and funding allocations, by institution, are presented in Figure~\ref{fig:mathPredictions}.
Apparent outliers (such as Coventry University or the University of Greenwich in the 4* plot) are among the smallest institutions by number of full-time equivalent (FTE) research staff.
There appears to be a pattern, however: weaker institutions are expected to do better, and stronger institutions are expected to do worse, than their actual published performance in the REF.

This shrinkage effect implies that some variation in assessed quality of outputs is not explained by journal identities alone.
It indicates that there is variation in quality within at least some journals, and that high-ranked institutions tend to publish more of the high-quality papers in such journals.

In terms of summary measures, the median index of dissimilarity for Mathematical Sciences is \(15.5\%\) and the median required redistribution of monetary reward is \(8.9\%\); the posterior distributions of these statistics are plotted in Figure~\ref{fig:indicesRidges}.

\hypertarget{physics-1}{%
\subsection{Physics}\label{physics-1}}

\begin{figure}
\subfloat[Probability of 4*\label{fig:physLeague-1}]{\includegraphics{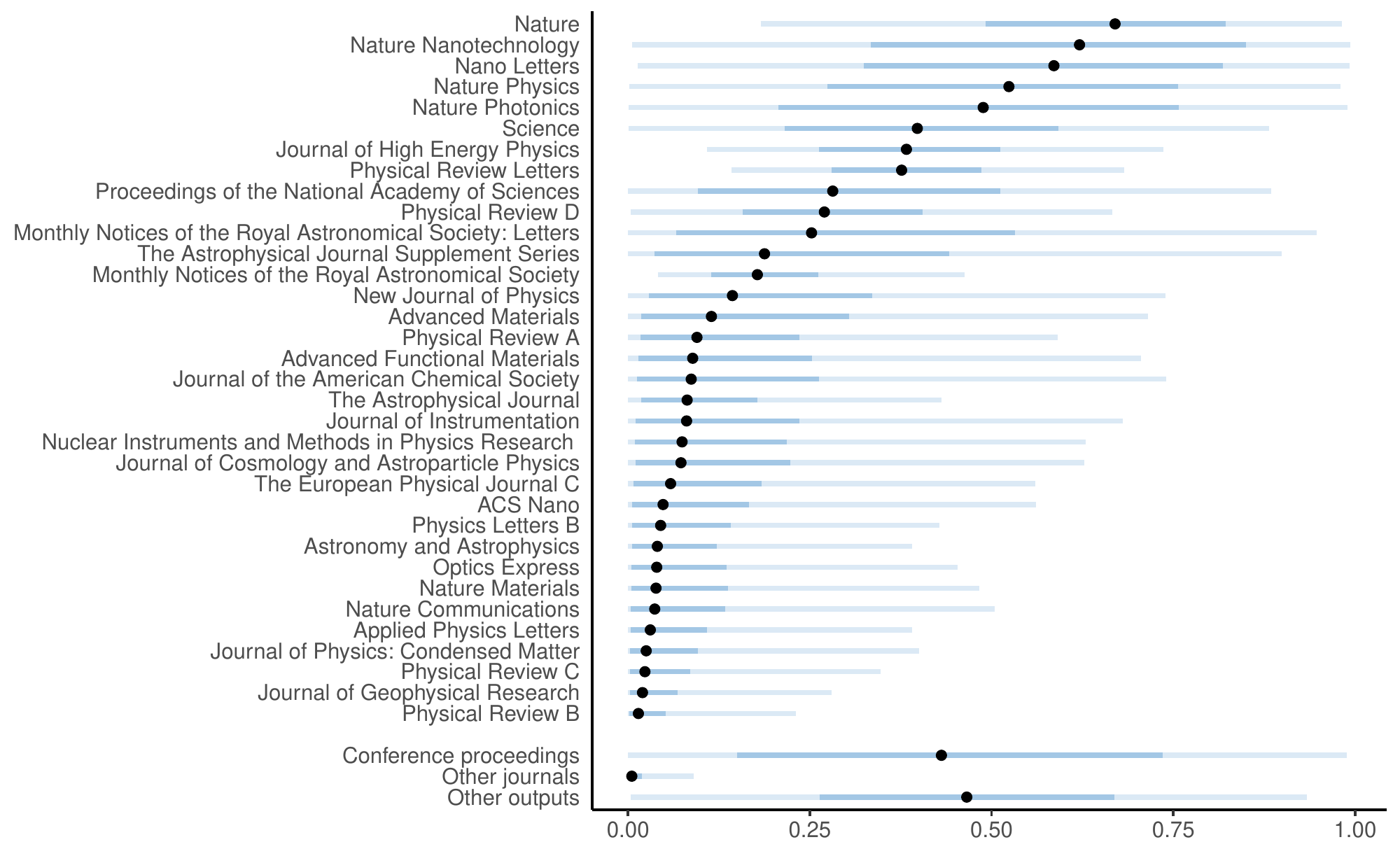} }\par\subfloat[Probability of 3* or 4*\label{fig:physLeague-2}]{\includegraphics{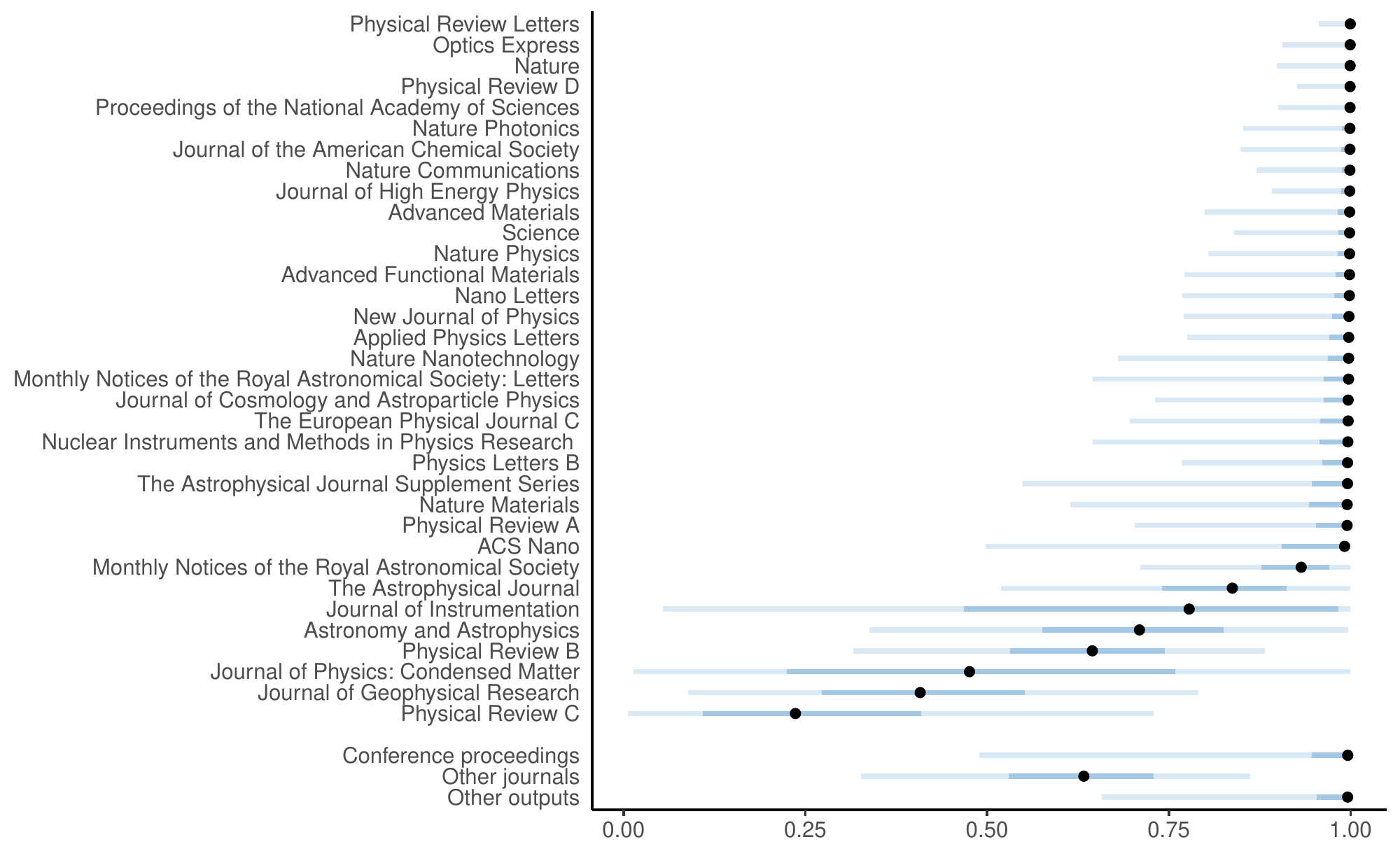} }\caption{Median estimated journal success probabilities of 4* ratings in Physics. Shaded line segments represent 50\% and 95\% posterior intervals. Named journals had 30 or more articles submitted in REF2014}\label{fig:physLeague}
\end{figure}

Posterior probabilities for the Physics sub-panel are presented in Figure~\ref{fig:physLeague}.
Journals from Nature Publishing Group have the highest estimated probabilities of attaining 4*, though no probabilities are near 100\%, perhaps owing to the relatively small number of 4* ratings awarded in this field generally.
The appearance of \emph{Physics Review Letters} above the \emph{Proceedings of the National Academy of Sciences (PNAS)} in the ranking might imply a preference by physicists in the review panel for physics-specific journals over general science ones.
In the international astrophysics community, \emph{Astrophysical Journal} might be considered more prestigious than \emph{Monthly Notices of the Royal Astronomical Society}, but the latter has a slightly higher estimated probability of 4*, which might be interpreted as a UK-centric bias (Ball 2019).
Relatively low success probabilities for \emph{Physics Review B} and \emph{C} could be attributed to an inter-journal dependence: namely, some works published in these journals also being announced in the highly-ranked \emph{Physical Review Letters}.

\begin{figure}
\subfloat[\% articles at 4*\label{fig:physPredictions-1}]{\includegraphics{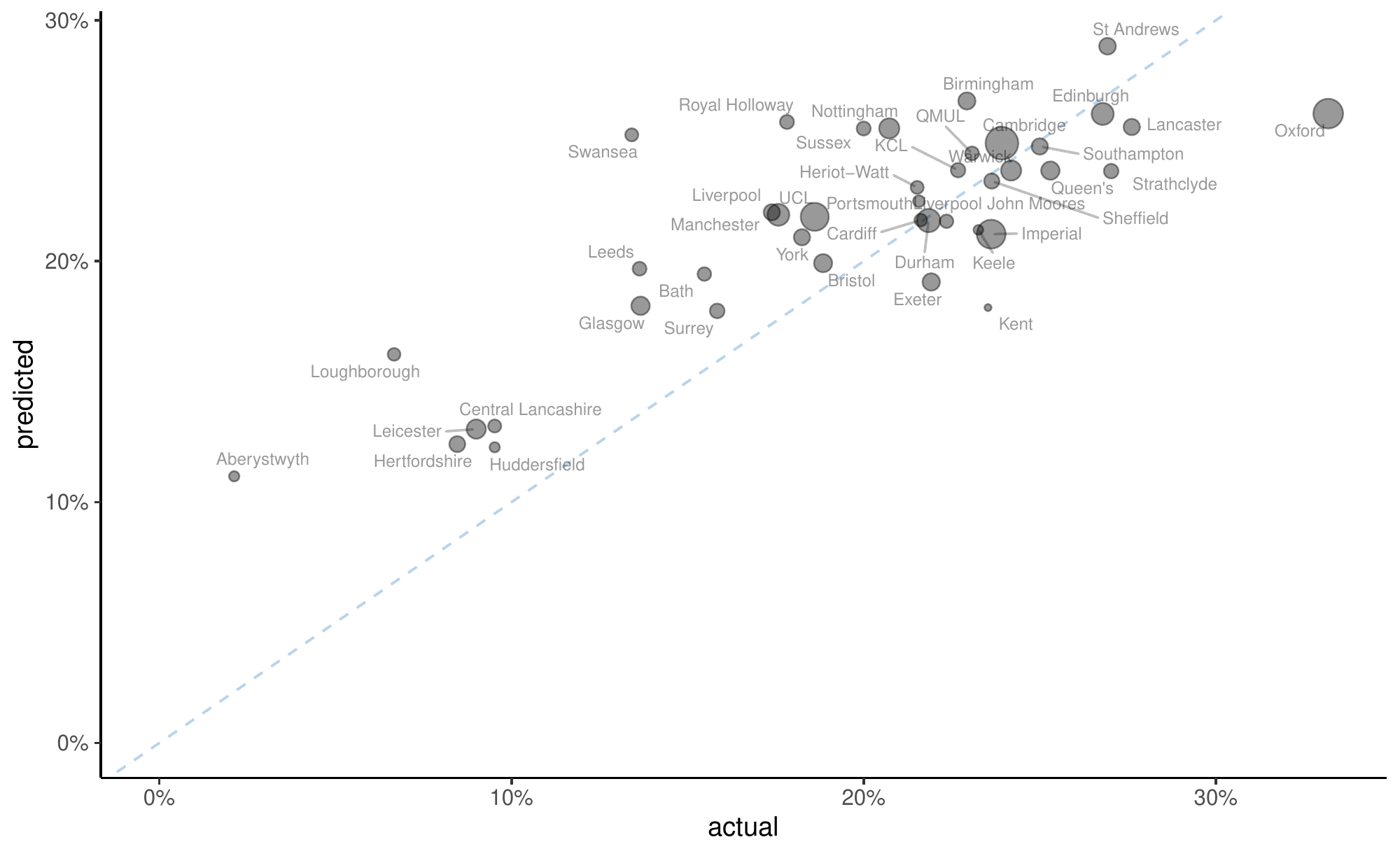} }\par\subfloat[funding allocation\label{fig:physPredictions-2}]{\includegraphics{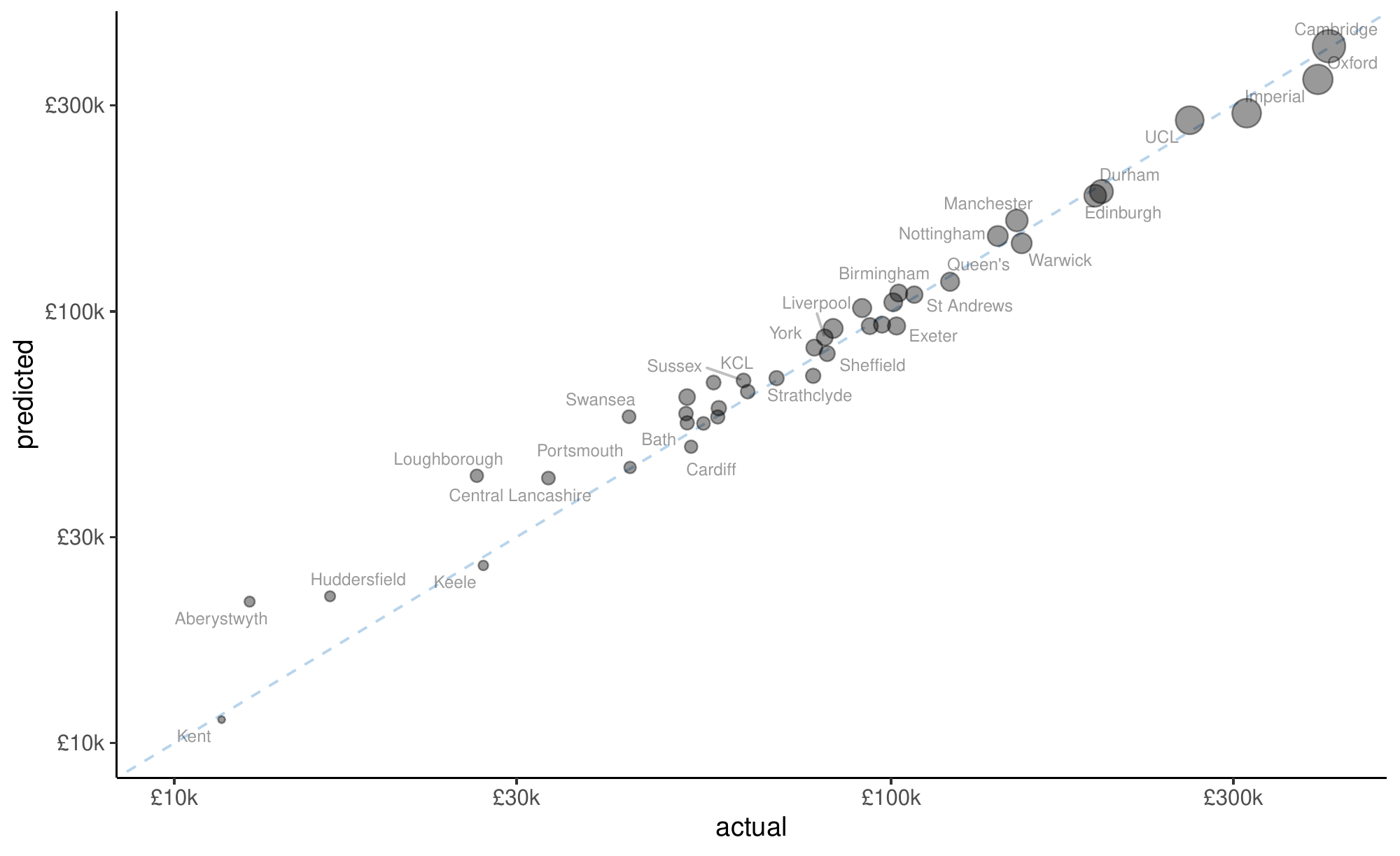} }\caption{Predictions versus observed REF2014 results for institutions submitting outputs to the Physics sub-panel, with point sizes proportional to number of FTE staff}\label{fig:physPredictions}
\end{figure}

As in Mathematical Sciences, a comparison of the predicted versus actual institutional REF results in Physics, shown in Figure~\ref{fig:physPredictions}, reveals a linear relationship, but an apparent shrinkage effect, implying some variation in assessed quality not explained by journal identities.
The performance of the University of Oxford, in particular, appears to be under-estimated by the model, suggesting that where there is variation of assessed quality within journals, the higher-quality outputs may be more likely to have been from Oxford researchers.

By summary measures, the median index of similarity in Physics is \(10.5\%\) and the median proportion of reallocated research funding is \(7.3\%\).

\hypertarget{chemistry-1}{%
\subsection{Chemistry}\label{chemistry-1}}

\begin{figure}
\subfloat[Probability of 4*\label{fig:chemLeague-1}]{\includegraphics[width=.8\linewidth]{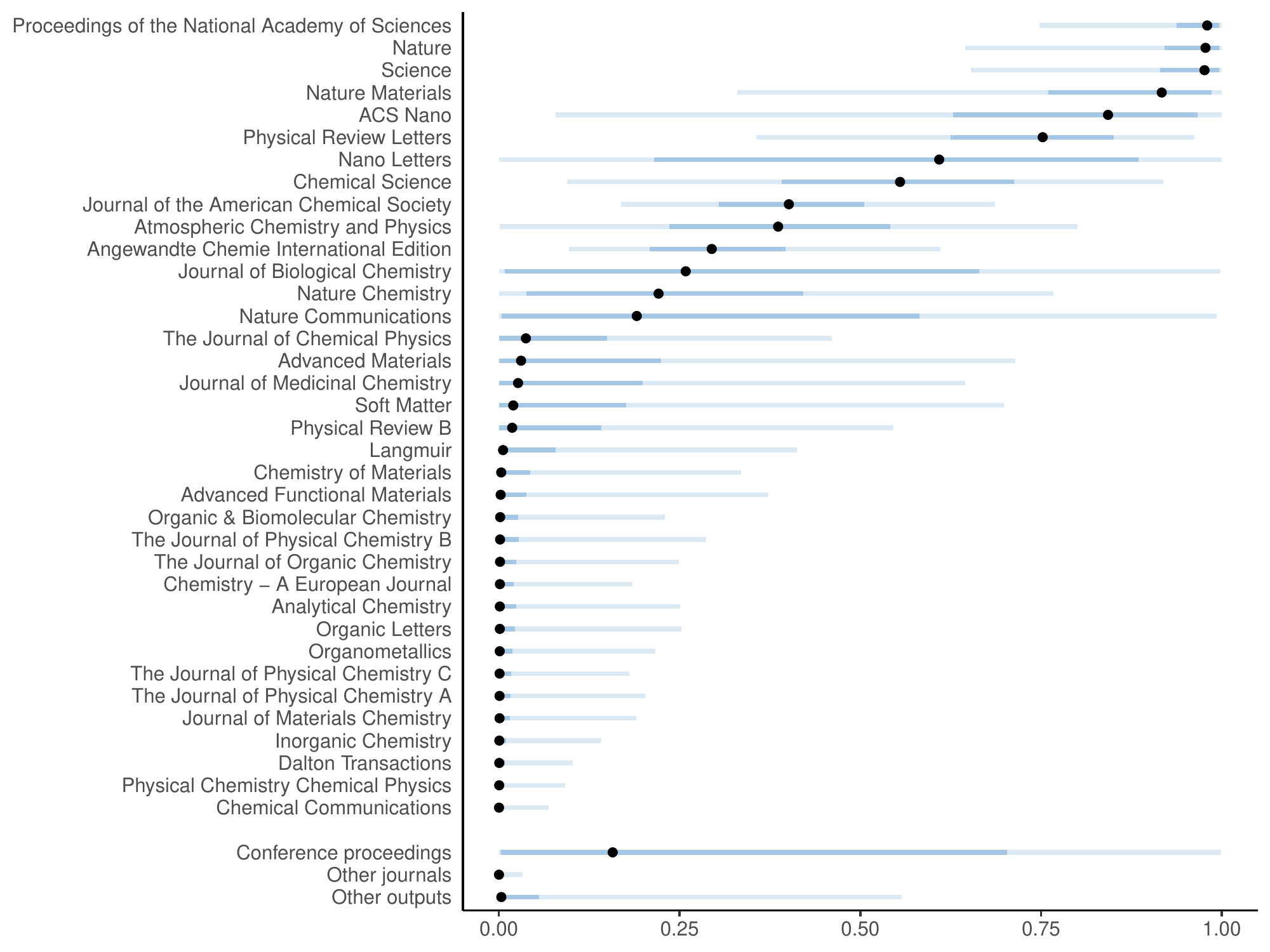} }\par\subfloat[Probability of 3* or 4*\label{fig:chemLeague-2}]{\includegraphics[width=.8\linewidth]{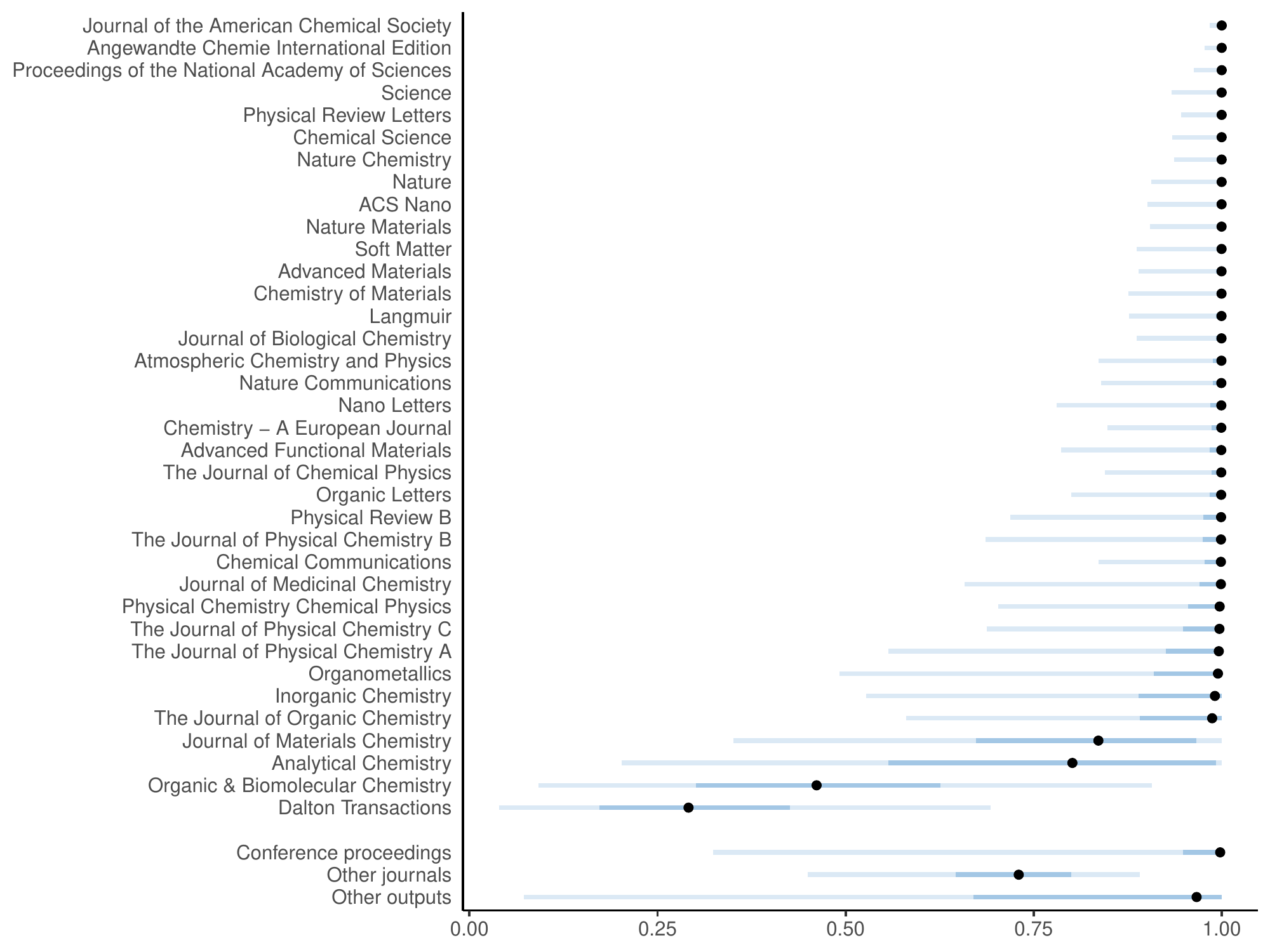} }\caption{Median estimated journal success probabilities of 4* ratings in Chemistry. Shaded line segments represent 50\% and 95\% posterior intervals. Named journals had 30 or more articles submitted in REF2014}\label{fig:chemLeague}
\end{figure}

Figure~\ref{fig:chemLeague} provides league tables of estimated journal REF success probabilities in Chemistry.
This field, unlike the others studied here, seems to be dominated by popular general science outlets, in \emph{PNAS}, \emph{Nature} and \emph{Science}, rather than dedicated chemistry journals.
There may be some dependence on types of articles published: some periodicals print different mixtures of `full' research papers and communications (letters).
\emph{Nature Chemistry} and \emph{Nature Communications} fall lower in the ranking than might be expected (Bugg 2019; Scott 2019).
This appears to be simply a result of work published in journals being submitted by several low-scoring institutions that were not awarded many 4* ratings in the REF.

\begin{figure}
\subfloat[\% articles at 4*\label{fig:chemPredictions-1}]{\includegraphics{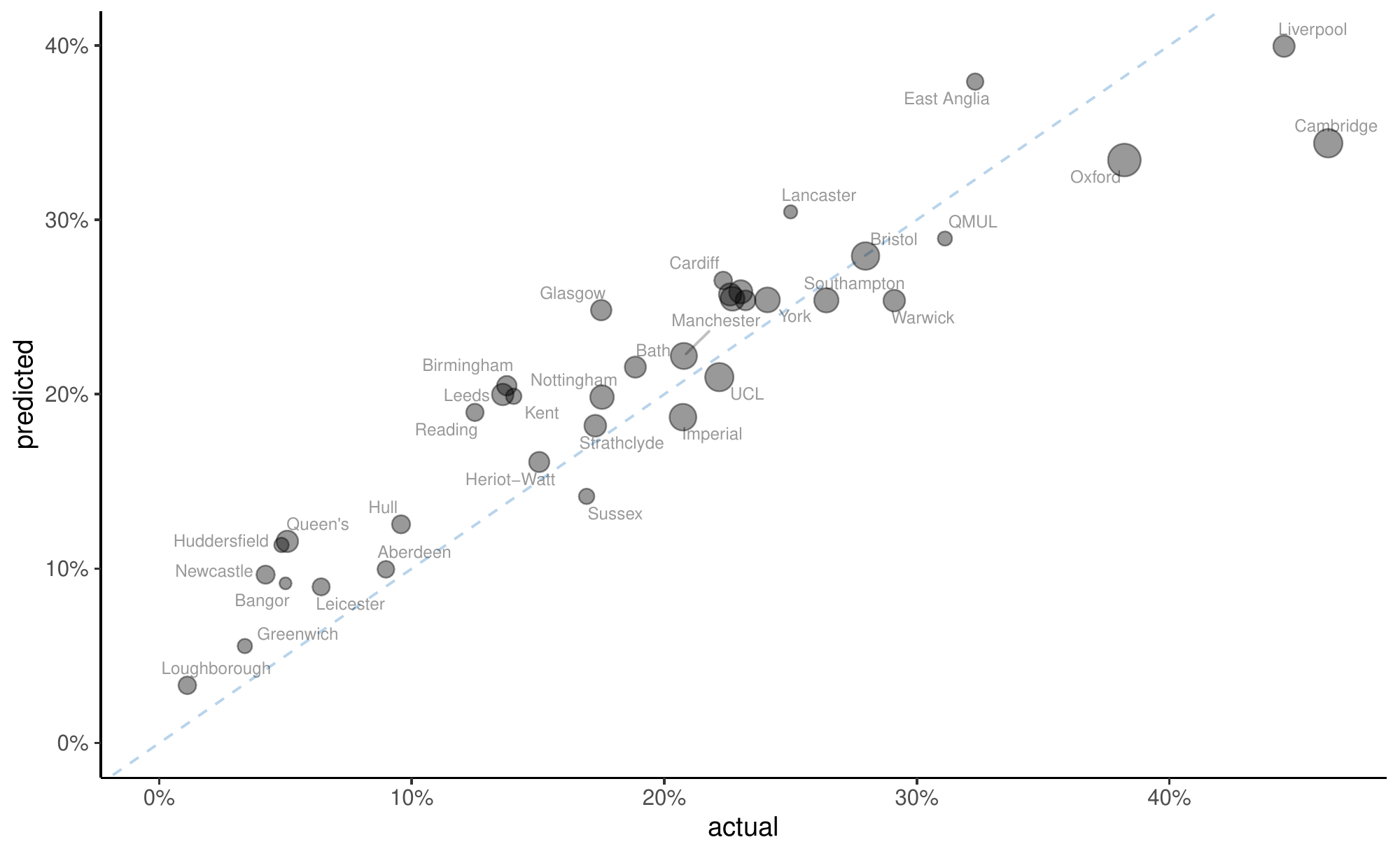} }\par\subfloat[funding allocation\label{fig:chemPredictions-2}]{\includegraphics{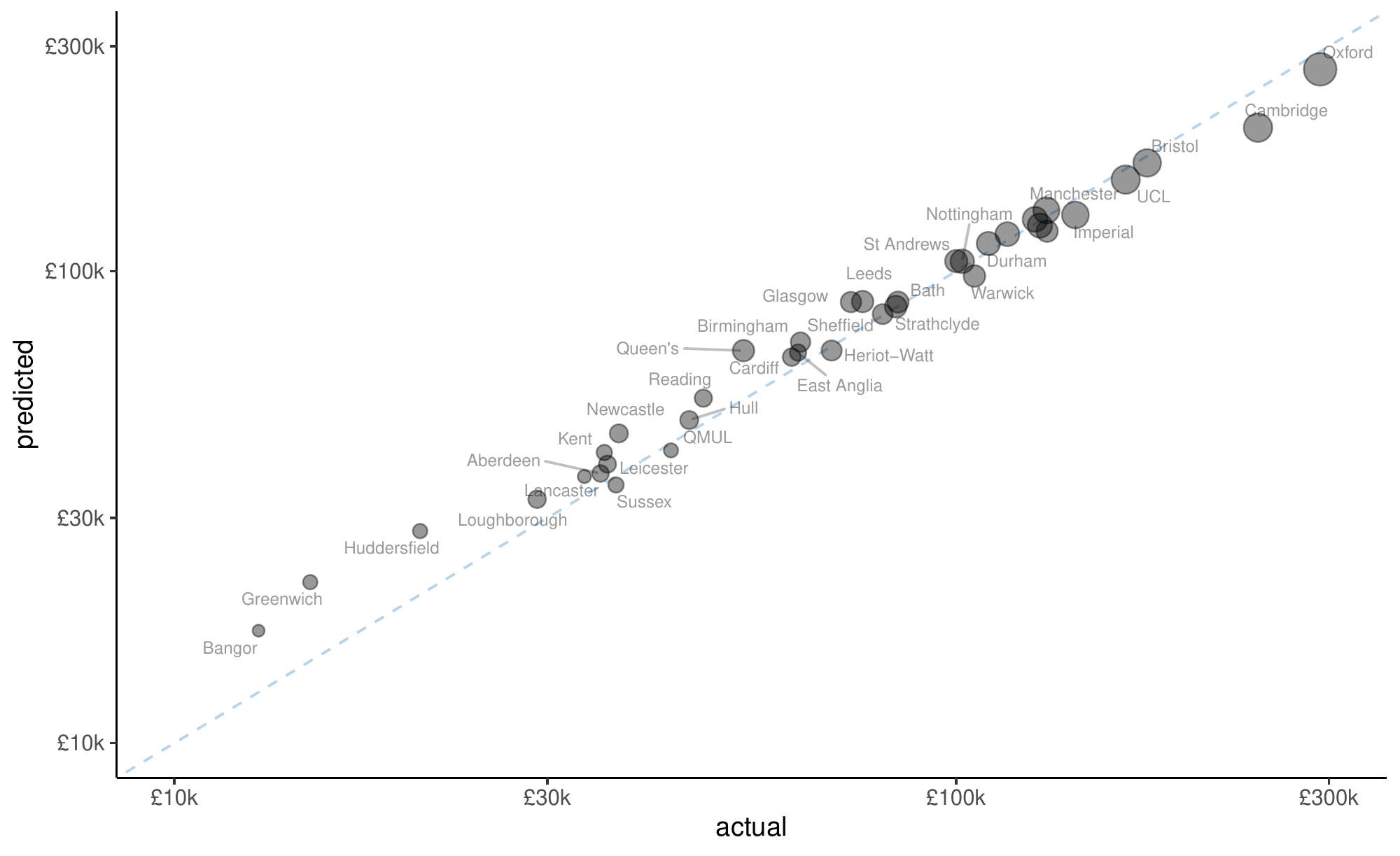} }\caption{Predictions versus observed REF2014 results for institutions submitting outputs to the Chemistry sub-panel, with point sizes proportional to number of FTE staff}\label{fig:chemPredictions}
\end{figure}

The expected versus actual institutional results are plotted in Figure~\ref{fig:chemPredictions}.
The pattern around the line of \(y=x\) is similar to that in the other sub-panels: broadly a linear relationship, but with lower-scoring institutions having higher predicted than actual results, and the converse for stronger institutions.
There are no noticeable outliers.

By summary measures, the median index of similarity in Chemistry is \(8.2\%\) and the median proportion of research funding that would need to be reallocated would be \(5.6\%\).
The posterior distributions are plotted in Figure~\ref{fig:indicesRidges}.
Performance, according to these metrics, appears similar to for other fields.

\hypertarget{comparison-with-journal-impact-factors}{%
\subsection{Comparison with Journal Impact Factors}\label{comparison-with-journal-impact-factors}}

Using data from Clarivate Analytics' \emph{Journal Citation Reports}, we can compare the latent journal REF effects with journal impact factors for the respective year.
For this article, we use the 2014 edition of \emph{Journal Citation Reports}, as this is based on citation data from the preceding two years.
(One could also consider the 2013 edition, though the results should not be too different.)

It may also be possible to compare with rival metrics, such as the CiteScore and Scimago Journal Rank (SJR), Scopus's versions of the impact factor and the Eigenfactor, respectively, however we do not make those comparisons here.

\hypertarget{economics-econometrics-2}{%
\subsection*{Economics \& Econometrics}\label{economics-econometrics-2}}
\addcontentsline{toc}{subsection}{Economics \& Econometrics}

\begin{figure}
\subfloat[Journal impact factor (2014)\label{fig:impactfactor-1}]{\includegraphics[width=0.5\linewidth]{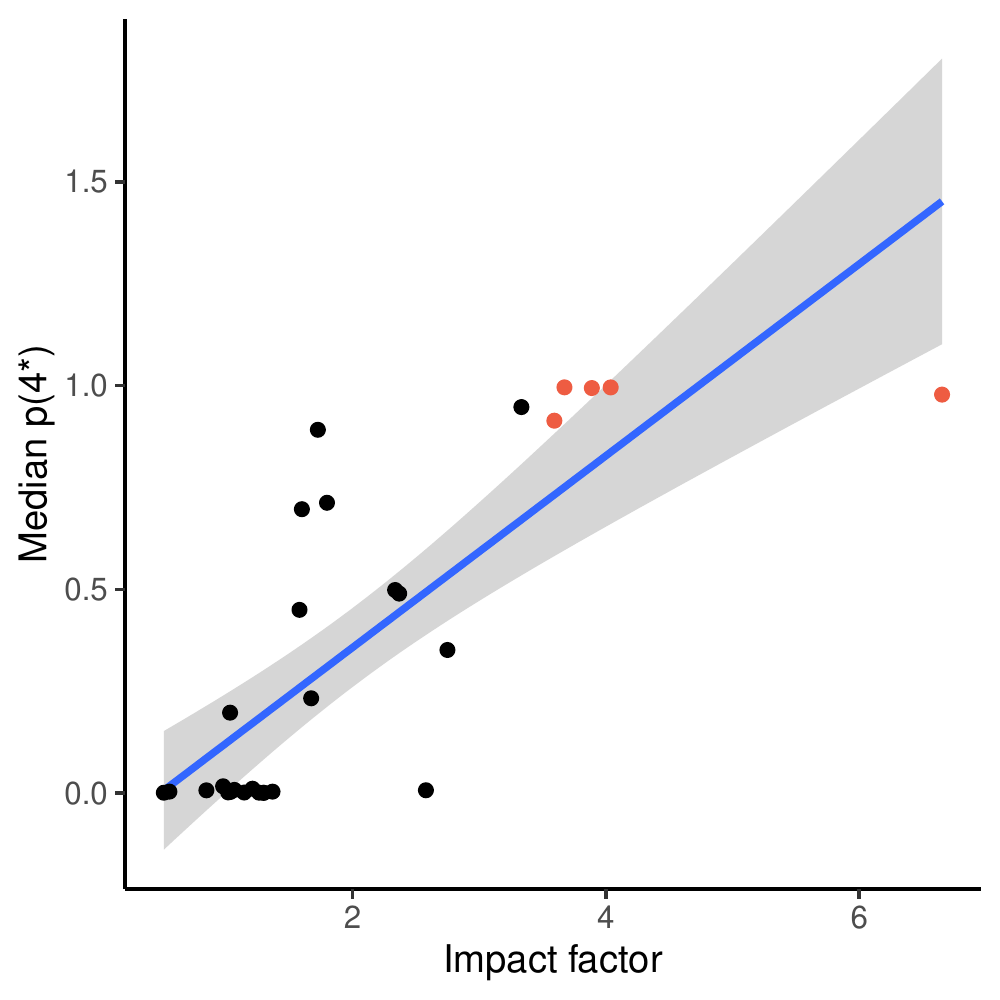} }\subfloat[Eigenfactor (2014)\label{fig:impactfactor-2}]{\includegraphics[width=0.5\linewidth]{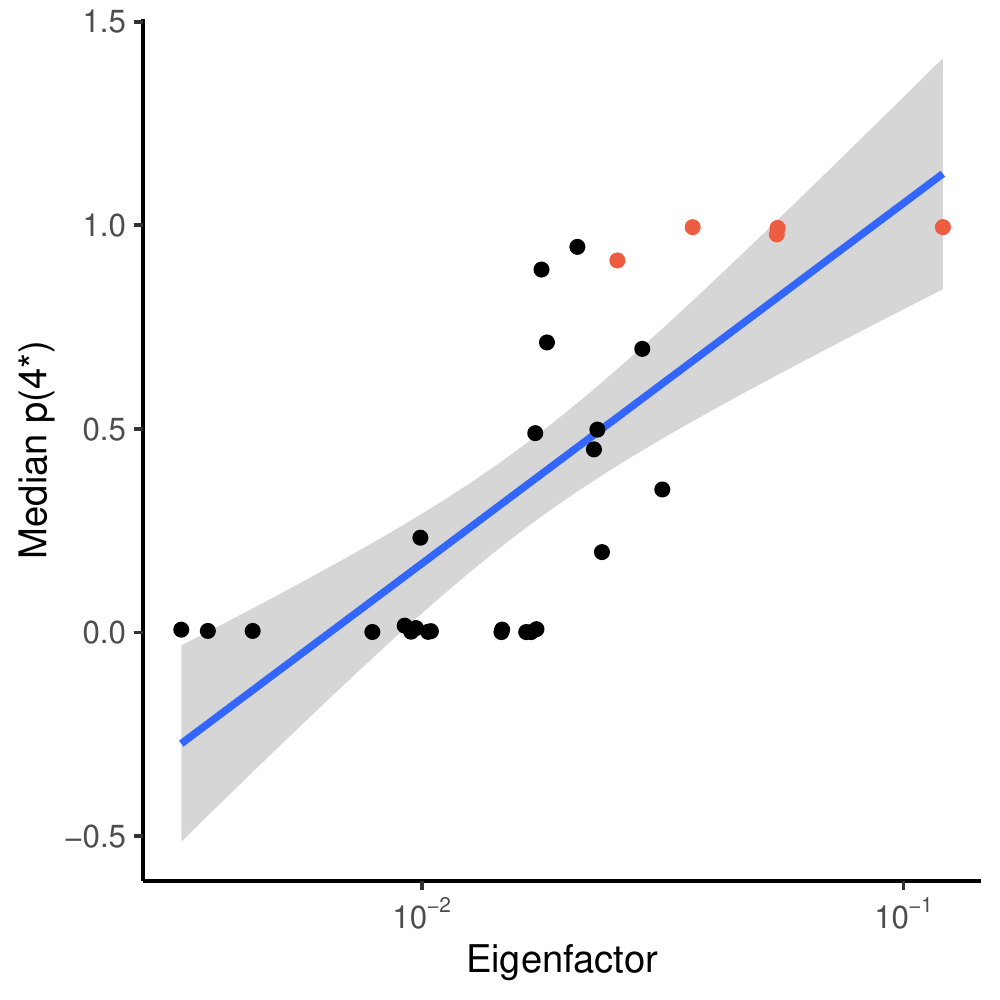} }\caption{Comparison of Economics and Econometrics journals' estimated probabilities of attaining 4* in the REF, versus Clarivate journal citation metrics, with line of best fit. So-called 'top 5' journals are highlighted in red}\label{fig:impactfactor}
\end{figure}

Comparisons are plotted in Figure~\ref{fig:impactfactor}.
Note the logarithmic scale for the Eigenfactor score.
Broadly speaking, there is a (weak) positive correlation between both citation metrics and the probability of attaining 4* in the REF.
Evidence for the supposed dominance of the `top 5' economics journals is mixed.
Whilst these periodicals are indeed highly ranked by journal impact factor, Eigenfactor and apparent REF effect, they do not completely dominate the top five spots, so their reputation must depend on other factors or perhaps be undeserved.
Moreover, as economists have explicitly known of the `top 5' designation for years, it may be a self-fulfilling prophecy.

\hypertarget{mathematical-sciences-2}{%
\subsection*{Mathematical Sciences}\label{mathematical-sciences-2}}
\addcontentsline{toc}{subsection}{Mathematical Sciences}

\begin{figure}
\subfloat[Journal impact factor (2014)\label{fig:mathsJIF-1}]{\includegraphics[width=0.5\linewidth]{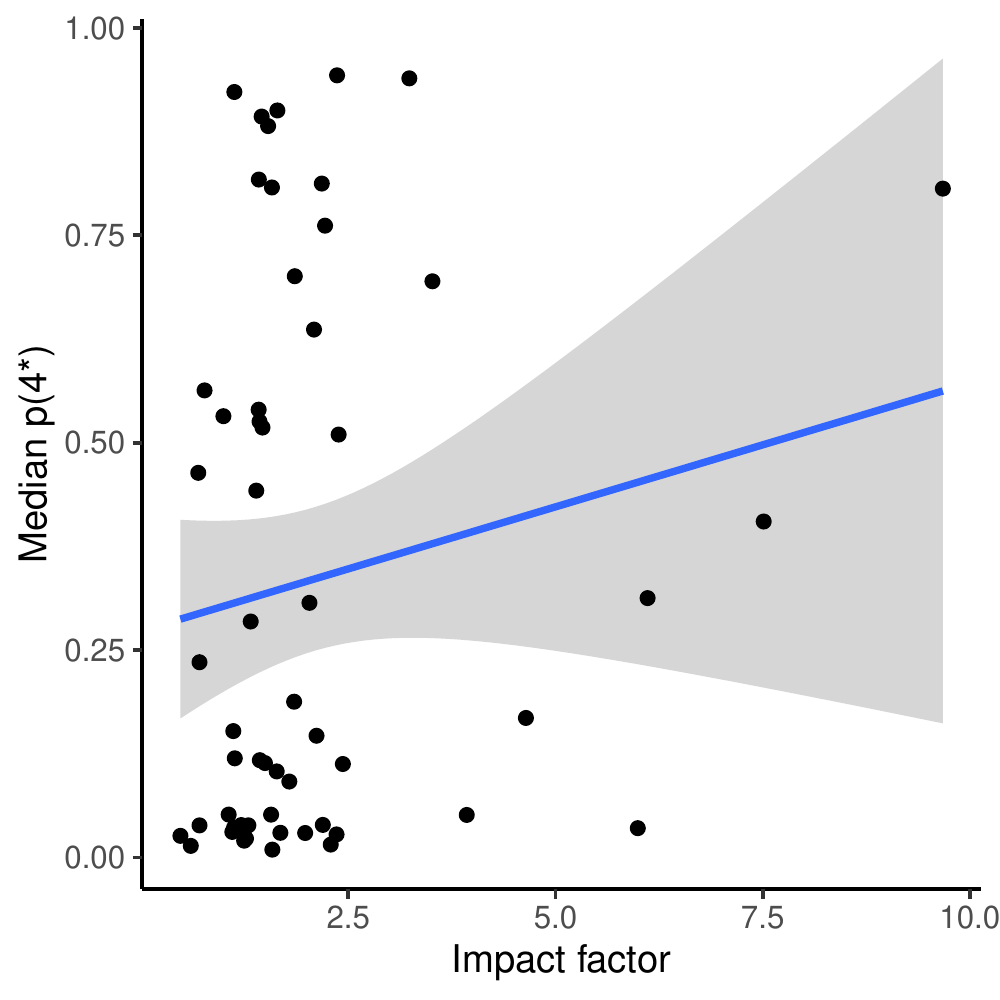} }\subfloat[Eigenfactor (2014)\label{fig:mathsJIF-2}]{\includegraphics[width=0.5\linewidth]{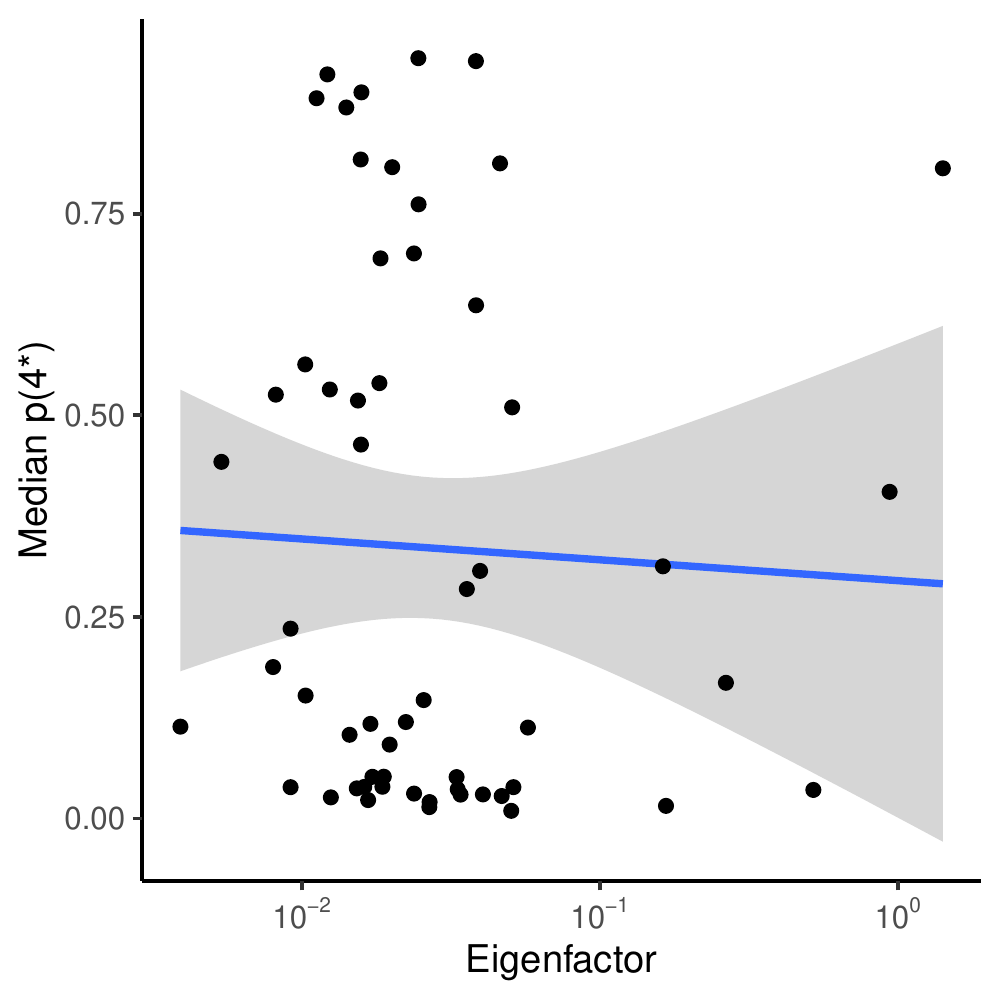} }\caption{Comparison of Mathematical Sciences journals' estimated probabilities of attaining 4* in the REF, versus Clarivate journal citation metrics, with line of best fit}\label{fig:mathsJIF}
\end{figure}

In Mathematical Sciences, however, there is almost no correlation between journal impact factor and the estimated probability of 4* in the REF; see Figure~\ref{fig:mathsJIF}.
This phenomenon could partly be explained by mathematical journals generally receiving lower impact factors; mathematics papers tend to have short reference lists and take longer to be noticed, when compared with publications in microbiology and other applied disciplines, so the journal impact factor (roughly speaking, counting citations over two years) is an especially poor metric for mathematics work.
Most mathematics journals here had an impact factor of around 1 or 2, so most of the variation between those scores might be attributed to random noise---see the left hand side of Figure~\ref{fig:mathsJIF}a.

\hypertarget{physics-2}{%
\subsection*{Physics}\label{physics-2}}
\addcontentsline{toc}{subsection}{Physics}

\begin{figure}
\subfloat[Journal impact factor (2014)\label{fig:physicsJIF-1}]{\includegraphics[width=0.5\linewidth]{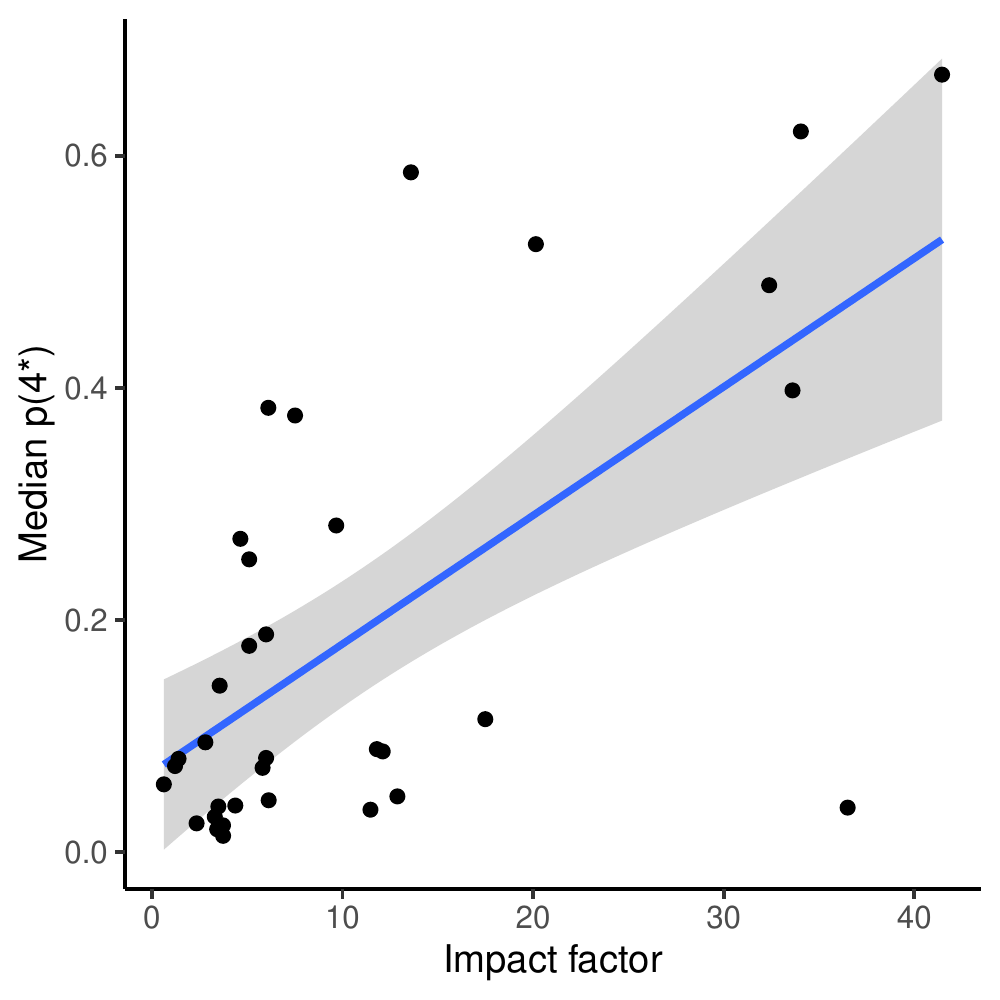} }\subfloat[Eigenfactor (2014)\label{fig:physicsJIF-2}]{\includegraphics[width=0.5\linewidth]{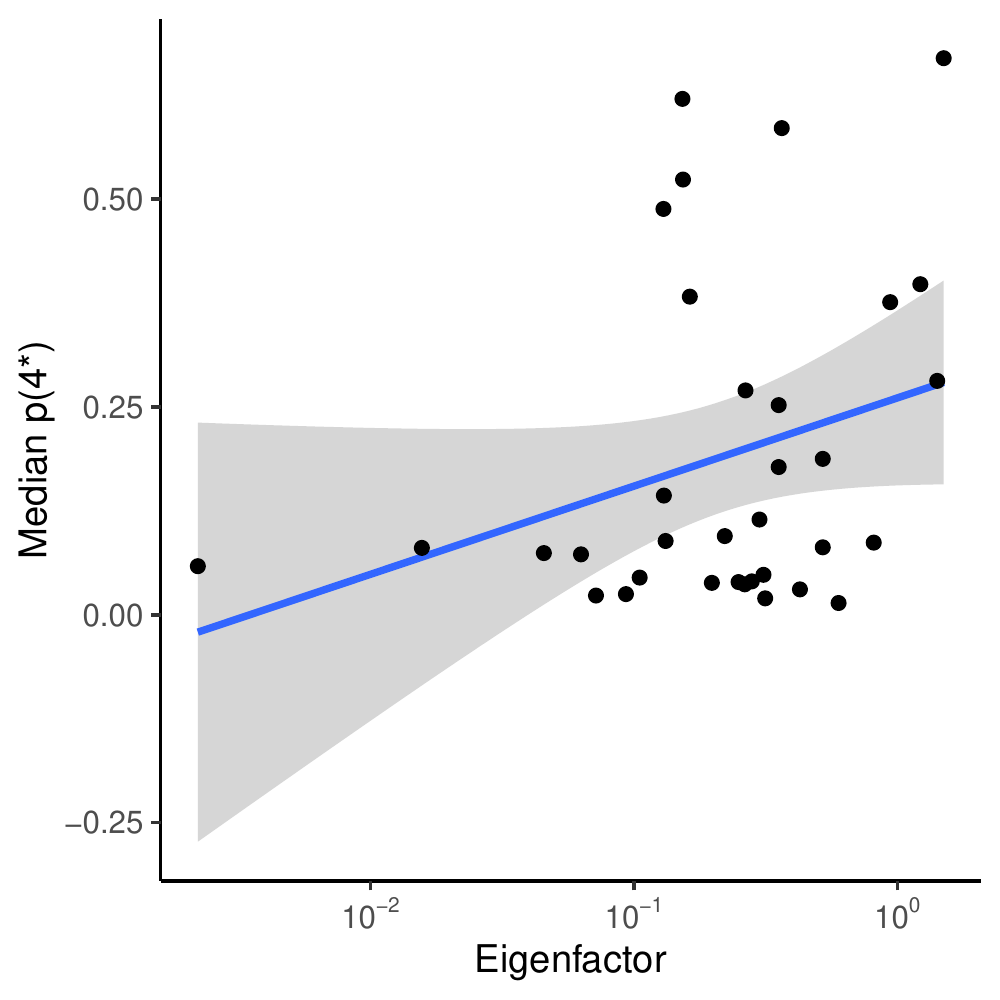} }\caption{Comparison of Physics journals' estimated probabilities of attaining 4* in the REF, versus Clarivate journal citation metrics, with line of best fit}\label{fig:physicsJIF}
\end{figure}

A positive correlation is present between Clarivate citation metrics and estimated probability of 4* for Physics as illustrated in Figure~\ref{fig:physicsJIF}.

\hypertarget{chemistry-2}{%
\subsection*{Chemistry}\label{chemistry-2}}
\addcontentsline{toc}{subsection}{Chemistry}

\begin{figure}
\subfloat[Journal impact factor (2014)\label{fig:chemistryJIF-1}]{\includegraphics[width=0.5\linewidth]{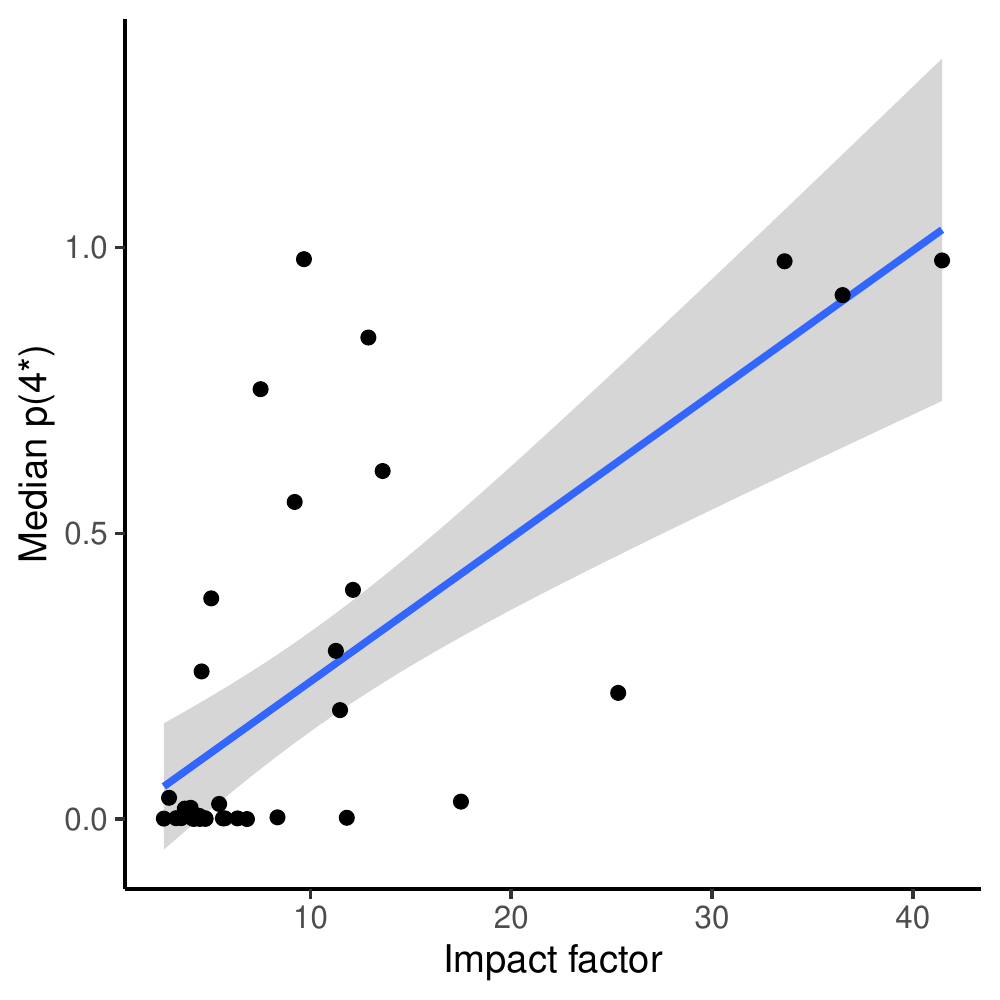} }\subfloat[Eigenfactor (2014)\label{fig:chemistryJIF-2}]{\includegraphics[width=0.5\linewidth]{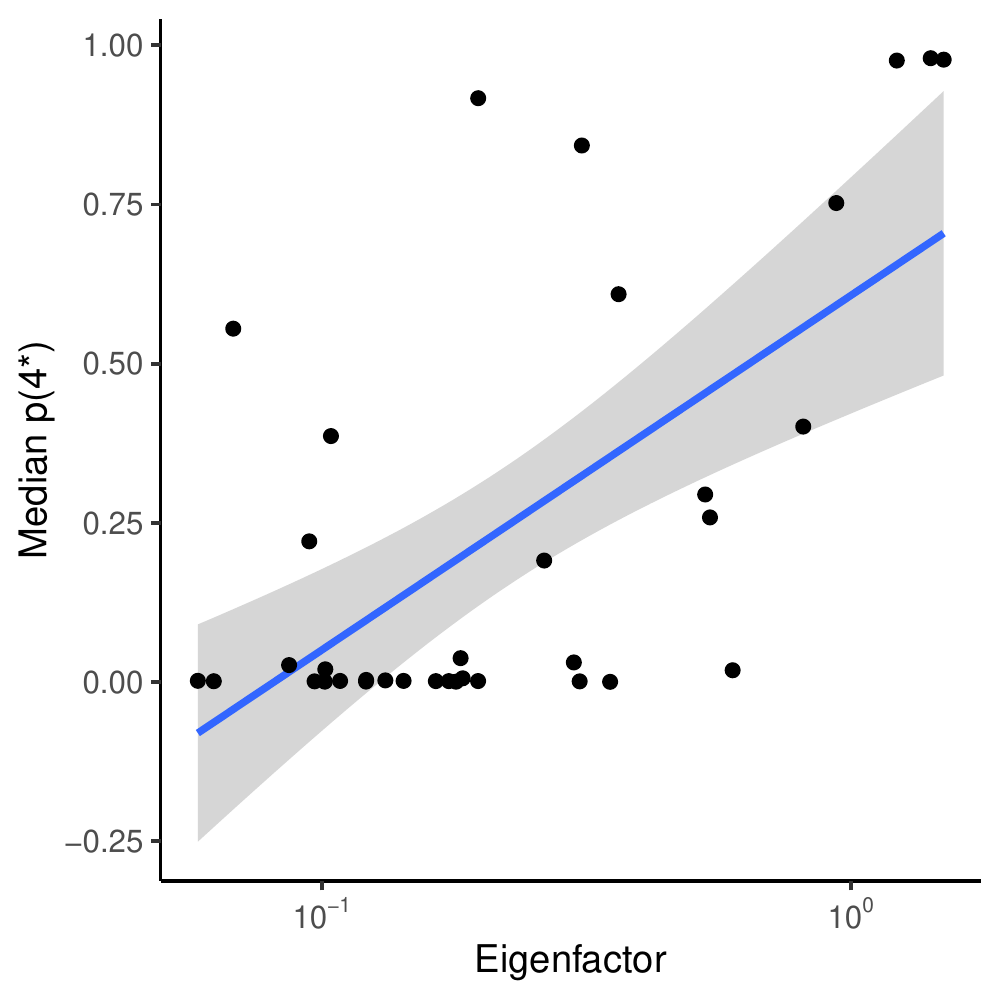} }\caption{Comparison of Chemistry journals' estimated probabilities of attaining 4* in the REF, versus Clarivate journal citation metrics, with line of best fit}\label{fig:chemistryJIF}
\end{figure}

The story for Chemistry is hard to interpret because so many journals have median estimated 4* probabilities close to zero.
But the top three journals by impact factor were also estimated to have the highest chances of their articles attaining 4* in the REF.
See Figure~\ref{fig:chemistryJIF}.

\hypertarget{discussion}{%
\section{Discussion}\label{discussion}}

This paper has explored the relationship between published REF2014 results and the journals in which institutions published research outputs submitted for REF2014 assessment.
The results are informative in various ways, including:

\begin{itemize}
\item
  implied rankings of the main journals from which work was submitted in each REF2014 sub-panel, together with measures of uncertainty on such rankings; and
\item
  for each REF2014 sub-panel studied, measurement of the \emph{maximum} extent to which REF2014 outcomes can be explained (retrospectively) by the identities of the journals from which work was submitted by each institution.
\end{itemize}

One reassuring aspect of the journal rankings derived for the four disciplines studied here is that they broadly agree with the informed opinions (informally elicited) of senior Warwick academics in those disciplines.
That is to say, for the main `named' journals in each field, the estimates and uncertainty intervals for the journals' probabilities of attaining 4* ratings in REF2014 made sense in the minds of the experts who were consulted.
Had the opposite been found, it would have been a strong reason to distrust the statistical methodology used here.

Our analysis of four disciplines found that in each of them there is---as expected---a strong or very strong relationship between the composition of journals seen in an institution's REF2014 submission and its published REF2014 Outputs profile results.

Naïvely, one might infer that the REF could therefore be replaced---at least for some disciplines---by a more automated `algorithmic' assessment that assigns quality ratings based on the journal in which each piece of research is published, rather than on an expert panel's reading of the work itself.
However, such an interpretation would not be justified.
As well as the potential for such an algorithmic approach to produce undesirable changes in behaviour, it is important to emphasise two aspects of our analysis.
Specifically:

\begin{enumerate}
\def\labelenumi{\arabic{enumi}.}
\item
  The analysis performed here is \emph{retrospective}, not predictive. The question asked, in each discipline, was effectively: if we imagine that the REF2014 panel based its assessments on journal identities alone, then what set of `journal quality' scores would yield the best match with the actual published REF2014 results?
  How good would such a `best match' be?
  The implied `journal quality' scores in our analysis came directly from the REF2014 \emph{results}; they were not known in advance by the REF panel, nor were they based on any explanatory covariates other than the journal identities themselves.
\item
  Although strong correlation was found between REF outcomes and aggregated `journal quality' scores (see Figures~\ref{fig:econPredictions}, \ref{fig:mathPredictions}, \ref{fig:physPredictions} and \ref{fig:chemPredictions}), there is a clear \emph{pattern} of deviation from that relationship, for each of the disciplines studied here.
  The `top' institutions are seen typically to do better in REF2014 than their aggregated `journal quality' scores would suggest; and conversely institutions at the other end of the scale tend to do worse, relative to purely journal-based scores.
  This indicates that REF assessment panels are in fact doing more than simply using journal identity to determine research quality.
  This finding is unsurprising: the published remit of REF panels is to \emph{read} the submitted research and evaluate its quality against clearly stated criteria.
  With that in mind, it is fully to be expected that a diligent REF panel will distinguish the `best' papers in each journal from those papers that are more ordinary.
\end{enumerate}

It could perhaps still be argued that the relationship between journal-based scores and REF outcomes is sufficiently strong that deviations from it could be ignored, in the interest of reducing the overall cost of the REF exercise.
But the clear \emph{pattern} of deviation described in point 2 implies that the resulting redistribution (of research funding, but also prestige) would systematically disadvantage those institutions where predominantly top-quality research is done.
While such redistribution of funds might represent a fairly modest fraction of the national funding total, its effects would systematically be concentrated in a few institutions at opposite ends of the scale.

Furthermore, the notion of judging work based on the container in which it is published, rather than on its own merits, seems to miss the point of research assessment entirely.
As Traag and Waltman (2019) points out, by relying on metrics, even those which correlate strongly with peer review results, `the goal of fostering ``high quality'' science may become displaced by the goal of obtaining a high metric' and have unintended consequences such as `favouring problematic research methods'.

More pragmatically, there is nothing to say that the esteem of academic journals in 2014 will remain constant until 2021.
Editors and authors change and publications can go defunct or start anew in such a long period.
Mryglod et al.~(2015a, 2015b) already showed that one research assessment exercise cannot necessarily be used to predict the next.

Perhaps the most interesting avenue for future research would be to apply these methods to \emph{all} subject areas in the REF and determine which fields are most beholden to the effect of journals on institutional rankings.
Data for all 36 units of assessment in the REF are readily available, and it should be straightforward to apply the methods developed here to those other fields.
With the 2021 REF approaching, this could be a topic of interest to many in academia, publishing and research assessment.
As seen in Table~\ref{tab:outputs}, however, subjects in the hard sciences tend to submit to scholarly journals more than other fields, such as the arts, who may produce books or artefacts, so the methodology would need to be adapted carefully for such areas, if indeed it can be applied at all.

\hypertarget{references}{%
\section*{References}\label{references}}
\addcontentsline{toc}{section}{References}

\hypertarget{refs}{}
\leavevmode\hypertarget{ref-Baccini2016}{}%
Baccini, Alberto, and Giuseppe De Nicolao. 2016. ``Do They Agree? Bibliometric Evaluation Versus Informed Peer Review in the Italian Research Assessment Exercise.'' \emph{Scientometrics} 108 (3): 1651--71. \url{https://doi.org/10.1007/s11192-016-1929-y}.

\leavevmode\hypertarget{ref-Balbuena2018}{}%
Balbuena, Lloyd D. 2018. ``The UK Research Excellence Framework and the Matthew Effect: Insights from Machine Learning.'' Edited by Lutz Bornmann. \emph{PLOS ONE} 13 (11): e0207919. \url{https://doi.org/10.1371/journal.pone.0207919}.

\leavevmode\hypertarget{ref-Ball2019}{}%
Ball, Robin C. 2019. Personal communication.

\leavevmode\hypertarget{ref-Brown1986}{}%
Brown, Philip J., and Clive D. Payne. 1986. ``Aggregate Data, Ecological Regression, and Voting Transitions.'' \emph{Journal of the American Statistical Association} 81 (394): 452--60. \url{https://doi.org/10.1080/01621459.1986.10478290}.

\leavevmode\hypertarget{ref-Bugg2019}{}%
Bugg, Tim. 2019. Personal communication.

\leavevmode\hypertarget{ref-Carpenter2017}{}%
Carpenter, Bob, Andrew Gelman, Matthew D. Hoffman, Daniel Lee, Ben Goodrich, Michael Betancourt, Marcus Brubaker, Jiqiang Guo, Peter Li, and Allen Riddell. 2017. ``Stan: A Probabilistic Programming Language.'' \emph{Journal of Statistical Software} 76 (1). \url{https://doi.org/10.18637/jss.v076.i01}.

\leavevmode\hypertarget{ref-rcrossref}{}%
Chamberlain, Scott, Hao Zhu, Najko Jahn, Carl Boettiger, and Karthik Ram. 2019. \emph{rcrossref: Client for Various 'CrossRef' APIs}. \url{https://CRAN.R-project.org/package=rcrossref}.

\leavevmode\hypertarget{ref-Cho1998}{}%
Cho, Wendy K. Tam. 1998. ``Iff the Assumption Fits\ldots{}: A Comment on the King Ecological Inference Solution.'' \emph{Political Analysis} 7: 143--63. \url{https://doi.org/10.1093/pan/7.1.143}.

\leavevmode\hypertarget{ref-Colquhoun2014}{}%
Colquhoun, David, and Andrew Plested. 2014. ``Why You Should Ignore Altmetrics and Other Bibliometric Nightmares.'' \emph{DC's Improbable Science}. \url{http://www.dcscience.net/2014/01/16/why-you-should-ignore-altmetrics-and-other-bibliometric-nightmares/}.

\leavevmode\hypertarget{ref-Duane1987}{}%
Duane, Simon, A.D. Kennedy, Brian J. Pendleton, and Duncan Roweth. 1987. ``Hybrid Monte Carlo.'' \emph{Physics Letters B} 195 (2): 216--22. \url{https://doi.org/10.1016/0370-2693(87)91197-x}.

\leavevmode\hypertarget{ref-Duncan1953}{}%
Duncan, Otis Dudley, and Beverly Davis. 1953. ``An Alternative to Ecological Correlation.'' \emph{American Sociological Review} 18 (6): 665. \url{https://doi.org/10.2307/2088122}.

\leavevmode\hypertarget{ref-Duncan1955}{}%
Duncan, Otis Dudley, and Beverly Duncan. 1955. ``A Methodological Analysis of Segregation Indexes.'' \emph{American Sociological Review} 20 (2): 210. \url{https://doi.org/10.2307/2088328}.

\leavevmode\hypertarget{ref-Else2015}{}%
Else, Holly. 2015. ``Research Funding Formula Tweaked After REF 2014 Results.'' \emph{Times Higher Education}. \url{https://www.timeshighereducation.com/news/research-funding-formula-tweaked-after-ref-2014-results/2018685.article}.

\leavevmode\hypertarget{ref-Flaxman2015}{}%
Flaxman, Seth R., Yu-Xiang Wang, and Alexander J. Smola. 2015. ``Who Supported Obama in 2012?'' In \emph{Proceedings of the 21st ACM SIGKDD International Conference on Knowledge Discovery and Data Mining - KDD '15}. ACM Press. \url{https://doi.org/10.1145/2783258.2783300}.

\leavevmode\hypertarget{ref-Flaxman2016}{}%
Flaxman, Seth, Dougal Sutherland, Yu-Xiang Wang, and Yee Whye Teh. 2016. ``Understanding the 2016 US Presidential Election Using Ecological Inference and Distribution Regression with Census Microdata,'' November. \url{http://arxiv.org/abs/1611.03787v1}.

\leavevmode\hypertarget{ref-BiasedUrn}{}%
Fog, Agner. 2015. \emph{BiasedUrn: Biased Urn Model Distributions}. \url{https://CRAN.R-project.org/package=BiasedUrn}.

\leavevmode\hypertarget{ref-Franceschet2011a}{}%
Franceschet, Massimo, and Antonio Costantini. 2011. ``The First Italian Research Assessment Exercise: A Bibliometric Perspective.'' \emph{Journal of Informetrics} 5 (2): 275--91. \url{https://doi.org/10.1016/j.joi.2010.12.002}.

\leavevmode\hypertarget{ref-Freedman1999}{}%
Freedman, D. A., M. Ostland, M. R. Roberts, and S. P. Klein. 1999. ``Reply to G. King.'' \emph{Journal of the American Statistical Association} 94 (445): 355--257. \url{https://doi.org/10.1080/01621459.1999.10473849}.

\leavevmode\hypertarget{ref-Goodman1953}{}%
Goodman, Leo A. 1953. ``Ecological Regressions and Behavior of Individuals.'' \emph{American Sociological Review} 18 (6): 663. \url{https://doi.org/10.2307/2088121}.

\leavevmode\hypertarget{ref-Goodman1959}{}%
---------. 1959. ``Some Alternatives to Ecological Correlation.'' \emph{American Journal of Sociology} 64 (6): 610--25. \url{https://doi.org/10.1086/222597}.

\leavevmode\hypertarget{ref-Groen-Xu2017}{}%
Groen-Xu, Moqi, Pedro Teixeira, Thomas Voigt, and Bernhard Knapp. 2017. ``Short-Termism in Science: Evidence from the UK Research Excellence Framework.'' \emph{SSRN Electronic Journal}. \url{https://doi.org/10.2139/ssrn.3083692}.

\leavevmode\hypertarget{ref-Heckman2018}{}%
Heckman, James, and Sidharth Moktan. 2018. ``Publishing and Promotion in Economics: The Tyranny of the Top Five.'' Technical report. National Bureau of Economic Research; National Bureau of Economic Research. \url{https://doi.org/10.3386/w25093}.

\leavevmode\hypertarget{ref-Hill2017}{}%
Hill, Steven. 2017. ``A Paper Exploring REF, Funding and Journal Metrics.'' \url{http://stevenhill.org.uk/a-paper-exploring-REF-funding-and-journal-metrics/}.

\leavevmode\hypertarget{ref-Hirsch2005}{}%
Hirsch, J. E. 2005. ``An Index to Quantify an Individuals Scientific Research Output.'' \emph{Proceedings of the National Academy of Sciences} 102 (46): 16569--72. \url{https://doi.org/10.1073/pnas.0507655102}.

\leavevmode\hypertarget{ref-Hole2017}{}%
Hole, Arne Risa. 2017. ``Ranking Economics Journals Using Data from a National Research Evaluation Exercise.'' \emph{Oxford Bulletin of Economics and Statistics} 79 (5): 621--36. \url{https://doi.org/10.1111/obes.12185}.

\leavevmode\hypertarget{ref-Hong2013}{}%
Hong, Yili. 2013. ``On Computing the Distribution Function for the Poisson Binomial Distribution.'' \emph{Computational Statistics \& Data Analysis} 59 (March): 41--51. \url{https://doi.org/10.1016/j.csda.2012.10.006}.

\leavevmode\hypertarget{ref-Jannarone1990}{}%
Jannarone, Robert J., Kai F. Yu, and James E. Laughlin. 1990. ``Easy Bayes Estimation for Rasch-Type Models.'' \emph{Psychometrika} 55 (3): 449--60. \url{https://doi.org/10.1007/bf02294760}.

\leavevmode\hypertarget{ref-Jump2013}{}%
Jump, Paul. 2013. ``'Game' of One-Fifth? Part-Time Contracts Rise in Run-up to REF.'' \emph{Times Higher Education}, no. 2120 (September): 6. \url{https://www.timeshighereducation.com/news/twenty-per-cent-contracts-rise-in-run-up-to-ref/2007670.article}.

\leavevmode\hypertarget{ref-King1997}{}%
King, Gary. 2013. \emph{A Solution to the Ecological Inference Problem: Reconstructing Individual Behavior from Aggregate Data}. Princeton University Press.

\leavevmode\hypertarget{ref-King1999}{}%
King, Gary, Ori Rosen, and Martin A. Tanner. 1999. ``Binomial-Beta Hierarchical Models for Ecological Inference.'' \emph{Sociological Methods \& Research} 28 (1): 61--90. \url{https://doi.org/10.1177/0049124199028001004}.

\leavevmode\hypertarget{ref-Koya2017}{}%
Koya, Kushwanth, and Gobinda Chowdhury. 2017. ``Metric-Based Vs Peer-Reviewed Evaluation of a Research Output: Lesson Learnt from UK's National Research Assessment Exercise.'' Edited by Pablo Dorta-González. \emph{PLOS ONE} 12 (7): e0179722. \url{https://doi.org/10.1371/journal.pone.0179722}.

\leavevmode\hypertarget{ref-Kuha2011}{}%
Kuha, Jouni, and David Firth. 2011. ``On the Index of Dissimilarity for Lack of Fit in Loglinear and Log-Multiplicative Models.'' \emph{Computational Statistics \& Data Analysis} 55 (1): 375--88.

\leavevmode\hypertarget{ref-Loeffler2019}{}%
Loeffler, David. 2019. Personal communication.

\leavevmode\hypertarget{ref-MacRoberts2017}{}%
MacRoberts, Michael H., and Barbara R. MacRoberts. 2017. ``The Mismeasure of Science: Citation Analysis.'' \emph{Journal of the Association for Information Science and Technology} 69 (3): 474--82. \url{https://doi.org/10.1002/asi.23970}.

\leavevmode\hypertarget{ref-Mansfield2016}{}%
Mansfield, Neil J. 2016. ``Ranking of Design Journals Based on Results of the UK Research Excellence Framework: Using REF as Referee.'' \emph{The Design Journal} 19 (6): 903--19. \url{https://doi.org/10.1080/14606925.2016.1216212}.

\leavevmode\hypertarget{ref-Marques2017}{}%
Marques, Marcelo, Justin JW Powell, Mike Zapp, and Gert Biesta. 2017. ``How Does Research Evaluation Impact Educational Research? Exploring Intended and Unintended Consequences of Research Assessment in the United Kingdom, 1986--2014.'' \emph{European Educational Research Journal} 16 (6): 820--42. \url{https://doi.org/10.1177/1474904117730159}.

\leavevmode\hypertarget{ref-McCullagh1989}{}%
McCullagh, P., and John A. Nelder. 1989. \emph{Generalized Linear Models}. Second Edition. Taylor \& Francis Ltd.

\leavevmode\hypertarget{ref-xkcd2008}{}%
Monroe, Randall. 2008. ``xkcd: Purity.'' \url{https://xkcd.com/435/}. \url{https://xkcd.com/435/}.

\leavevmode\hypertarget{ref-Mryglod2015a}{}%
Mryglod, Olesya, Ralph Kenna, Yurij Holovatch, and Bertrand Berche. 2015a. ``Predicting Results of the Research Excellence Framework Using Departmental \(h\)-Index.'' \emph{Scientometrics} 102 (3): 2165--80. \url{https://doi.org/10.1007/s11192-014-1512-3}.

\leavevmode\hypertarget{ref-Mryglod2015b}{}%
---------. 2015b. ``Predicting Results of the Research Excellence Framework Using Departmental H-Index: Revisited.'' \emph{Scientometrics} 104 (3): 1013--7. \url{https://doi.org/10.1007/s11192-015-1567-9}.

\leavevmode\hypertarget{ref-Oswald2007}{}%
Oswald, Andrew J. 2007. ``An Examination of the Reliability of Prestigious Scholarly Journals: Evidence and Implications for Decision-Makers.'' \emph{Economica} 74 (293): 21--31. \url{https://doi.org/10.1111/j.1468-0335.2006.00575.x}.

\leavevmode\hypertarget{ref-Oswald2019}{}%
---------. 2019. Personal communication.

\leavevmode\hypertarget{ref-Poisson1837}{}%
Poisson, S. D. 1837. \emph{Recherches Sur La Probabilité Des Jugements En Matière Criminelle et En Matière Civile}. Paris: Bachelier.

\leavevmode\hypertarget{ref-Reed2017}{}%
Reed, Mark, and Simon Kerridge. 2017. ``How Much Was an Impact Case Study Worth in the UK Research Excellence Framework?'' \emph{Fast Track Impact} 1 (January): 47--49. \url{https://www.fasttrackimpact.com/single-post/2017/02/01/How-much-was-an-impact-case-study-worth-in-the-UK-Research-Excellence-Framework}.

\leavevmode\hypertarget{ref-REF2014}{}%
REF. 2015. ``REF Frequently Asked Questions.'' \url{https://www.ref.ac.uk/2014/faq/}.

\leavevmode\hypertarget{ref-REF2019}{}%
---------. 2019. ``What Is the REF?'' \url{https://www.ref.ac.uk/about/what-is-the-ref/}. \url{https://www.ref.ac.uk/about/what-is-the-ref/}.

\leavevmode\hypertarget{ref-Rosen2001}{}%
Rosen, Ori, Wenxin Jiang, Gary King, and Martin A. Tanner. 2001. ``Bayesian and Frequentist Inference for Ecological Inference: The R×C Case.'' \emph{Statistica Neerlandica} 55 (2): 134--56. \url{https://doi.org/10.1111/1467-9574.00162}.

\leavevmode\hypertarget{ref-Rosenman2019}{}%
Rosenman, Evan. 2019. ``Some New Results for Poisson Binomial Models,'' July. \url{https://arxiv.org/abs/1907.09053}.

\leavevmode\hypertarget{ref-Rosenman2018}{}%
Rosenman, Evan, and Nitin Viswanathan. 2018. ``Using Poisson Binomial GLMs to Reveal Voter Preferences,'' February. \url{https://arxiv.org/abs/1802.01053v1}.

\leavevmode\hypertarget{ref-Scott2019}{}%
Scott, Peter. 2019. Personal communication.

\leavevmode\hypertarget{ref-Sgroi2019}{}%
Sgroi, Daniel. 2019. Personal communication.

\leavevmode\hypertarget{ref-Shah1973}{}%
Shah, B. K. 1973. ``On the Distribution of the Sum of Independent Integer Valued Random Variables.'' \emph{The American Statistician} 27 (3): 123--24. \url{https://www.jstor.org/stable/2683639}.

\leavevmode\hypertarget{ref-Stan2018}{}%
Stan Development Team. 2018. ``RStan: The R Interface to Stan.'' \url{http://mc-stan.org/}.

\leavevmode\hypertarget{ref-Stockhammer2017}{}%
Stockhammer, Engelbert, Quirin Dammerer, and Sukriti Kapur. 2017. ``The Research Excellence Framework 2014, Journal Ratings and the Marginalization of Heterodox Economics.'' Research report 1715. Post-Keynesian Economics Society. \url{http://postkeynesian.net/working-papers/1715/}.

\leavevmode\hypertarget{ref-Szabo2016}{}%
Szabó, Zoltan, Bharath Sriperumbudur, Barnabás Póczos, and Arthur Gretton. 2014. ``Learning Theory for Distribution Regression.'' \emph{Journal of Machine Learning Research} 17 (152): 1--40. \url{http://arxiv.org/abs/http://arxiv.org/abs/1411.2066v4}.

\leavevmode\hypertarget{ref-Traag2019}{}%
Traag, V. A., and L. Waltman. 2019. ``Systematic Analysis of Agreement Between Metrics and Peer Review in the UK REF.'' \emph{Palgrave Communications} 5 (1). \url{https://doi.org/10.1057/s41599-019-0233-x}.

\leavevmode\hypertarget{ref-Varin2016}{}%
Varin, Cristiano, Manuela Cattelan, and David Firth. 2016. ``Statistical Modelling of Citation Exchange Between Statistics Journals.'' \emph{Journal of the Royal Statistical Society: Series A (Statistics in Society)} 179 (1): 1--63.

\leavevmode\hypertarget{ref-Wakefield2005}{}%
Wakefield, Jon. 2005. ``Ecological Inference for 2×2 Tables (with Discussion).'' \emph{Journal of the Royal Statistical Society: Series A (Statistics in Society)} 167 (3): 385--445. \url{https://doi.org/10.1111/j.1467-985x.2004.02046.x}.

\leavevmode\hypertarget{ref-Wang1993}{}%
Wang, Yuan H. 1993. ``On the Number of Successes in Independent Trials.'' \emph{Statistica Sinica} 3 (2): 295--312.

\leavevmode\hypertarget{ref-Wilsdon2015}{}%
Wilsdon, James, Liz Allen, Eleonora Belfiore, Philip Campbell, Stephen Curry, Steven Hill, Richard Jones, et al. 2015. ``The Metric Tide: Report of the Independent Review of the Role of Metrics in Research Assessment and Management.'' \url{https://responsiblemetrics.org/the-metric-tide/}; Unpublished. \url{https://doi.org/10.13140/rg.2.1.4929.1363}.

\leavevmode\hypertarget{ref-Yan2017}{}%
Yan, Zizhong. 2017. ``How Does the REF Panel Perceive Journals? A New Approach to Estimating Ordinal Response Model with Censored Outcomes.'' PhD thesis, Department of Economics, University of Warwick. \url{http://wrap.warwick.ac.uk/91064}.

\hypertarget{appendix-appendix}{%
\appendix}

\hypertarget{appendix}{%
\section{Appendix}\label{appendix}}

\begin{table}

\caption{\label{tab:econJournals}Distribution of Economics and Econometrics REF2014 submissions by containing journal (named titles contained $\geq 20$ submissions)}
\centering
\begin{tabular}[t]{lrr}
\toprule
Volume title & Outputs & \%\\
\midrule
American Economic Review & 104 & 4.0\\
The Economic Journal & 103 & 4.0\\
Journal of Econometrics & 93 & 3.6\\
Journal of Economic Theory & 81 & 3.1\\
Games and Economic Behavior & 78 & 3.0\\
\addlinespace
Econometrica & 68 & 2.6\\
Journal of the European Economic Association & 65 & 2.5\\
Review of Economic Studies & 63 & 2.4\\
Economics Letters & 62 & 2.4\\
Review of Economics and Statistics & 58 & 2.2\\
\addlinespace
Journal of Public Economics & 57 & 2.2\\
European Economic Review & 51 & 2.0\\
Economic Theory & 48 & 1.8\\
Journal of Development Economics & 47 & 1.8\\
Journal of Economic Dynamics and Control & 44 & 1.7\\
\addlinespace
Journal of Economic Behavior \& Organization & 42 & 1.6\\
Journal of Monetary Economics & 42 & 1.6\\
Econometric Theory & 35 & 1.3\\
Journal of International Economics & 35 & 1.3\\
Journal of Health Economics & 33 & 1.3\\
\addlinespace
Journal of Money, Credit and Banking & 32 & 1.2\\
Quarterly Journal of Economics & 29 & 1.1\\
International Economic Review & 28 & 1.1\\
Oxford Bulletin of Economics and Statistics & 28 & 1.1\\
Canadian Journal of Economics & 25 & 1.0\\
\addlinespace
Journal of Applied Econometrics & 24 & 0.9\\
Oxford Economic Papers & 24 & 0.9\\
Journal of Banking \& Finance & 23 & 0.9\\
Journal of Political Economy & 22 & 0.8\\
Conference proceedings & 2 & 0.1\\
\addlinespace
Other journals & 944 & 36.3\\
Other outputs & 210 & 8.1\\
\bottomrule
\end{tabular}
\end{table}

\begin{table}

\caption{\label{tab:physicsJournals}Distribution of Physics REF2014 submissions by containing journal (named titles contained $\geq 30$ submissions)}
\centering
\begin{tabular}[t]{lrr}
\toprule
Volume title & Outputs & \%\\
\midrule
Physical Review Letters & 1227 & 19.0\\
Monthly Notices of the Royal Astronomical Society & 678 & 10.5\\
The Astrophysical Journal & 393 & 6.1\\
Physical Review D & 281 & 4.4\\
Physical Review B & 242 & 3.8\\
\addlinespace
Journal of High Energy Physics & 226 & 3.5\\
Nature & 207 & 3.2\\
Astronomy and Astrophysics & 196 & 3.0\\
Physics Letters B & 189 & 2.9\\
Science & 175 & 2.7\\
\addlinespace
Applied Physics Letters & 123 & 1.9\\
Nature Physics & 96 & 1.5\\
Physical Review A & 92 & 1.4\\
The European Physical Journal C & 87 & 1.3\\
Nature Communications & 85 & 1.3\\
\addlinespace
Journal of Geophysical Research & 81 & 1.3\\
Optics Express & 81 & 1.3\\
Proceedings of the National Academy of Sciences & 81 & 1.3\\
New Journal of Physics & 76 & 1.2\\
Nano Letters & 70 & 1.1\\
\addlinespace
Nature Materials & 65 & 1.0\\
Advanced Materials & 58 & 0.9\\
Physical Review C & 55 & 0.9\\
Journal of the American Chemical Society & 52 & 0.8\\
Journal of Cosmology and Astroparticle Physics & 46 & 0.7\\
\addlinespace
Nature Photonics & 46 & 0.7\\
Monthly Notices of the Royal Astronomical Society: Letters & 45 & 0.7\\
Nuclear Instruments and Methods in Physics Research & 45 & 0.7\\
Journal of Instrumentation & 41 & 0.6\\
Nature Nanotechnology & 37 & 0.6\\
\addlinespace
Advanced Functional Materials & 35 & 0.5\\
ACS Nano & 30 & 0.5\\
Journal of Physics: Condensed Matter & 30 & 0.5\\
The Astrophysical Journal Supplement Series & 30 & 0.5\\
Conference proceedings & 18 & 0.3\\
\addlinespace
Other journals & 1075 & 16.7\\
Other outputs & 52 & 0.8\\
\bottomrule
\end{tabular}
\end{table}

\clearpage

\begin{longtable}[t]{lrr}
\caption{\label{tab:mathsJournals}Distribution of Mathematical Sciences REF2014 submissions by containing journal (named titles contained $\geq 30$ submissions)}\\
\toprule
Volume title & Outputs & \%\\
\midrule
Journal of Fluid Mechanics & 254 & 3.6\\
Physical Review Letters & 209 & 3.0\\
Journal of High Energy Physics & 159 & 2.3\\
Communications in Mathematical Physics & 140 & 2.0\\
Proceedings of the Royal Society A & 126 & 1.8\\
\addlinespace
Advances in Mathematics & 116 & 1.7\\
Journal of Physics A & 112 & 1.6\\
Physical Review E & 110 & 1.6\\
Physical Review D & 107 & 1.5\\
Journal of Algebra & 83 & 1.2\\
\addlinespace
Proceedings of the London Mathematical Society & 70 & 1.0\\
The Annals of Probability & 70 & 1.0\\
Journal of Functional Analysis & 66 & 0.9\\
Nonlinearity & 66 & 0.9\\
Transactions of the American Mathematical Society & 66 & 0.9\\
\addlinespace
The Annals of Applied Probability & 61 & 0.9\\
Biometrika & 57 & 0.8\\
SIAM Journal on Numerical Analysis & 55 & 0.8\\
Journal of the London Mathematical Society & 54 & 0.8\\
International Mathematics Research Notices & 52 & 0.7\\
\addlinespace
Mathematische Annalen & 50 & 0.7\\
Archive for Rational Mechanics and Analysis & 49 & 0.7\\
JRSS Series B (Statistical Methodology) & 48 & 0.7\\
Crelles Journal & 47 & 0.7\\
JRSS Series C (Applied Statistics) & 47 & 0.7\\
\addlinespace
Physics of Fluids & 45 & 0.6\\
Journal of Mathematical Physics & 44 & 0.6\\
Annals of Mathematics & 42 & 0.6\\
SIAM Journal on Applied Mathematics & 42 & 0.6\\
Probability Theory and Related Fields & 41 & 0.6\\
\addlinespace
SIAM Journal on Scientific Computing & 41 & 0.6\\
Stochastic Processes and their Applications & 41 & 0.6\\
Journal of Differential Equations & 40 & 0.6\\
Proceedings of the National Academy of Sciences & 40 & 0.6\\
Bulletin of the London Mathematical Society & 39 & 0.6\\
\addlinespace
Duke Mathematical Journal & 38 & 0.5\\
Journal of Pure and Applied Algebra & 38 & 0.5\\
SIAM Journal on Mathematical Analysis & 38 & 0.5\\
Geometric and Functional Analysis & 37 & 0.5\\
Journal of Mathematical Biology & 37 & 0.5\\
\addlinespace
Journal of the American Statistical Association & 37 & 0.5\\
The Astrophysical Journal & 37 & 0.5\\
Bulletin of Mathematical Biology & 36 & 0.5\\
Inventiones mathematicae & 36 & 0.5\\
The Annals of Statistics & 35 & 0.5\\
\addlinespace
Mathematische Zeitschrift & 34 & 0.5\\
Physica D & 34 & 0.5\\
European Journal of Operational Research & 33 & 0.5\\
Nuclear Physics B & 33 & 0.5\\
Biometrics & 32 & 0.5\\
\addlinespace
Journal of Computational Physics & 32 & 0.5\\
Journal of Theoretical Biology & 32 & 0.5\\
Compositio Mathematica & 31 & 0.4\\
Journal of Statistical Physics & 31 & 0.4\\
Electronic Journal of Probability & 30 & 0.4\\
\addlinespace
Geometry \& Topology & 30 & 0.4\\
The Annals of Applied Statistics & 30 & 0.4\\
Conference proceedings & 17 & 0.2\\
Other journals & 3291 & 47.1\\
Other outputs & 246 & 3.5\\
\bottomrule
\end{longtable}

\clearpage

\begin{longtable}[t]{lrr}
\caption{\label{tab:chemistryJournals}Distribution of Chemistry REF2014 submissions by containing journal (named titles contained $\geq 30$ submissions)}\\
\toprule
Volume title & Outputs & \%\\
\midrule
Journal of the American Chemical Society & 690 & 14.7\\
Angewandte Chemie International Edition & 472 & 10.0\\
Chemical Communications & 258 & 5.5\\
Proceedings of the National Academy of Sciences & 142 & 3.0\\
Chemistry - A European Journal & 138 & 2.9\\
\addlinespace
Physical Chemistry Chemical Physics & 126 & 2.7\\
Nature Chemistry & 119 & 2.5\\
The Journal of Chemical Physics & 116 & 2.5\\
Physical Review Letters & 112 & 2.4\\
Chemical Science & 94 & 2.0\\
\addlinespace
Science & 90 & 1.9\\
The Journal of Physical Chemistry C & 80 & 1.7\\
Dalton Transactions & 76 & 1.6\\
Inorganic Chemistry & 71 & 1.5\\
The Journal of Organic Chemistry & 70 & 1.5\\
\addlinespace
Organic Letters & 68 & 1.4\\
Chemistry of Materials & 57 & 1.2\\
Organic \& Biomolecular Chemistry & 55 & 1.2\\
Advanced Materials & 54 & 1.1\\
Analytical Chemistry & 53 & 1.1\\
\addlinespace
Nature & 53 & 1.1\\
Langmuir & 51 & 1.1\\
Journal of Materials Chemistry & 50 & 1.1\\
ACS Nano & 49 & 1.0\\
The Journal of Physical Chemistry A & 49 & 1.0\\
\addlinespace
Nature Materials & 47 & 1.0\\
Atmospheric Chemistry and Physics & 44 & 0.9\\
Soft Matter & 43 & 0.9\\
The Journal of Physical Chemistry B & 43 & 0.9\\
Organometallics & 41 & 0.9\\
\addlinespace
Physical Review B & 41 & 0.9\\
Advanced Functional Materials & 38 & 0.8\\
Journal of Medicinal Chemistry & 37 & 0.8\\
Journal of Biological Chemistry & 34 & 0.7\\
Nano Letters & 32 & 0.7\\
\addlinespace
Nature Communications & 31 & 0.7\\
Conference proceedings & 2 & 0.0\\
Other journals & 1064 & 22.6\\
Other outputs & 8 & 0.2\\
\bottomrule
\end{longtable}

\begin{figure}
\subfloat[Economics and Econometrics\label{fig:alphaDensity-1}]{\includegraphics[width=.45\linewidth]{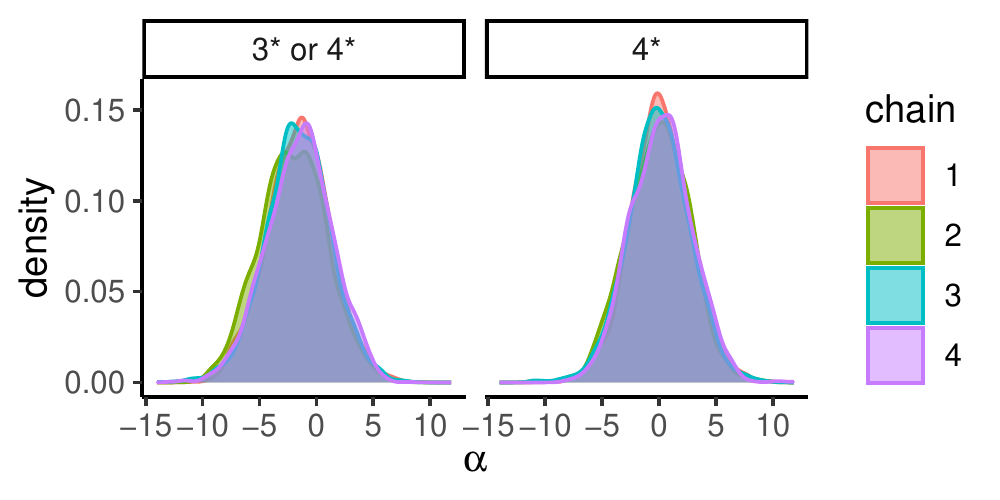} }\subfloat[Mathematical Sciences\label{fig:alphaDensity-2}]{\includegraphics[width=.45\linewidth]{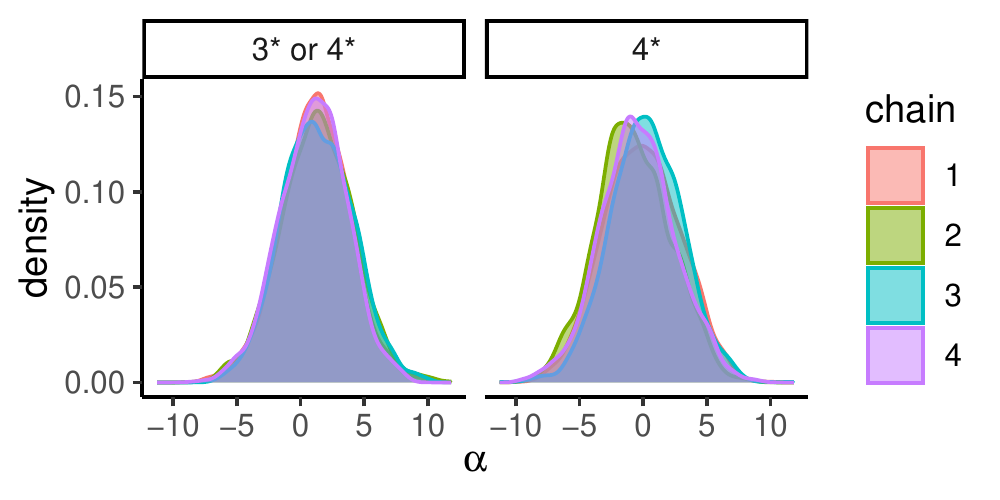} }\newline\subfloat[Physics\label{fig:alphaDensity-3}]{\includegraphics[width=.45\linewidth]{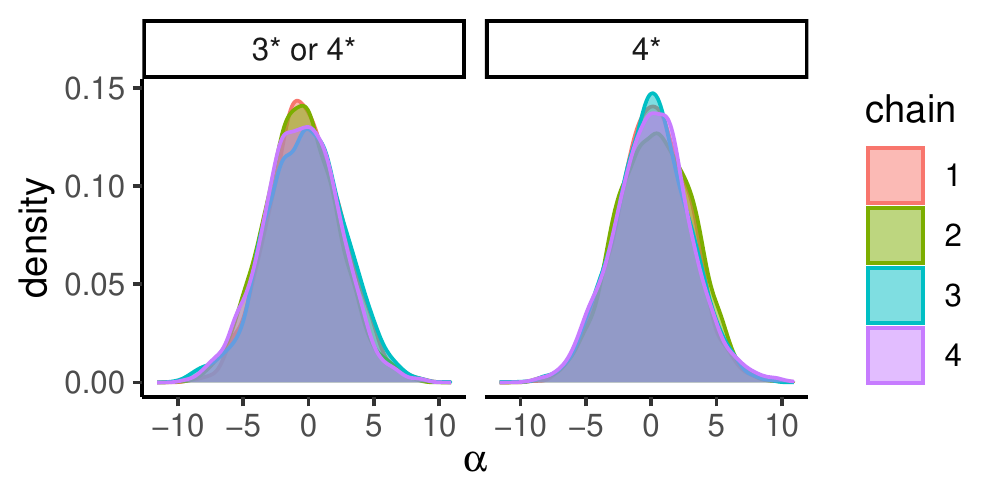} }\subfloat[Chemistry\label{fig:alphaDensity-4}]{\includegraphics[width=.45\linewidth]{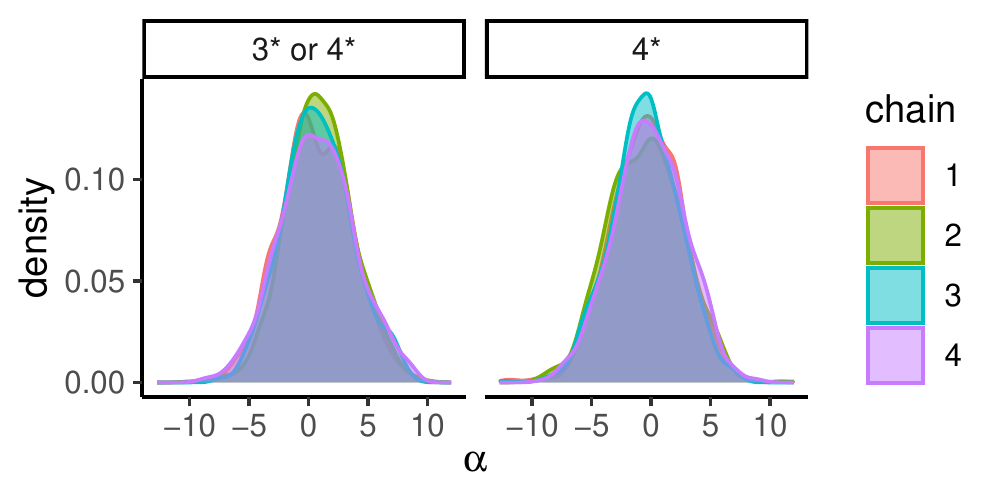} }\caption{Marginal density of \(\alpha\) hyper-parameter for four chains of Hamiltonian Monte Carlo, run on 4* and 3*+ profiles for each field. The prior for \(\alpha\) is a normal distribution with mean zero and standard deviation 3}\label{fig:alphaDensity}
\end{figure}

\begin{figure}
\subfloat[Economics and Econometrics\label{fig:traceplots-1}]{\includegraphics[width=.75\linewidth]{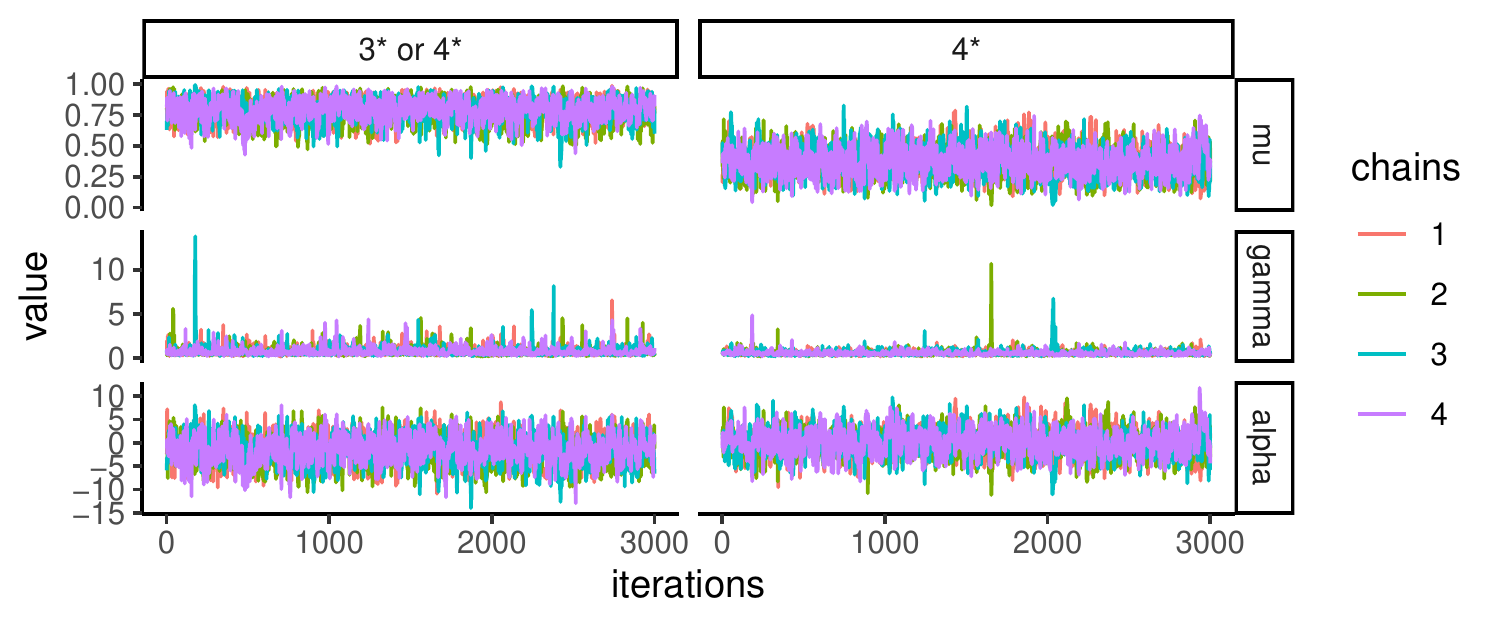} }\newline\subfloat[Mathematical Sciences\label{fig:traceplots-2}]{\includegraphics[width=.75\linewidth]{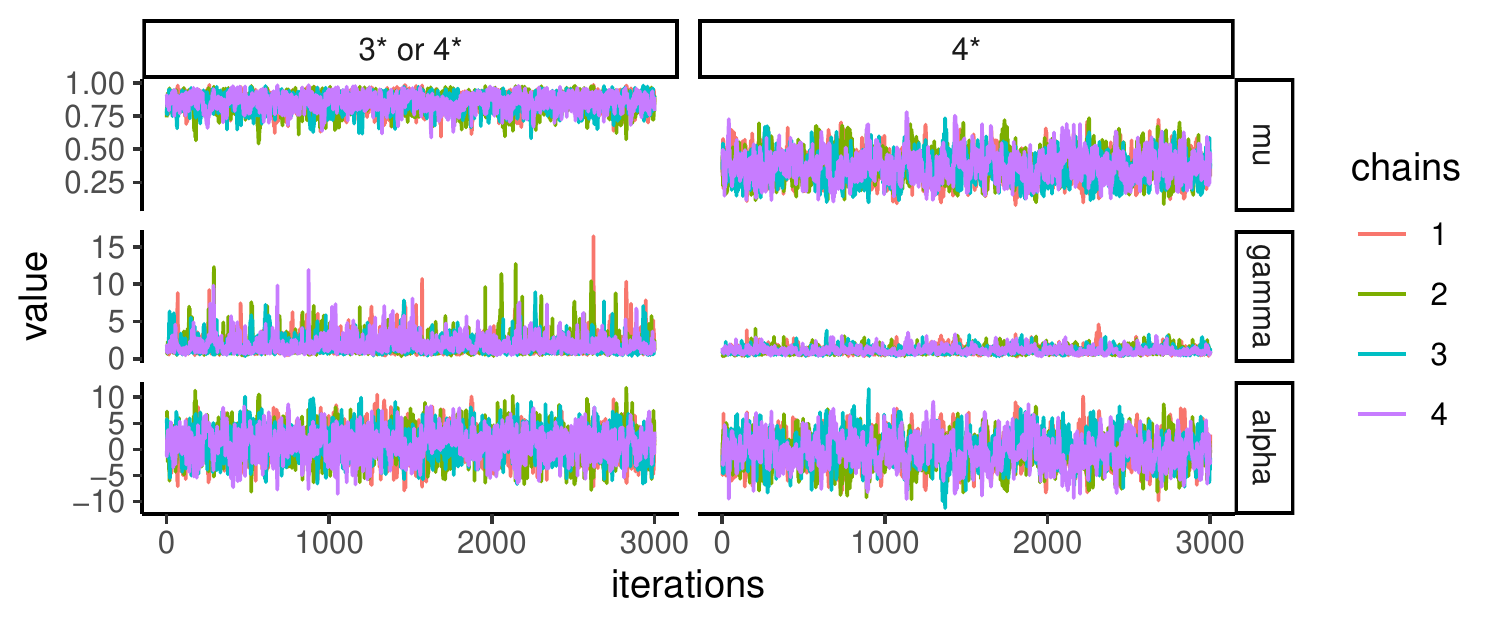} }\newline\subfloat[Physics\label{fig:traceplots-3}]{\includegraphics[width=.75\linewidth]{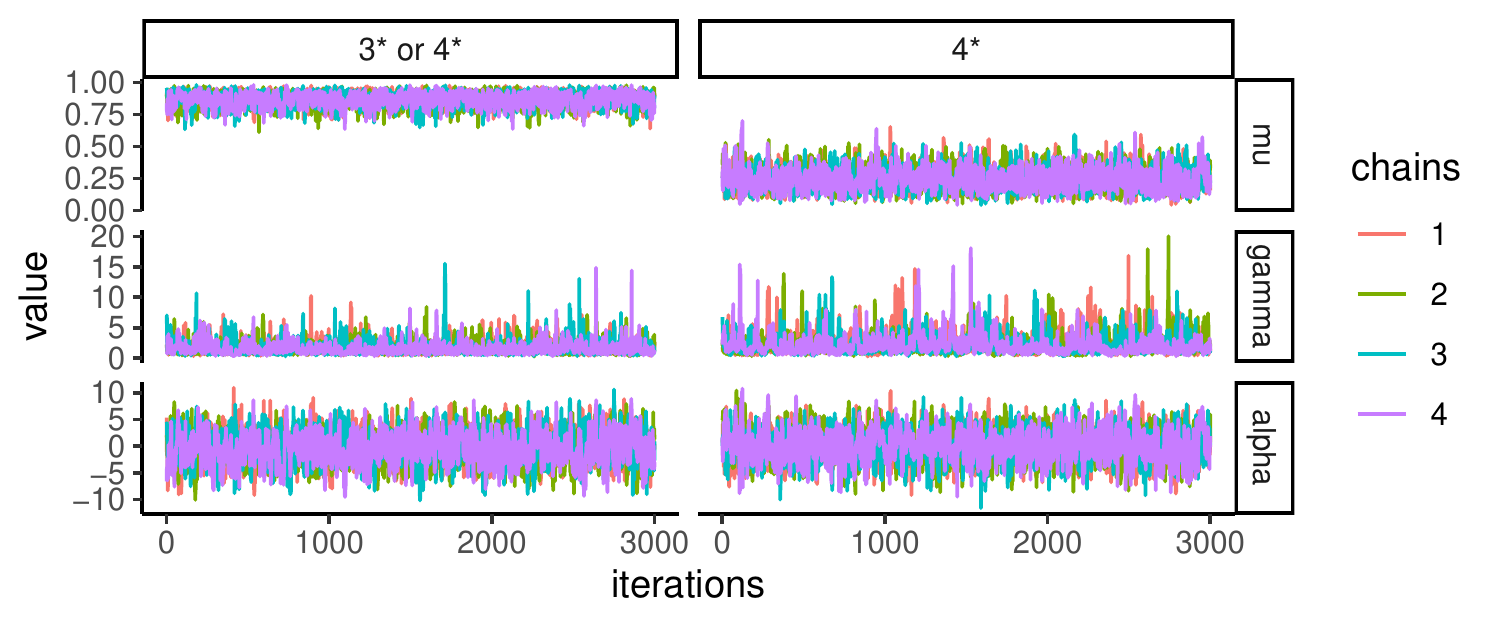} }\newline\subfloat[Chemistry\label{fig:traceplots-4}]{\includegraphics[width=.75\linewidth]{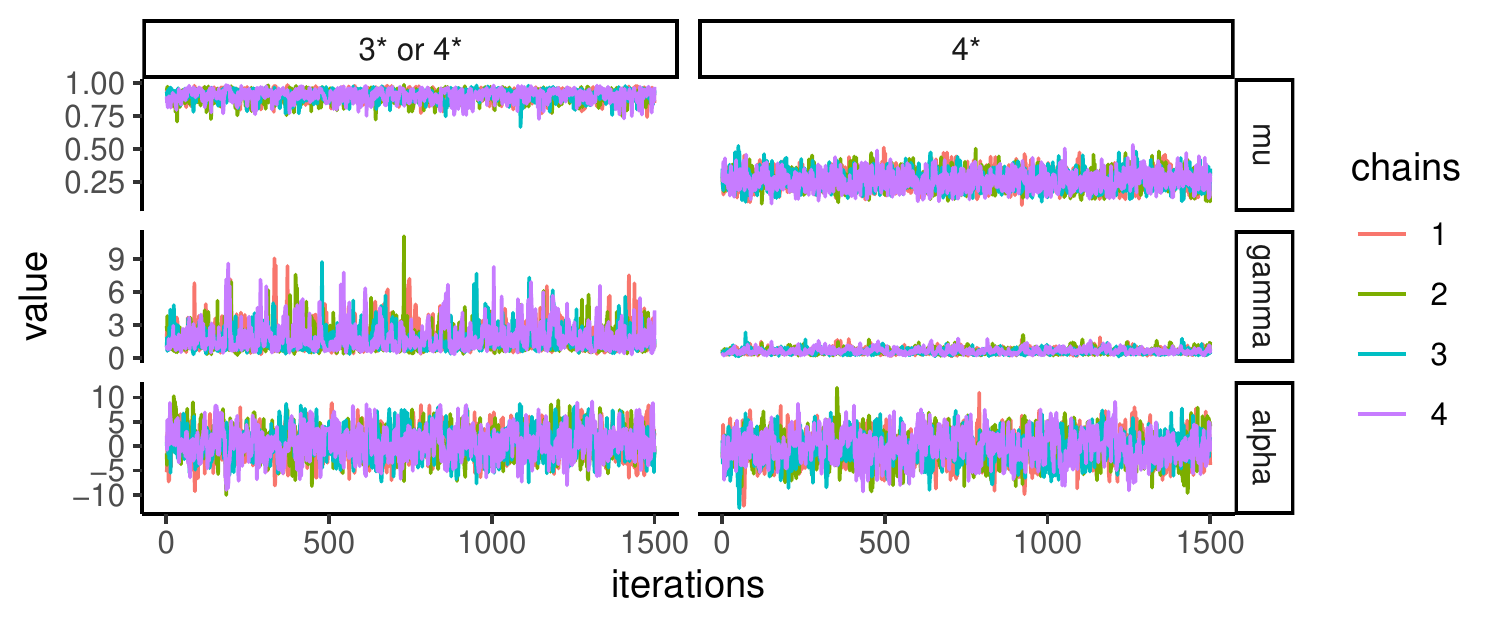} }\caption{Hamiltonian Monte Carlo trace plots for different parameters in the Poisson binomial model, run on 4* and 3*+ profiles for each field}\label{fig:traceplots}
\end{figure}

\end{document}